\documentclass[published]{JHEP3}
\accepted{July 31, 2002}
\keywords{Superstrings and Heterotic Strings, D-branes}
\received{June 7, 2002}
\JHEP{07(2002)071}

\skip\footins = 1\bigskipamount plus 2pt minus 4pt
\usepackage{epsfig}
\newcommand\OSp{\mathop{\rm OSp}} 

\newcommand{\be}{\begin{equation}}
\newcommand{\ee}{\end{equation}}
\newcommand{\pd}{\partial}
\newcommand{\nnbb}{\nonumber\\}
\newcommand{\tr}{\mathop{\rm tr}}
\newcommand{\un}{$\mathop{\rm {}U}(n)\;$}
\newcommand{\alp}{{\alpha^\prime}}
\newcommand{\sump}{\displaystyle{\sum_{\rm perm}\tr}}
\newcommand{\sumpp}{\displaystyle{\sum_{\rm perm^{\prime}}\tr}}

\newcommand\diag{\mathop{\rm diag}}

\title{The open superstring 5-point amplitude revisited}

\author{Ricardo Medina\\
Instituto de Ci{\^e}ncias, Universidade Federal de Itajub\'{a}\\
Itajub\'a, Minas Gerais, Brazil\\
E-mail: \email{rmedina@unifei.edu.br}.}

\author{Fernando T. Brandt and Fabiano R. Machado \\
Instituto de F\'{\i}sica Universidade de S\~ao Paulo\\
S\~ao Paulo, SP, Brazil\\
E-mail: \email{fbrandt@usp.br}, \email{fabiano@fma.if.usp.br}.}

\abstract{We derive the complete five-gluon scattering amplitude at
tree level, within the context of Open Superstring theory. We find the
general expression in terms of kinematic factors, and also find its
complete expansion up to ${\cal O}({\alpha'}^3)$ terms.  We use our
scattering amplitude to test three non-equivalent ${\cal
O}({\alpha'}^3)$ effective lagrangians that have recently been matter
of some controversy.}

\begin{document}

\section{Introduction}

\looseness=1 A nice feature of String Theory, among many things, is
the fact that it reproduces known non-massive field theories in its
low energy limit ($\alpha' \rightarrow 0$). In fact, the String Theory
corrections to these field theories may be found perturbatively in
$\alpha'$ by means of scattering amplitude arguments, as was done
in~\cite{Gross:1986iv}, where the leading corrections to the
Yang-Mills\pagebreak[3] and the Einstein-Hilbert lagrangians were
found. Moreover, in~\cite{Fradkin:1985qd, Tseytlin}, the infinite
$\alpha'$ series describing the interaction between open massless
strings (corresponding to photons) was found to be the Born-Infeld
theory~\cite{Born:1934gh}, as long as the field strength $F_{\mu \nu}$
is kept constant.

Up to now, there hasn't been found a non-abelian generalization of
this last result and the only achievements that have been done so far
are strictly perturbative in $\alpha'$.  Indeed, besides the leading
Yang-Mills term, the structure of the non-abelian Born-Infeld
lagrangian is completely known only up to ${\cal O}(\alpha'^2)$
terms~\cite{Gross:1986iv,Tseytlin}. Results involving higher order
corrections have only been partial in the sense that a non-abelian
prescription that worked for ${\cal O}(\alpha'^2)$
terms~\cite{Tseytlin2} did not work at higher orders
(see~\cite{Bilal:2001} and references therein) or that only some terms
of a higher order correction in the effective lagrangian have been
found~\cite{Bilal:2001}, but not all.

So a good starting point to consider would be the determination of the
complete ${\cal O}(\alpha'^3)$ terms in the non-abelian Born-Infeld
lagrangian.  This problem was studied a long time ago
in~\cite{Kitazawa:1987xj} by means of the partial computation of the
five-point amplitude in Open Superstring theory. Recent attempts have
avoided this direct calculation, succeeding in arriving to an
effective lagrangian~\cite{Koerber:2001uu, Refolli:2001df}.  But
controversy emerged since the ${\cal O}(\alpha'^3)$ terms of these
effective lagrangians turned up to be
non-equivalent~\cite{Koerber:2001uu,Koerber:2001hk}.

In this paper we reconsider the scattering amplitude approach to the
effective lagrangian by calculating the complete open superstring
five-point amplitude.

Our paper is organized as follows. In section~\ref{review} we give a
brief review of $N$-gluon scattering amplitudes in the context of Open
Superstring theory (at tree level). In section~\ref{review2} we
shortly give the known three and four-gluon amplitudes, as derived
directly from the general $N$-gluon formula, presented in
section~\ref{review}. In section~\ref{low} we consider the low energy
gluon lagrangian responsible for the previous three and four-gluon
amplitudes (up to ${\cal O}({\alpha'}^2)$ terms) and we introduce the
${\cal O}({\alpha'}^3)$ lagrangian terms of~\cite{Koerber:2001uu}. In
section~\ref{main} we develop the main result of this paper, namely,
the five-gluon tree amplitude as derived from Open Superstring
theory. In the final section, we make an analysis of the different
existing versions of the lagrangian terms at order ${\cal
O}({\alpha'}^3)$, in light of our five-point amplitude result. Our
main conclusion is that we find complete agreement with the one
in~\cite{Koerber:2001uu}.  Appendix~\ref{kinematic} contains the
kinematic factors which appear in the five-gluon tree
amplitude. Appendix~\ref{A0A2A3} contains the explicit expressions for
the second and third order contributions, in $\alpha'$, to this
amplitude. Appendices~\ref{appC} and~\ref{appD} constitute a
fundamental support to all the effective lagrangian expressions
appearing in this work since they contain the Feynman rules and the
scattering amplitudes derived from these lagrangians, which match with
all the corresponding scattering amplitudes derived from Open
Superstring theory. We have treated these very involved calculations
using the \emph{Maple} version of the computer algebra package
HIP~\cite{hsieh:1992ti}.

While this paper was being written, the authors came across a recent
preprint~\cite{Collinucci:2002ac}, in which the ${\cal
O}({\alpha'}^3)$ lagrangian terms of~\cite{Koerber:2001uu} were
confirmed by studying $\alpha'$ deformations of $D=10$ Super
Yang-Mills theory up to order three, imposing supersymmetry order by
order. Our work has been done with complete independence and without
any reference to this preprint.

\section{Review of gluon tree amplitudes in Open Superstring theory}
\label{review}

Upon quantization of the superstring it turns out that Lorentz
invariance is violated unless $D=10$ and non massive states ($M^2 =
0$) are part of the infinite
spectrum~\cite{Green:1987sp,Polchinski:1998rq,Polchinski:1998rr}.
These states describe a vector particle $A^{\mu}$, and a spinor
particle $\psi^{\mu}$. If Chan-Paton factors are associated to the
ends of the open superstring, then this allows for the introduction of
a \un gauge group. In this case the couple $(A^{\mu}_a,\psi^{\mu}_a)$
is the one describing $D=10$ Supersymmetric Yang-Mills theory. Along
this paper we will refer to the $A^{\mu}_a$ vector particle as a
gluon.  The tree level scattering amplitude of $N$ gluons with
polarizations $\zeta_1, \zeta_2, \ldots, \zeta_N$, external momenta
$k_1, k_2, \ldots, k_N$\footnote{The momenta $k_i$ are all assumed to
be \emph{inward} in the corresponding scattering process.}  and colors
$a_1, a_2, \ldots, a_N$, is calculated in Open Superstring theory
as~\cite{Green:1987sp,Polchinski:1998rq,Polchinski:1998rr}
\begin{equation} 
{\cal A}^{(N)} = i (2 \pi)^{10} \delta^{10} (k_1+ k_2 +\cdots +k_{N})
\cdot {\sumpp} (\lambda^{a_{1}} \lambda^{a_{2}} \ldots
\lambda^{a_{n}}) A(1,2, \ldots ,N) \,,
\label{general-amplitude}
\end{equation} 
where the sum $\sum_{\rm perm^{\prime}}$ is over \emph{all non-cyclic}
permutations of the sets $\{\zeta_1,k_1,a_1\}$,\linebreak
$\{\zeta_2,k_2,a_2\},\ldots, \{\zeta_N,k_N,a_N\}$, $\lambda^{a_{n}}$
are the \un generators satisfying the relations~(\ref{colmat}) and the
argument $1,2,\ldots,N$ of the Lorentz factor $A(1,2, \ldots ,N)$ is a
compact notation for $\zeta_1$, $k_1;\zeta_2$, $k_2;\ldots;\zeta_N$,
$k_N$. $A(1,2, \ldots ,N)$ corresponds to the $N$-particle scattering
amplitude of open superstrings which do not carry color indices and
which are placed among themselves in the specific ordering
$\{1,2,\ldots, N\}$ (modulo cyclic permutations).  For example, the
case of the ordering $\{1, 2,\ldots, N\}$ is shown in
figure~\ref{planar}. The first step of the calculations presented in
appendix~\ref{appD} is to express all the amplitudes in the form
given by eq.~(\ref{general-amplitude}).

\FIGURE[t]{\centerline{\epsfig{file=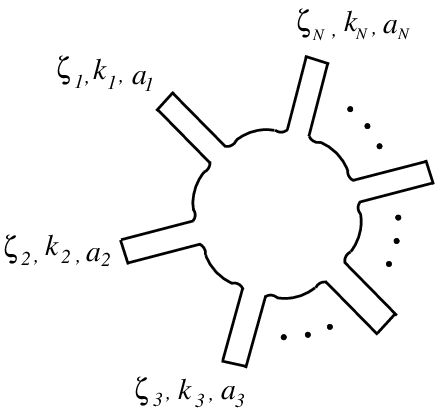, scale=1.1}\caption{The
planar open-string diagram.}\label{planar}}}

Using the vertex operator formalism, the following expression may be
derived for the corresponding amplitude~\cite{Green:1987sp}:
\begin{eqnarray}
A(1, 2, \ldots,N) & = & 2 \frac{g^{N-2}}{(2 \alpha')^{7N/4+2}}
 (x_{N-1}-x_1)(x_{N}-x_1)\times \nonumber \\ && {}\times \int dx_2
 \ldots dx_{N-2} \int d \theta_1 \ldots d \theta_{N-2} \prod_{i>j}^N |
 x_i - x_j - \theta_i \theta_j |^{2 \alpha' k_i \cdot k_j}\times
\nonumber \\& &
{} \times \int d \phi_1 \ldots d \phi_N e^{f_N(\zeta, k, \theta,
 \phi)}\,,
\label{N-amplitude}
\end{eqnarray}
where
\begin{equation} 
f_N(\zeta, k, \theta, \phi) = \sum_{i \neq j}^N
\frac{(\theta_i-\theta_j) \phi_i (\zeta_i \cdot k_j) (2
\alpha')^{11/4} -1/2 \phi_i \phi_j (\zeta_i \cdot \zeta_j) (2
\alpha')^{9/2}}{x_i-x_j-\theta_i \theta_j}\,.
\label{fN}
\end{equation} 
In~(\ref{N-amplitude}) and~(\ref{fN}) the $x_i$ are all bosonic
(commuting) variables, while the $\theta_i$, together with the
$\phi_i$ variables, are all fermionic (anticommuting) ones.  There are
some remarks about eqs.~(\ref{N-amplitude}) and~(\ref{fN}):
\begin{itemize}
\item Equation~(\ref{N-amplitude}) contains all consistent factors that
lead to the $N$-gluon tree amplitude in Yang-Mills theory, in the
limit $\alpha' \rightarrow 0$.\footnote{In eq.~(\ref{N-amplitude}) $g$
is the Yang-Mills coupling constant.}
\item The curious powers of $(2 \alpha')$ that appear in
eqs.~(\ref{N-amplitude}) and~(\ref{fN}) are just a consequence of a
dimensional analysis of the corresponding formula in~\cite[section
7.3]{Green:1987sp}, in which the convention $\alpha' = 1/2$ is
used. As will be seen later, the $\alpha'$ expansion of the amplitudes
contains only \emph{integer} powers of it.
\item The original amplitude $A(1, 2, 3, \ldots,N)$ is invariant under
$\OSp(1|2)$ transformations of all the $x_i$ and $\theta_i$ ($i=1,
\ldots, N$)~\cite{Green:1987sp}.  As a consequence of this symmetry,
the variables $x_1$, $x_{N-1}$, $x_N$, $\theta_{N-1}$ and $\theta_N$,
all behave like $free$ parameters (the final result does not depend on
them). Now, in eqs.~(\ref{N-amplitude}) and~(\ref{fN}), variables
$x_1$, $x_{N-1}$ and $x_N$ have been kept arbitrary, while
$\theta_{N-1}$ and $\theta_N$ have been set to $0$. 
\item To guarantee the specific ordering of the superstrings, $\{1, 2,
3, \ldots, N\}$, the condition
\begin{equation} 
x_1 < x_2 < x_3 <\ldots < x_N
\label{ordering}
\end{equation} 
should be imposed in eq.~(\ref{N-amplitude}).
\item Although not manifest, $A(1, 2, 3, \ldots, N)$ has the cyclic
property.\footnote{The formula in eq.~(\ref{N-amplitude}) comes
directly from the corresponding formula in~\cite[section
7.3]{Green:1987sp}, which does have the cyclic property.}
\item The external momenta $k_i$ and the polarizations $\zeta_i$
in~(\ref{N-amplitude}) satisfy
\begin{equation} \label{onshell}
\left. \begin{array}{lcl} i) \mbox{ Physical state condition:} & &
\zeta_1 \cdot k_1 = \zeta_2 \cdot k_2 = \ldots = \zeta_N \cdot k_N = 0
\\[3pt] ii) \mbox{ On-shell condition:} & & k_1^2 = k_2^2 = \ldots =
k_n^2 = 0
\end{array} \right\} 
\end{equation} 
\end{itemize}
Our convention for the Minkowski metric is the following:
\begin{equation}
\eta_{\mu \nu} = \diag(-, +, +, \ldots , +)\,.
\label{metric}
\end{equation}

\section{Review of three and four-gluon tree amplitudes}
\label{review2}

As an application of the formula in eq.~(\ref{N-amplitude}) we have
the well known three and four-point amplitudes. In the first case it becomes
\begin{eqnarray}
\lefteqn{A(1, 2, 3)  =  2 \frac{g}{(2 \alpha')^{29/4}}
  (x_2-x_1)(x_3-x_1)\times}\ &&
\nonumber \\ &  &\times
\int d \theta_1 | x_2 - x_1 - \theta_2 \theta_1 |^{2 \alpha' k_2\cdot
k_1}
| x_3 - x_1 - \theta_3 \theta_1 |^{2 \alpha' k_3\cdot k_1} |x_3 - x_2 -
 \theta_3 \theta_2 |^{2 \alpha' k_3\cdot k_2}\times 
\nonumber \\
 & &\times  \int d \phi_1 d \phi_2 d \phi_3 e^{f_3(\zeta, k, \theta,
\phi)}\,,
\label{A123}
\end{eqnarray}
\noindent 
In this case there is no $x$ integration ($x_1$, $x_2$ and $x_3$ are
arbitrary, with the only condition $x_1 < x_2 <x_3$) and we have
$\theta_2 = \theta_3 = 0$. Momentum conservation implies that $
k_2\cdot k_1 = k_3\cdot k_1 = k_3\cdot k_2 = 0$, so that $A(1, 2, 3)$
has a simple expression which, after substituted in ${\cal A}^{(3)}$
leads to~\cite{Green:1987sp,Polchinski:1998rr}
\begin{eqnarray}
{\cal A}^{(3)} & = & 2i g (2 \pi)^{10} \delta^{10} (k_1+k_2+k_3)\times
\label{Athree} \\ & & {}\times \left(   (\zeta_1 \cdot k_2)(\zeta_2
\cdot \zeta_3) + (\zeta_2 \cdot k_3)(\zeta_3 \cdot \zeta_1) + (\zeta_3
\cdot k_1)(\zeta_1 \cdot \zeta_2)   \right) \tr(\lambda^{a_1}
[\lambda^{a_2}, \lambda^{a_3}]) \,.
\nonumber
\end{eqnarray}
This is exactly the three-gluon tree amplitude, as derived from
Yang-Mills theory (see eq.~(\ref{A3a})): it has no superstring theory
corrections.  Now, for the four gluon tree
amplitude,~(\ref{N-amplitude}) becomes
\begin{eqnarray}
\lefteqn{A(1, 2, 3, 4) = 2 \frac{g^{2}}{(2 \alpha')^{9}}
  (x_3-x_1)(x_4-x_1)\times} \!\! &&
\nonumber \\ && {}\times \! \int_{x_1}^{x_3}\!\! dx_2 \!\int\! d
\theta_1 d \theta_2 | x_2 - x_1 - \theta_2 \theta_1 |^{2 \alpha' k_2
\cdot k_1}
| x_3 - x_1 - \theta_3 \theta_1 |^{2 \alpha' k_3 \cdot k_1} | x_4 -
x_1 - \theta_4 \theta_1 |^{2 \alpha' k_4 \cdot k_1} \! \times
\nonumber \\ && {}\times | x_3 - x_2 - \theta_3 \theta_2 |^{2 \alpha'
k_3 \cdot k_2} | x_4 - x_2 - \theta_4 \theta_2 |^{2 \alpha' k_4 \cdot
k_2} | x_4 - x_3 - \theta_4 \theta_3 |^{2 \alpha' k_4 \cdot k_3}
\times \nonumber \\ && {}\times \int d \phi_1 d \phi_2 d \phi_3 d
\phi_4 e^{f_4(\zeta, k, \theta, \phi)}\,.
\label{A1234}
\end{eqnarray}
In this case the only $x$ integration is carried over $x_2$ (where
$x_1 < x_2 < x_3 < x_4$) and we also have $\theta_3 = \theta_4 =
0$. In order to simplify the calculations $x_1$, $x_3$ and $x_4$ are
typically chosen to be $0$, $1$ and $+ \infty$. After doing the
calculations and introducing the Mandelstam variables,
\begin{equation}
s = - 2 k_1 \cdot k_2\,,\qquad
t = -2 k_1 \cdot k_4\,,\qquad
u = - 2 k_1 \cdot k_3\,,
\label{Mandel}
\end{equation}
$A(1, 2, 3, 4)$ can be obtained in a closed expression, which, after
substituted in ${\cal A}^{(4)}$ leads
to~\cite{Green:1987sp,Polchinski:1998rr}
\begin{eqnarray}
{\cal A}^{(4)} & = & 8i g^2 (\alpha')^2 (2 \pi)^{10} \delta^{10}(k_1 +
k_2 +
k_3 +k_4)\times \nonumber \\&&{}\times  \Biggl[ \frac{\Gamma(- \alpha' s)
 \Gamma(- \alpha' t)}{\Gamma(1- \alpha' s - \alpha' t)}
\{ \tr(\lambda^{a_1} \lambda^{a_2} \lambda^{a_3} \lambda^{a_4}) +
\tr(\lambda^{a_1} \lambda^{a_4} \lambda^{a_3} \lambda^{a_2}) \}+
 \nonumber \\ & & 
\hphantom{{}\times  \Biggl[} 
+\frac{\Gamma(- \alpha' t) \Gamma(- \alpha' u)}{\Gamma(1- \alpha' t -
\alpha' u)} \{ \tr(\lambda^{a_1} \lambda^{a_4} \lambda^{a_2}
\lambda^{a_3}) +
\tr(\lambda^{a_1} \lambda^{a_3} \lambda^{a_2} \lambda^{a_4}) \}+
 \nonumber \\ & &
\hphantom{{}\times  \Biggl[} 
 + \frac{\Gamma(- \alpha' u) \Gamma(- \alpha' s)}{\Gamma(1-
\alpha' u - \alpha' s)} \{ \tr(\lambda^{a_1} \lambda^{a_3}
\lambda^{a_4} \lambda^{a_2}) +
\tr(\lambda^{a_1} \lambda^{a_2} \lambda^{a_4} \lambda^{a_3}) \}
 \Biggr] +
\nonumber \\&&{}\times  K(\zeta_1, k_1; \zeta_2, k_2;
 \zeta_3, k_3; \zeta_4, k_4)\,,
\label{A4}
\end{eqnarray}
where the kinematic factor $K(\zeta_1, k_1; \zeta_2, k_2; \zeta_3,
k_3; \zeta_4, k_4)$ is given by
\begin{eqnarray}
K & = & - \frac{1}{4} \Bigl[ ts(\zeta_1 \cdot \zeta_3)(\zeta_2 \cdot
  \zeta_4) + su(\zeta_2 \cdot \zeta_3)(\zeta_1 \cdot \zeta_4) +
  ut(\zeta_1 \cdot \zeta_2)(\zeta_3 \cdot \zeta_4) \Bigr] +
\nonumber
\\&&{}+ \frac{1}{2} s \Bigl[ (\zeta_1 \cdot k_4)(\zeta_3 \cdot
  k_2)(\zeta_2 \cdot \zeta_4) + (\zeta_2 \cdot k_3)(\zeta_4 \cdot
  k_1)(\zeta_1 \cdot \zeta_3) + 
\nonumber \\&&
\hphantom{{}+ \frac{1}{2} s \Bigl[}
+(\zeta_1 \cdot k_3)(\zeta_4 \cdot
  k_2)(\zeta_2 \cdot \zeta_3) 
+ (\zeta_2 \cdot k_4)(\zeta_3 \cdot k_1)(\zeta_1 \cdot \zeta_4)
  \Bigr] +
\nonumber\\ &&
{}+ \frac{1}{2} t \Bigl[ (\zeta_2 \cdot k_1)(\zeta_4 \cdot
  k_3)(\zeta_3 \cdot \zeta_1) + (\zeta_3 \cdot k_4)(\zeta_1 \cdot
  k_2)(\zeta_2 \cdot \zeta_4) +
\nonumber \\&&
\hphantom{{}+ \frac{1}{2} t \Bigl[}
+ (\zeta_2 \cdot k_4)(\zeta_1 \cdot k_3)(\zeta_3 \cdot \zeta_4) +
  (\zeta_3 \cdot k_1)(\zeta_4 \cdot k_2)(\zeta_2 \cdot \zeta_1)
  \Bigr] + 
\nonumber\\ &&
{}+\frac{1}{2} u \Bigl[ (\zeta_1 \cdot k_2)(\zeta_4 \cdot
  k_3)(\zeta_3 \cdot \zeta_2) +  
 (\zeta_3
  \cdot k_4)(\zeta_2 \cdot k_1)(\zeta_1 \cdot \zeta_4) + 
\nonumber \\&&
\hphantom{{}+\frac{1}{2} u \Bigl[} 
+  (\zeta_1 \cdot k_4)(\zeta_2 \cdot k_3)(\zeta_3 \cdot \zeta_4) +
  (\zeta_3 \cdot k_2)(\zeta_4 \cdot k_1)(\zeta_1 \cdot \zeta_2)
  \Bigr] 
\label{K}
\end{eqnarray}
An important thing about ${\cal A}^{(4)}$, and that does not happen in
${\cal A}^{(3)}$, is the fact that it contains an infinite number of
higher order corrections in $\alpha'$. The coefficients of this
$\alpha'$ expansion can all be determined in terms of the Riemann Zeta
function, evaluated in integer values. For example, the first term
involving Gamma functions in~(\ref{A4}) has the following $\alpha'$
expansion:
\begin{equation} 
\frac{\Gamma(- \alpha' s) 
\Gamma(- \alpha' t)} {\Gamma(1- \alpha' s - \alpha' t)} =
 \frac{1}{{\alpha'}^2 s t} - \frac{\pi^2}{6} - \zeta(3) (s+t) \alpha'
 + {\cal O}({\alpha'}^2)\,.
\label{gammas}
\end{equation}
After substituting this last expression and similar ones for the other
terms involving Gamma functions in~(\ref{A4}), the leading term (i.e.,
order zero in $\alpha'$) in ${\cal A}^{(4)}$ is nothing else than the
Yang-Mills four-gluon tree amplitude; the first superstring theory
contributions to ${\cal A}^{(4)}$ begin at order two in $\alpha'$ and
they all have a common factor $\pi^2$; the next ones occur at order
three in $\alpha'$ and they all have a common factor $\zeta(3)$, and
so on.

\section{Effective lagrangian up to ${\cal O}(\alpha'^2)$ and beyond}
\label{low}

\subsection{Effective lagrangian up to ${\cal O}(\alpha'^2)$}

Knowledge of the on-shell scattering amplitudes of four gluons is
enough to determine the effective lagrangian up to ${\cal
O}(\alpha'^2)$. This has been known for a long
time~\cite{Gross:1986iv,Tseytlin} to be:
\begin{eqnarray}
{\cal L}_{(0, 2)} &=&{}- \frac{1}{4} tr\left(F_{{\mu}_1 {\mu}_2}
F^{{\mu}_1 {\mu}_2} \right) +
\nonumber \\&&{}+ \left(2 \pi \alpha'
\right)^2 g^2 \tr \biggl( \frac{1}{24}
F_{{\mu}_1}^{\;\,{\mu}_2}\,F_{{\mu}_2}^{\;\,{\mu}_3}
\,F_{{\mu}_3}^{\;\,{\mu}_4}\,F_{{\mu}_4}^{\;\,{\mu}_1} +
\frac{1}{12}\,F_{{\mu}_1}^{\;\,{\mu}_2}\,F_{{\mu}_2}^{\;\,{\mu}_3}
\,F_{{\mu}_4}^{\;\,{\mu}_1}\,F_{{\mu}_3}^{\;\,{\mu}_4}-
\nonumber \\&&
\hphantom{{}+ \left(2 \pi \alpha' \right)^2 g^2 \tr \biggl(} 
- \frac{1}{48}\,F_{{\mu}_1}^{\;\,{\mu}_2}\,F_{{\mu}_2}^{\;\,{\mu}_1}
\,F_{{\mu}_3}^{\;\,{\mu}_4}\,F_{{\mu}_4}^{\;\,{\mu}_3} -
\frac{1}{96}\,F_{{\mu}_1}^{\;\,{\mu}_2}\,F_{{\mu}_3}^{\;\,{\mu}_4}
\,F_{{\mu}_2}^{\;\,{\mu}_1}\,F_{{\mu}_4}^{\;\,{\mu}_3}\biggr)
\label{L2}
\end{eqnarray}
where the field strength is defined as
\be
F_{\mu\nu} = \pd_\mu\, A_\nu - \pd_\nu\, A_\mu -
i\,g\,[A_\mu,\,A_\nu]\,.
\label{F}
\ee
Here, the index ${(0,2)}$ has been used to denote that this lagrangian
contains terms of order $0$ and $2$ in $\alpha'$. No order $1$ terms
in $\alpha'$ appear.

\subsection{${\cal O}(\alpha'^3)$ contributions to the effective
  lagrangian} 

As was argued in the previous section, the ${\cal A}^{(4)}$ amplitude
receives, beyond ${\cal O}(\alpha'^2)$, ${\cal O}(\alpha'^3)$ and
higher order contributions. In fact, using the $\alpha'$ expansion for
the Gamma functions in ${\cal A}^{(4)}$ (see eq.~(\ref{A4})) some new
${\cal O}(\alpha'^3)$ terms may be derived in the effective
lagrangian. These would be the $D^2 F^4$ terms, as may be understood
by dimensional analysis. Since ${\cal A}^{(3)}$ does not contain any
$\alpha'$ terms at all, no $D^4 F^3$ terms will be present in the
effective lagrangian.  So, the remaining ${\cal O}(\alpha'^3)$ terms
can only be of $F^5$ type.  To derive them it would be necessary to
compute the five-gluon tree amplitude, ${\cal A}^{(5)}$, up to order
three in $\alpha'$, which is going to be done in section~\ref{main}.

At this point, it is important to say, as was already mentioned in the
introduction, that in the literature there can be found three
non-equivalent versions of the ${\cal O}(\alpha'^3)$ terms of the
effective lagrangian (see references~\cite{Koerber:2001uu}
and~\cite{Koerber:2001hk} for a detailed comparison of them). The
first of them~\cite{Kitazawa:1987xj} was derived a long time ago by
means of the computation of the five-gluon amplitude in Open
Superstring theory. The other two versions are recent. One of them is
derived from the effective action of $N=4$ super Yang-Mills in
$D=4$~\cite{Refolli:2001df} and the other uses deformations of a
particular kind of solutions of Yang-Mills
theory~\cite{Koerber:2001uu} in any even dimensional spacetime. We
will see, in the next sections, that our detailed calculation of the
five gluon tree amplitude ${\cal A}^{(5)}$ matches completely the last
of these three versions, due to Koerber and Sevrin. Their effective
lagrangian, at order ${\cal O}(\alpha'^3)$, is the
following~\cite{Koerber:2001uu}:\footnote{There are two differences in
the conventions used by the authors of~\cite{Koerber:2001uu} and ours:
{\renewcommand{\theenumi}{\roman{enumi}}
\begin{enumerate} 
\item They use anti-hermitian matrices as generators of the \un Lie
algebra, while we use hermitian matrices. This implies that in our
formulas in equations~(\ref{L2}),~(\ref{F}) and~(\ref{L3}), there are
some signals and $i$ factors that do not appear in their effective
lagrangian.
\item They introduce the coupling constant $g$ only as a global factor
$-1/g^2$ in the lagrangian, while we introduce it in the definition of
$F_{\mu \nu}$, in eq.~(\ref{F}), and also as powers of it multiplying
each individual term in the lagrangian.\end{enumerate}}}
\begin{eqnarray}
\lefteqn{{\cal L}_{(3)} =-(2\,\alpha')^3\,2\,\zeta(3)\times}&&
\label{L3}
\\ && 
\times\tr \biggl[
i\,g^3\,F_{{\mu}_1}^{\;\,{\mu}_2}\,F_{{\mu}_2}^{\;\,{\mu}_3}
\,F_{{\mu}_3}^{\;\,{\mu}_4}\,F_{{\mu}_5}^
{\;\,{\mu}_1}\,F_{{\mu}_4}^{\;\,{\mu}_5}+ 
i\,g^3\,F_{{\mu}_1}^{\;\,{\mu}_2}\,F_{{\mu}_4}^{\;\,{\mu}_5}
\,F_{{\mu}_2}^{\;\,{\mu}_3}\,F_{{\mu}_5}^{\;\,{\mu}_1}
\,F_{{\mu}_3}^{\;\,{\mu}_4}-
\nonumber \\&&
\hphantom{\times\tr \biggl[}
-  \frac{i\,g^3}{2}
\,F_{{\mu}_1}^{\;\,{\mu}_2}\,F_{{\mu}_2}^{\;\,{\mu}_3}
\,F_{{\mu}_4}^{\;\,{\mu}_5}\,F_{{\mu}_3}^{\;\,{\mu}_1}\,F_{{\mu}_5}^{\;\,{\mu}_4}
-g^2\,F_{{\mu}_1}^{\;\,{\mu}_2}\left(D^{{\mu}_1}
\,F_{{\mu}_3}^{\;\,{\mu}_4}\right)\left(D_{{\mu}_5}
\,F_{{\mu}_2}^{\;\,{\mu}_3}\right)\,F_{{\mu}_4}^{\;\,{\mu}_5}+
\nonumber \\ &&
\hphantom{\times\tr \biggl[}
+  \frac{g^2}{2}
\left(D^{{\mu}_1}\,F_{{\mu}_2}^{\;\,{\mu}_3}\right)
\left(D_{{\mu}_1}\,F_{{\mu}_3}^{\;\,{\mu}_4}\right)
\,F_{{\mu}_5}^{\;\,{\mu}_2}\,F_{{\mu}_4}^{\;\,{\mu}_5}
+\frac{g^2}{2}\left(D^{{\mu}_1}\,F_{{\mu}_2}^{\;\,{\mu}_3}\right)
\,F_{{\mu}_5}^{\;\,{\mu}_2}\left(D_{{\mu}_1}
\,F_{{\mu}_3}^{\;\,{\mu}_4}\right)\,F_{{\mu}_4}^{\;\,{\mu}_5}-
\nonumber \\&&
\hphantom{\times\tr \biggl[}
-  \frac{g^2}{8}
\left(D^{{\mu}_1}\,F_{{\mu}_2}^{\;\,{\mu}_3}\right)
\,F_{{\mu}_4}^{\;\,{\mu}_5}\left(D_{{\mu}_1}
\,F_{{\mu}_3}^{\;\,{\mu}_2}\right)\,F_{{\mu}_5}^{\;\,{\mu}_4} + g^2\,
\left(D_{{\mu}_5}\,F_{{\mu}_1}^{\;\,{\mu}_2}\right)
\,F_{{\mu}_3}^{\;\,{\mu}_4}\left(D^{{\mu}_1}
\,F_{{\mu}_2}^{\;\,{\mu}_3}\right)\,F_{{\mu}_4}^{\;\,{\mu}_5}\biggr]\,,
\nonumber
\end{eqnarray}
where, besides~(\ref{F}), the covariant derivative acting on some
field $\phi$ is defined as
\be
D_{\mu}\,\phi = \pd_\mu\,\phi - i\,g\,[A_\mu,\,\phi]\,.
\label{Dphi}
\ee

\section{Five-gluon tree amplitude in Open Superstring theory}
\label{main}

In this section we develop the main result of this paper, namely, the
five gluon tree amplitude, as derived from the Open Superstring
theory.  Our formula in eq.~(\ref{N-amplitude}), in the case of five
gluons becomes
\begin{eqnarray} 
A(1, 2, 3, 4, 5) & = & 2 \frac{g^3}{(2 \alpha')^{43/4}} (x_4-x_1)
(x_5-x_1)\times
\nonumber \\&&{}\times  \int_{x_1}^{x_4} d x_3 \int_{x_1}^{x_3} d x_2
 \int
d \theta_1 d \theta_2 d \theta_3 \prod_{i>j}^5 | x_i - x_j - \theta_i
\theta_j |^{2 \alpha' k_i \cdot k_j}\times
\nonumber \\&&{}\times  \int d \phi_1 d \phi_2 d \phi_3 d \phi_4 d
 \phi_5
e^{f_5(\zeta, k, \theta, \phi)}\,,
\label{A12345} 
\end{eqnarray} 
where $\theta_4= \theta_5 =0$.  Starting from this formula and
arriving to a final answer, where $A(1, 2, 3, 4, 5)$ is an explicit
expression of $\alpha'$, the momenta $k_i$ and the polarizations
$\zeta_i$, is a huge task. To save the reader from the very lengthy
calculations involved in this task, we have decided to present the
explicit expression for $A(1, 2, 3, 4, 5)$, up to ${\cal
O}(\alpha'^3)$ terms, in subsection~\ref{5-point} and the details of
the calculations in the following subsections.

\subsection{Explicit expression for the 5-point amplitude}
\label{5-point}

Up to ${\cal O}(\alpha'^3)$ terms, the final result for $A(1, 2, 3, 4,
5)$ has the following form:
\begin{equation} 
A(1, 2, 3, 4, 5) = A^{(0)}(1, 2, 3, 4, 5) + A^{(2)}(1, 2, 3, 4, 5)
\cdot {\alpha'}^2 + A^{(3)}(1, 2, 3, 4, 5) \cdot {\alpha'}^3 + {\cal
O}(\alpha'^4)\,,
\label{A5final}
\end{equation} 
where the terms $A^{(0)}(1, 2, 3, 4, 5)$, $A^{(2)}(1, 2, 3, 4, 5)$ and
$A^{(3)}(1, 2, 3, 4, 5)$ do not depend on $\alpha'$. The first of
these three terms is nothing else than the Yang-Mills five-gluon tree
amplitude, whose expression is the following:\footnote{The complete
Yang-Mills five-gluon tree amplitude is constructed from~(\ref{A(0)})
using the same rule given in eq.~(\ref{general-amplitude}).}
\begin{eqnarray}
\lefteqn{A^{(0)}(1, 2, 3, 4, 5) =} && 
\nonumber\\ && 2 g^3
\Biggl[   (\zeta_1 \cdot \zeta_2)(\zeta_3 \cdot \zeta_4)\times
\nonumber\\ &&
\hphantom{2 g^3 \Biggl[} 
\times \biggl\{   (\zeta_5 \cdot k_1) \alpha_{23}
\left\{ \frac{1}{\rho \alpha_{12}} + \frac{1}{\rho \alpha_{23}} +
\frac{1}{\alpha_{34} \phi} + \frac{1}{\alpha_{23} \phi}
+ \frac{1}{\alpha_{34} \alpha_{12}} \right\} -
\nonumber \\ & & 
\hphantom{2 g^3 \Biggl[\times \biggl\{}
- (\zeta_5\cdot k_2) \alpha_{13}
\left\{ \frac{1}{\alpha_{34} \alpha_{12}} 
+ \frac{1}{\rho \alpha_{12}}
\right\} + 
\nonumber\\ &&
\hphantom{2 g^3 \Biggl[\times \biggl\{}
+(\zeta_5 \cdot k_3) \biggl( \alpha_{24} \left\{ \frac{1}{\alpha_{34}
\alpha_{12}}   + \frac{1}{\alpha_{34} \phi}
 \right\} +
\nonumber \\ & & 
\hphantom{2 g^3 \Biggl[\times \biggl\{+(\zeta_5 \cdot k_3) \biggl(}
+ \alpha_{23}
\left\{ \frac{1}{\rho \alpha_{12}} + \frac{1}{\rho \alpha_{23}} +
\frac{1}{\alpha_{34} \phi} + \frac{1}{\alpha_{23} \phi} +
\frac{1}{\alpha_{34} \alpha_{12}} \right\}    \biggr)
\biggr\}+
\nonumber \\ & & 
\hphantom{2 g^3 \Biggl[} 
+ (\zeta_1 \cdot \zeta_3)(\zeta_2 \cdot
 \zeta_4)
\biggl\{   -(\zeta_5 \cdot k_1) \alpha_{23} \left\{
\frac{1}{\alpha_{23} \rho} + \frac{1}{\alpha_{23} \phi} \right\} +
\nonumber \\ & & 
\hphantom{2 g^3 \Biggl[+ (\zeta_1 \cdot \zeta_3)(\zeta_2 \cdot
    \zeta_4)\biggl\{  }  
+ (\zeta_5 \cdot k_2)
\left( \alpha_{34} \left\{ \frac{1}{\alpha_{34} \phi} \right\} 
- \alpha_{23}
\left\{ \frac{1}{\alpha_{23} \rho}  + \frac{1}{\alpha_{23} \phi} \right\}
\right)-
\nonumber\\ &&
\hphantom{2 g^3 \Biggl[+ (\zeta_1 \cdot \zeta_3)(\zeta_2 \cdot
    \zeta_4)\biggl\{  }  
- (\zeta_5 \cdot k_3) \alpha_{12}
 \left\{ \frac{1}{\rho \alpha_{12}} \right\} \biggr\}+
\nonumber \\ & &  
\hphantom{2 g^3 \Biggl[} 
+ (\zeta_1 \cdot \zeta_4)(\zeta_2 \cdot
 \zeta_3)
\biggl\{   (\zeta_5 \cdot k_1)   \left(  
\alpha_{34} \left\{ \frac{1}{\alpha_{23} \phi} + \frac{1}{\alpha_{34} \phi}
\right\}
- \alpha_{13} \left\{ \frac{1}{\alpha_{23} \rho} \right\} \right) -
\nonumber\\ & &
\hphantom{2 g^3 \Biggl[+ (\zeta_1 \cdot \zeta_4)(\zeta_2 \cdot  \zeta_3)
\biggl\{ } 
 -(\zeta_5 \cdot k_2) \frac{\alpha_{13}}{\alpha_{23} \rho}
+ (\zeta_5 \cdot k_3) \alpha_{12} \left\{ \frac{1}{\alpha_{23} \rho} +
\frac{1}{\alpha_{12} \rho} \right\}    \biggr\}+
\nonumber \\ & &  
\hphantom{2 g^3 \Biggl[} 
+ (\zeta_2 \cdot \zeta_3)
\biggl\{   (\zeta_5 \cdot k_1)   \biggl(  
(\zeta_1 \cdot k_2)(\zeta_4 \cdot k_3)   \left\{ \frac{1}{\rho
\alpha_{12}} + \frac{1}{\rho \alpha_{23}} + \frac{1}{\alpha_{34} \phi}
+ \frac{1}{\alpha_{23} \phi} + \frac{1}{\alpha_{34} \alpha_{12}} \right\}-
\nonumber\\ & & 
\hphantom{2 g^3 \Biggl[+ (\zeta_2 \cdot \zeta_3)\biggl\{   (\zeta_5
    \cdot k_1)   \biggl(  }  
- (\zeta_1 \cdot k_3)(\zeta_4 \cdot k_2)
\left\{ \frac{1}{\alpha_{23} \rho}  + \frac{1}{\alpha_{23} \phi} \right\}
\biggr)+
\nonumber \\ & & 
\hphantom{2 g^3 \Biggl[+ (\zeta_2 \cdot \zeta_3)\biggl\{ }
+ (\zeta_5 \cdot k_2)
\biggl(   (\zeta_1 \cdot k_3)(\zeta_4 \cdot k_1) 
\frac{1}{\alpha_{23} \rho} +
\nonumber\\ &&
\hphantom{2 g^3 \Biggl[+ (\zeta_2 \cdot \zeta_3)\biggl\{ + (\zeta_5
    \cdot k_2)\biggl(} 
+ (\zeta_1 \cdot k_3)(\zeta_4 \cdot k_3) \left\{
\frac{1}{\alpha_{23} \alpha_{34}} + \frac{1}{\rho \alpha_{23}}
- \frac{\alpha_{24}}{\alpha_{23} \phi \alpha_{34}} \right\} +
\nonumber \\ & &
\hphantom{2 g^3 \Biggl[+ (\zeta_2 \cdot \zeta_3)\biggl\{ + (\zeta_5
    \cdot k_2)\biggl(} 
 + (\zeta_1 \cdot k_2)(\zeta_4 \cdot k_3)
\left\{ \frac{1}{\rho \alpha_{12}} {+} \frac{1}{\rho \alpha_{23}} {+}
\frac{1}{\alpha_{34} \phi} {+} \frac{1}{\alpha_{23} \phi}
{+} \frac{1}{\alpha_{34} \alpha_{12}} \right\} +
\nonumber \\ & & 
\hphantom{2 g^3 \Biggl[+ (\zeta_2 \cdot \zeta_3)\biggl\{ + (\zeta_5
    \cdot k_2)\biggl(} 
+ (\zeta_1
 \cdot k_4)(\zeta_4 \cdot k_3)
\left\{ \frac{1}{\alpha_{23} \phi}
 + \frac{1}{\alpha_{34} \phi} \right\} \biggl) -
\nonumber \\ & & 
\hphantom{2 g^3 \Biggl[+ (\zeta_2 \cdot \zeta_3)\biggl\{}
- (\zeta_5 \cdot k_3) \biggl(   (\zeta_1 \cdot k_2)(\zeta_4
 \cdot k_1) 
\left\{ \frac{1}{\alpha_{23} \rho} + \frac{1}{\alpha_{12} \rho}
\right\} +
\nonumber\\ &&
\hphantom{2 g^3 \Biggl[+ (\zeta_2 \cdot \zeta_3)\biggl\{- (\zeta_5
    \cdot k_3) \biggl(} 
+ (\zeta_1 \cdot k_3)(\zeta_4 \cdot k_2) \left\{ \frac{1}{\alpha_{23}
  \rho} + \frac{1}{\alpha_{23} \phi} \right\}+
\nonumber \\ & & 
\hphantom{2 g^3 \Biggl[+ (\zeta_2 \cdot \zeta_3)\biggl\{- (\zeta_5
    \cdot k_3) \biggl(} 
+ (\zeta_1 \cdot k_2)(\zeta_4 \cdot k_2)
\left\{ \frac{1}{\alpha_{23} \alpha_{12}} + \frac{1}{\phi \alpha_{23}}
- \frac{\alpha_{13}}{\alpha_{23} \rho \alpha_{12}} \right\}+
\nonumber\\ &&
\hphantom{2 g^3 \Biggl[+ (\zeta_2 \cdot \zeta_3)\biggl\{- (\zeta_5
    \cdot k_3) \biggl(} 
+ (\zeta_1 \cdot k_4)(\zeta_4 \cdot k_2)
 \frac{1}{\alpha_{23} \phi} \biggr) \biggr\}+
\nonumber \\ & &  
\hphantom{2 g^3 \Biggl[}
+ (\zeta_1 \cdot \zeta_4)
\biggl\{ -   (\zeta_5 \cdot k_1)   \biggl(  
(\zeta_2 \cdot k_3)(\zeta_3 \cdot k_4)   \left\{
\frac{1}{\alpha_{23} \phi} + \frac{1}{\alpha_{34} \phi} \right\}
- (\zeta_2 \cdot k_4)(\zeta_3 \cdot k_2)
\frac{1}{\alpha_{23} \phi} +
\nonumber \\ & & 
\hphantom{2 g^3 \Biggl[+ (\zeta_1 \cdot \zeta_4)\biggl\{ -   (\zeta_5
    \cdot k_1)   \biggl(  }  
+ (\zeta_2 \cdot
 k_4)(\zeta_3 \cdot k_4)
\frac{1}{\alpha_{34} \phi}
- (\zeta_2 \cdot k_1)(\zeta_3 \cdot k_4)
 \frac{1}{\alpha_{12} \alpha_{34}} \biggr)+
\nonumber \\ & & 
\hphantom{2 g^3 \Biggl[+ (\zeta_1 \cdot \zeta_4)\biggl\{}
+ (\zeta_5 \cdot k_4)
\biggl(   (\zeta_2 \cdot k_1)(\zeta_3 \cdot k_2)   \left\{
\frac{1}{\alpha_{23} \rho} + \frac{1}{\alpha_{12} \rho} \right\} + (\zeta_2
\cdot k_1)(\zeta_3 \cdot k_1)
\frac{1}{\alpha_{12} \rho} -
\nonumber \\ & & 
\hphantom{2 g^3 \Biggl[+ (\zeta_1 \cdot \zeta_4)\biggl\{+ (\zeta_5
    \cdot k_4)\biggl(} 
- (\zeta_2 \cdot
 k_3)(\zeta_3 \cdot k_1)  \frac{1}{\alpha_{23} \rho} \biggr)
 \biggr\} +
\nonumber \\ & & 
\hphantom{2 g^3 \Biggl[}
+ \left(  
\mbox{cyclic permutations of indexes (1,2,3,4,5)} \right) \Biggr],
\label{A(0)}
\end{eqnarray}
where $\alpha_{ij} = k_i \cdot k_j$, $\rho = \alpha_{12} + \alpha_{13}
+ \alpha_{23}$ and $\phi = \alpha_{23} + \alpha_{24} + \alpha_{34}$.
$A^{(2)}(1,2,3,4,5)$ and $A^{(3)}(1,2,3,4,5)$ have similar expressions
to the one shown in~(\ref{A(0)}) for $A^{(0)}(1,2,3,4,5)$. Their
explicit expressions are given in appendix~\ref{A0A2A3}.  As expected,
in~(\ref{A5final}) there is no linear term in $\alpha'$.

\subsection{Fixing some of the free parameters in $A(1, 2, 3, 4, 5)$}

In this subsection we begin the derivation of each of the terms in
formula eq.~(\ref{A5final}). As a starting point we
consider~(\ref{A12345}). Our only aim in this subsection is to
simplify a little this last formula, after fixing $\theta_4 =0$,
$\theta_5=0$, and $x_5 = + \infty$ ($x_1$ and $x_4$ will be fixed to
$0$ and $1$, respectively, in subsection~\ref{Grasstheta}). For this
purpose it is necessary to expand the $exp(f_5(\zeta, k,\theta,
\phi))$ term in eq.~(\ref{A12345}) and to do some of the Grassmann
integrations.  To start with, in eq.~(\ref{fN}) we have that
\begin{equation} 
f_5(\zeta, k, \theta, \phi) = \sum_{i \neq j}^5
\frac{(\theta_i-\theta_j) \phi_i (\zeta_i \cdot k_j) (2
\alpha')^{11/4} -1/2 \phi_i \phi_j (\zeta_i \cdot \zeta_j) (2
\alpha')^{9/2}}{x_i-x_j-\theta_i \theta_j}\,,
\label{f5}
\end{equation} 
so $f_5(\zeta, k,\theta, \phi)$ may be split into two sums, which can
be expanded in powers of $1/x_5$ as follows:
\begin{equation}
\sum_{i \neq j}^5 \frac{(\theta_i-\theta_j) \phi_i (\zeta_i \cdot
k_j)} {x_i - x_j - \theta_i \theta_j} = A + \frac{B}{x_5} + {\cal
O}\left(\frac{1}{{x_5}^2}\right)
\label{S1}
\end{equation}
and
\begin{equation}
\frac{1}{2} \sum_{i \neq j}^5 \frac{\phi_i \phi_j (\zeta_i \cdot
\zeta_j)} {x_i - x_j - \theta_i \theta_j} = C + \frac{D}{x_5} + {\cal
O}\left(\frac{1}{{x_5}^2}\right).
\label{S2}
\end{equation}
After fixing $\theta_4 =0$ and $\theta_5=0$, the bosonic terms $A$,
$B$, $C$ and $D$ in this last formula are given by
\begin{eqnarray}
A & = & \left[ \theta_1 \left( \frac{\zeta_1 \cdot k_2}{x_1-x_2}
\right.  \right. + \frac{\zeta_1 \cdot k_3}{x_1-x_3} \left. +
\frac{\zeta_1 \cdot k_4}{x_1-x_4} \right)
- \theta_2 \frac{\zeta_1 \cdot k_2}{x_1-x_2}
\left. - \theta_3 \frac{\zeta_1 \cdot k_3}{x_1-x_3} \right] \phi_1
+ \nonumber \\&&{}+ \left[ - \theta_1 \frac{\zeta_2 \cdot
 k_1}{x_2-x_1} \right.
+ \theta_2 \left( \frac{\zeta_2 \cdot k_1}{x_2-x_1} \right.  +
\frac{\zeta_2 \cdot k_3}{x_2-x_3} \left. + \frac{\zeta_2 \cdot
k_4}{x_2-x_4} \right)
\left. - \theta_3 \frac{\zeta_2 \cdot k_3}{x_2-x_3} \right] \phi_2
+ \nonumber \\&&{}+ \left[ - \theta_1 \frac{\zeta_3 \cdot
 k_1}{x_3-x_1} \right.
- \theta_2 \frac{\zeta_3 \cdot k_2}{x_3-x_2}
+ \theta_3 \left( \frac{\zeta_3 \cdot k_1}{x_3-x_1} \right.  +
\frac{\zeta_3 \cdot k_2}{x_3-x_2} \left. \left. + \frac{\zeta_3 \cdot
k_4}{x_3-x_4} \right) \right] \phi_3
-\nonumber \\&&{}- \left[ \theta_1 \frac{\zeta_4 \cdot k_1}{x_4-x_1}
 \right.
+ \theta_2 \frac{\zeta_4 \cdot k_2}{x_4-x_2} \left. + \theta_3
\frac{\zeta_4 \cdot k_3}{x_4-x_3} \right] \phi_4\,,
\label{A}
\\
B & = & -\theta_1 (\zeta_1 \cdot k_5) \phi_1 -\theta_2 (\zeta_2 \cdot
k_5)
\phi_2 -\theta_3 (\zeta_3 \cdot k_5) \phi_3 
-\nonumber \\&&{}- \Bigl[
   \theta_1 (\zeta_5 \cdot k_1)
+ \theta_2 (\zeta_5 \cdot k_2) + \theta_3 (\zeta_5 \cdot k_3)
  \Bigr] \phi_5\,,
\label{B}
\\
C & = & \frac{\phi_2 \phi_1 (\zeta_2 \cdot \zeta_1)}{x_2-x_1- \theta_2
\theta_1}+ \frac{\phi_3 \phi_1 (\zeta_3 \cdot \zeta_1)}{x_3-x_1-
\theta_3 \theta_1}+ \frac{\phi_4 \phi_1 (\zeta_4 \cdot
\zeta_1)}{x_4-x_1}+ \frac{\phi_3 \phi_2 (\zeta_3 \cdot
\zeta_2)}{x_3-x_2- \theta_3 \theta_2}
+\nonumber \\&&{}+ \frac{\phi_4 \phi_2 (\zeta_4 \cdot
 \zeta_2)}{x_4-x_2}
+ \frac{\phi_4 \phi_3 (\zeta_4 \cdot \zeta_3)}{x_4-x_3},
\label{C}
\\
D &=& \phi_5 \phi_1 (\zeta_5 \cdot \zeta_1) + \phi_5 \phi_2 (\zeta_5
\cdot \zeta_2) + \phi_5 \phi_3 (\zeta_5 \cdot \zeta_3) + \phi_5 \phi_4
(\zeta_5 \cdot \zeta_4)\,.
\label{D}
\end{eqnarray}
Now, multiplying separately the factors containing $x_5$ in the
$\prod$ term of~(\ref{A12345}), using the on-shell
condition~(\ref{onshell}) and using the sums given in eqs.~(\ref{S1})
and~(\ref{S2}) for the $f_5(\zeta, k,\theta, \phi)$ term in the
exponential, we have that
\begin{eqnarray}
\lefteqn{A(1, 2, 3, 4, 5) = 2 g^3 (2 \alpha') (x_4-x_1) (x_5-x_1)
\times}&& 
\label{A12345-1} \\&&{}\times  \int_{x_1}^{x_4} d x_3
\int_{x_1}^{x_3} d x_2 \int d \theta_1 d \theta_2 d \theta_3
\prod_{i>j}^4 | x_i - x_j - \theta_i \theta_j |^{2 \alpha' k_i \cdot
k_j} 
\times\nonumber \\&&{}\times  \left[   1- (2 \alpha') \Big(
  (k_1 \cdot k_5) x_1 + (k_2 \cdot k_5) x_2 +(k_3 \cdot k_5)
x_3 + (k_4 \cdot k_5) x_4   \Big)\frac{1}{x_5} + {\cal O}
\left(\frac{1}{x_5^2}\right)   \right] 
\times\nonumber \\&&{}\times 
\int d \phi_1 d \phi_2 d \phi_3 d \phi_4 d \phi_5 \left\{ \left[
\frac{AC^2}{2} + (2 \alpha') \frac{A^3}{6} \right] \right.  
+\nonumber
\\ & & 
\hphantom{\times 
\int d \phi_1 d \phi_2 d \phi_3 d \phi_4 d \phi_5 \left\{
\left[\right.\right.}  
\left. + \left[ \frac{2ACD+BC^2}{2}+ (2 \alpha') \frac{3A^2
BC+A^3 D}{6} \right] \frac{1}{x_5} + {\cal O}
\left(\frac{1}{x_5^2}\right) \right\}.
\nonumber
\end{eqnarray}
In this last formula, the term $[ \frac{AC^2}{2} + (2 \alpha')
\frac{A^3}{6}]$ gives no contribution after the Grassmann integration
in $d \phi_5$ is done, since neither $A$ nor $C$ contain the $\phi_5$
variable. So, after choosing $x_5 \rightarrow + \infty$,
eq.~(\ref{A12345-1}) becomes
\begin{eqnarray}
A(1, 2, 3, 4, 5) & = & 2 g^3 (2 \alpha') (x_4 - x_1) \times
\nonumber \\ &&
{} \times  \int_{x_1}^{x_4} d x_3
\int_{x_1}^{x_3} d x_2 \int d \theta_1 d \theta_2 d \theta_3
\prod_{i>j}^4 |
x_i - x_j - \theta_i \theta_j |^{2 \alpha' k_i \cdot k_j} \times
\nonumber \\
&&{}\times  \int d \phi_1 d \phi_2 d \phi_3 d \phi_4 d \phi_5
\left\{ \frac{2ACD+BC^2}{2}+ (2 \alpha') \frac{3A^2 BC+A^3 D}{6}
\right\}.\qquad\ \ 
\label{A12345-2}
\end{eqnarray}   
Although $\theta_4$ appears in the $\prod$ term of this formula, it is
assumed to be $0$. So the only free parameters in eq.~(\ref{A12345-2})
are $x_1$ and $x_4$ (satisfying $x_4 > x_1$).

\subsection{Calculating the $\phi$ Grassmann integration}

In this subsection we integrate all the $\phi_i$ Grassmann variables
appearing in~(\ref{A12345-2}).
{\renewcommand{\theenumi}{\roman{enumi}}
\begin{enumerate} 
\item As a first step we substitute the expressions of $A$, $B$, $C$
and $D$, given respectively in~(\ref{A}),~(\ref{B}),~(\ref{C})
and~(\ref{D}), in the first term of the $\phi_i$ Grassmann integration
in formula~(\ref{A12345-2}). After some steps we arrive to
\begin{eqnarray}
\int d \phi_1 d \phi_2 d \phi_3 d \phi_4 d \phi_5 \left\{
\frac{2ACD+BC^2}{2} \right\} & = & -S_{12345} \theta_1 + S_{21345}
\theta_2
+ S_{32145} \theta_3 
\nonumber \\ & & - (T_{12345} + T_{21345} +
 T_{32145}) \theta_1 \theta_2 \theta_3\,,\qquad
\label{int1}
\end{eqnarray}
where
\begin{eqnarray}
\lefteqn{S_{12345}  =}&&
\label{S12345}
\\  &&
- \left( \frac{\zeta_1 \cdot k_2}{x_2 - x_1} \right.
\left. + \frac{\zeta_1 \cdot k_3}{x_3 - x_1} + \frac{\zeta_1 \cdot
k_4}{x_4 - x_1} \right) \times
\nonumber\\ &&{}\times
\left(\frac{(\zeta_3 \cdot \zeta_2)(\zeta_5
\cdot \zeta_4)}{x_3 - x_2} \right.
- \frac{(\zeta_4 \cdot \zeta_2)(\zeta_5 \cdot \zeta_3)}{x_4 - x_2}
\left. + \frac{(\zeta_4 \cdot \zeta_3)(\zeta_5 \cdot \zeta_2)}{x_4 -
x_3} \right)
-\nonumber \\&&{}- \frac{\zeta_2 \cdot k_1}{x_2 - x_1}
\left(-\frac{(\zeta_3 \cdot \zeta_1)(\zeta_5 \cdot \zeta_4)}{x_3 -
x_1} \right.  + \frac{(\zeta_4 \cdot \zeta_1)(\zeta_5 \cdot
\zeta_3)}{x_4 - x_1} \left. - \frac{(\zeta_4 \cdot \zeta_3)(\zeta_5
\cdot \zeta_1)}{x_4 - x_3} \right)
-\nonumber \\&&{}- \frac{\zeta_3 \cdot k_1}{x_3 - x_1}
\left(\frac{(\zeta_2 \cdot \zeta_1)(\zeta_5 \cdot \zeta_4)}{x_2 - x_1}
\right.
- \frac{(\zeta_4 \cdot \zeta_1)(\zeta_5 \cdot \zeta_2)}{x_4 - x_1}
\left. + \frac{(\zeta_4 \cdot \zeta_2)(\zeta_5 \cdot \zeta_1)}{x_4 -
x_2} \right)
-\nonumber \\&&{}- \frac{\zeta_4 \cdot k_1}{x_4 - x_1}
\left(-\frac{(\zeta_2 \cdot \zeta_1)(\zeta_5 \cdot \zeta_3)}{x_2 -
x_1} \right.  + \frac{(\zeta_3 \cdot \zeta_1)(\zeta_5 \cdot
\zeta_2)}{x_3 - x_1} \left. - \frac{(\zeta_3 \cdot \zeta_2)(\zeta_5
\cdot \zeta_1)}{x_3 - x_2} \right)
+\nonumber \\&&{}+ (\zeta_5 \cdot k_1)
\left(\frac{(\zeta_2 \cdot \zeta_1)(\zeta_4 \cdot \zeta_3)}{(x_2 -
x_1)(x_4 - x_3)} \right.
- \frac{(\zeta_3 \cdot \zeta_1)(\zeta_4 \cdot \zeta_2)}{(x_3 -
x_1)(x_4 - x_2)}
\left. + \frac{(\zeta_4 \cdot \zeta_1)(\zeta_3 \cdot \zeta_2)}{(x_4 -
x_1)(x_3 - x_2)} \right)
\nonumber\end{eqnarray}
and
\begin{eqnarray}
T_{12345} & = & \frac{(\zeta_3 \cdot \zeta_2)(\zeta_5 \cdot
\zeta_4)}{(x_3 - x_2)^2} \left( \frac{\zeta_1 \cdot k_2}{x_2 - x_1}
\right.  \left. + \frac{\zeta_1 \cdot k_3}{x_3 - x_1} + \frac{\zeta_1
\cdot k_4}{x_4 - x_1} \right)
-\nonumber \\&&{}- \frac{(\zeta_3 \cdot \zeta_2)(\zeta_5 \cdot
 \zeta_1)}{(x_3 - x_2)^2}
\frac{\zeta_4 \cdot k_1}{x_4 - x_1}
- \frac{(\zeta_4 \cdot \zeta_1)(\zeta_3 \cdot \zeta_2)}{(x_3 - x_2)^2}
\frac{\zeta_5 \cdot k_1}{x_4 - x_1}\,.
\label{T12345}
\end{eqnarray}
The coefficients $S_{21345}$, $S_{32145}$, $T_{21345}$ and $T_{32145}$
appearing in~(\ref{int1}) are to be derived using the corresponding
expressions for $S_{12345}$ and $T_{12345}$ and interchanging the
corresponding indexes.

\item As a second step we now substitute the expressions of $A$, $B$,
$C$ and $D$ in the second term of the $\phi_i$ Grassmann integration
in~(\ref{A12345-2}). Again, after some steps we arrive to
\begin{eqnarray}
\lefteqn{\int d \phi_1 d \phi_2 d \phi_3 d \phi_4 d \phi_5
  \left\{\frac{3A^2 BC+A^3 D}{6} \right\}  = }\qquad&&
\label{int2}
\\ &&
-\biggl[ U_{12345} \frac{\zeta_2 \cdot \zeta_1}{x_2 - x_1} 
+ U_{13245} \frac{\zeta_3 \cdot \zeta_1}{x_3 - x_1}
+ U_{23145} \frac{\zeta_3 \cdot \zeta_2}{x_3 - x_2} +
\nonumber \\&&
\hphantom{-\biggl[}
+ 
 V_{12345} \frac{\zeta_4 \cdot \zeta_1}{x_4 - x_1}
+ V_{21345} \frac{\zeta_4 \cdot \zeta_2}{x_4 - x_2}
+ V_{32145} \frac{\zeta_4 \cdot \zeta_3}{x_4 - x_3} 
+\nonumber \\&& 
\hphantom{-\biggl[}
+ W_{1234} (\zeta_5 \cdot \zeta_1) + W_{2134} (\zeta_5 \cdot \zeta_2)
+ W_{3124} (\zeta_5 \cdot \zeta_3) +
Z_{1234}
 (\zeta_5 \cdot \zeta_4)   \biggr]
\theta_1 \theta_2 \theta_3 \,,
\nonumber\end{eqnarray}
where
\begin{eqnarray}
U_{12345} & = & - \left[ \frac{\zeta_3 \cdot k_2}{x_3 - x_2} \cdot
\frac{\zeta_4 \cdot k_3}{x_4 - x_3}   + \left( \frac{\zeta_3
\cdot k_2}{x_3 - x_2} {+} \frac{\zeta_3 \cdot k_1}{x_3 - x_1} 
 {+} \frac{\zeta_3 \cdot k_4}{x_3 - x_4} \right)
 \frac{\zeta_4 \cdot k_2}{x_4 - x_2} \right] (\zeta_5 \cdot k_1)
{+} \nonumber \\&&{}+ \left[ \frac{\zeta_3 \cdot k_1}{x_3 - x_1} \cdot
 \frac{\zeta_4 \cdot k_3}{x_4 - x_3} 
+ \left( \frac{\zeta_3 \cdot k_2}{x_3 - x_2} {+} \frac{\zeta_3 \cdot
k_1}{x_3 - x_1}  {+} \frac{\zeta_3 \cdot k_4}{x_3 - x_4}
\right)
 \frac{\zeta_4 \cdot k_1}{x_4 - x_1} \right] (\zeta_5 \cdot k_2)
{-} \nonumber \\&&{}- 
\left[ \frac{\zeta_3 \cdot k_1}{x_3 - x_1} \cdot \frac{\zeta_4 \cdot
k_2}{x_4 - x_2} \right.
-
\left. \frac{\zeta_3 \cdot k_2}{x_3 - x_2} \cdot \frac{\zeta_4 \cdot
k_1}{x_4 - x_1} \right] (\zeta_5 \cdot k_3),
\label{U12345}
\\
V_{12345} & = & - \biggl[    \left( \frac{\zeta_2 \cdot
k_1}{x_2 - x_1} + \frac{\zeta_2 \cdot k_3}{x_2 - x_3} 
 + \frac{\zeta_2 \cdot k_4}{x_2 - x_4} \right) \left(
\frac{\zeta_3 \cdot k_2}{x_3 - x_2} + \frac{\zeta_3 \cdot k_1}{x_3 -
x_1} 
 + \frac{\zeta_3 \cdot k_4}{x_3 - x_4} \right) -
\nonumber \\
 & & \hphantom{ - \biggl[}
-  \frac{\zeta_2 \cdot k_3}{x_2 - x_3} \cdot \frac{\zeta_3
  \cdot k_2}{x_3 - x_2} \biggr]
(\zeta_5 \cdot k_1) -
\label{V12345}
 \\&&{}- \left[   
\left( \frac{\zeta_3 \cdot k_2}{x_3 - x_2} {+} \frac{\zeta_3 \cdot
k_1}{x_3 - x_1}  {+} \frac{\zeta_3 \cdot k_4}{x_3 - x_4}
\right) \frac{\zeta_2 \cdot k_1}{x_2 - x_1} {+}  \frac{\zeta_2
\cdot k_3}{x_2 - x_3} \cdot \frac{\zeta_3 \cdot k_1}{x_3 -
x_1} \right] (\zeta_5 \cdot k_2) -
\nonumber \\&&{}- \left[  
\frac{\zeta_2 \cdot k_1}{x_2 - x_1} \cdot \frac{\zeta_3 \cdot k_2}{x_3
- x_2} + \left( \frac{\zeta_2 \cdot k_1}{x_2 - x_1} {+} \frac{\zeta_2
\cdot k_3}{x_2 - x_3} {+} \frac{\zeta_2 \cdot k_4}{x_2 - x_4} \right)
\frac{\zeta_3 \cdot k_1}{x_3 - x_1} \right] (\zeta_5 \cdot k_3)\,,
\nonumber
\\
W_{1234} & = & \left[  \frac{\zeta_3 \cdot k_2}{x_3
- x_2} \cdot \frac{\zeta_4 \cdot k_3}{x_4 - x_3} +\! \left(
\frac{\zeta_3 \cdot k_2}{x_3 - x_2} + \frac{\zeta_3 \cdot k_1}{x_3 -
x_1}  + \frac{\zeta_3 \cdot k_4}{x_3 - x_4} \right)
 \frac{\zeta_4 \cdot k_2}{x_4 - x_2} \right] \frac{\zeta_2 \cdot
 k_1}{x_2 - x_1} +
\nonumber \\&&{}+ 
\left[   \frac{\zeta_3 \cdot k_1}{x_3 - x_1} \cdot
\frac{\zeta_4 \cdot k_3}{x_4 - x_3}
- \left( \frac{\zeta_3 \cdot k_2}{x_3 - x_2} + \frac{\zeta_3 \cdot
 k_1}{x_3 - x_1}
 + \frac{\zeta_3 \cdot k_4}{x_3 - x_4} \right)
 \frac{\zeta_4 \cdot k_1}{x_4 - x_1} \right] \times
\nonumber \\ & &
{}\times \left( \frac{\zeta_2 \cdot k_1}{x_2 - x_1} + \frac{\zeta_2
\cdot k_3}{x_2 - x_3} \right.
\left. + \frac{\zeta_2 \cdot k_4}{x_2 - x_4} \right) -
\nonumber \\ & 
 &{}-
\left[   \frac{\zeta_3 \cdot k_1}{x_3 - x_1} \cdot
\frac{\zeta_4 \cdot k_2}{x_4 - x_2}
-  \frac{\zeta_3 \cdot k_2}{x_3 - x_2} \cdot \frac{\zeta_4 \cdot
 k_1}{x_4 - x_1}
   \right] \frac{\zeta_2 \cdot k_3}{x_2 - x_3}
\label{W1234}
\end{eqnarray}
and
\begin{eqnarray}
Z_{1234} & = & \biggl[  \left( \frac{\zeta_2 \cdot
k_1}{x_2 - x_1} + \frac{\zeta_2 \cdot k_3}{x_2 - x_3}
 + \frac{\zeta_2 \cdot k_4}{x_2 - x_4} \right) \left(
\frac{\zeta_3 \cdot k_2}{x_3 - x_2} + \frac{\zeta_3 \cdot k_1}{x_3 -
x_1}  + \frac{\zeta_3 \cdot k_4}{x_3 - x_4} \right)-
\nonumber \\ & & 
\hphantom{ \biggl[ } 
-  \frac{\zeta_2 \cdot k_3}{x_2 - x_3} \cdot
\frac{\zeta_3 \cdot k_2}{x_3 - x_2} \biggr] \left( \frac{\zeta_1 \cdot
k_2}{x_1 - x_2} + \frac{\zeta_1 \cdot k_3}{x_1 - x_3} 
 + \frac{\zeta_1 \cdot k_4}{x_1 - x_4} \right) -
\label{Z1234}\\ && {}-
  \left[  
\left( \frac{\zeta_3 \cdot k_2}{x_3 - x_2} + \frac{\zeta_3 \cdot
k_1}{x_3 - x_1}  {+} \frac{\zeta_3 \cdot k_4}{x_3 - x_4}
\right) \frac{\zeta_2 \cdot k_1}{x_2 - x_1} {+}  \frac{\zeta_2
\cdot k_3}{x_2 - x_3} \cdot \frac{\zeta_3 \cdot k_1}{x_3 - x_1}
\right]
\frac{\zeta_1 \cdot k_2}{x_1 - x_2} {-}
\nonumber \\&&{}- 
\left[ \frac{\zeta_2 \cdot k_1}{x_2 - x_1} \cdot \frac{\zeta_3 \cdot
k_2}{x_3 - x_2} 
-\!
\left( \frac{\zeta_2 \cdot k_1}{x_2 - x_1} + \frac{\zeta_2 \cdot
k_3}{x_2 - x_3} \right.  \left. + \frac{\zeta_2 \cdot k_4}{x_2 - x_4}
\right)  \frac{\zeta_3 \cdot k_1}{x_3 - x_1} \right]
\frac{\zeta_1 \cdot k_3}{x_1 - x_3} \,.
\nonumber \end{eqnarray}

\item We summarize the results of this subsection by substituting
eqs.~(\ref{int1}) and~(\ref{int2}) in eq.~(\ref{A12345-2}). So the
expression for $A(1, 2, 3, 4, 5)$ turns into
\begin{eqnarray}
\lefteqn{ A(1, 2, 3, 4, 5)  = }&& 
\nonumber\\ &&
2 g^3 (2 \alpha') (x_4 - x_1) \int_{x_1}^{x_4}
d x_3 \int_{x_1}^{x_3} d x_2 \int d \theta_1 d \theta_2 d \theta_3
\prod_{i>j}^4 | x_i - x_j - \theta_i \theta_j |^{2 \alpha' k_i \cdot
k_j} \times
\nonumber \\ & &{}\times \biggl[   \Big\{  
-S_{12345} \theta_1 + S_{21345} \theta_2 + S_{32145} \theta_3
-(T_{12345}+T_{21345}+T_{32145}) \theta_1 \theta_2 \theta_3 \Big\}
-
\nonumber \\ & & 
\hphantom{{}\times \biggl[} 
- (2 \alpha') \biggl\{ U_{12345} \frac{\zeta_2 \cdot
 \zeta_1}{x_2 -
x_1}  + U_{13245} \frac{\zeta_3 \cdot \zeta_1}{x_3 - x_1} +
U_{23145} \frac{\zeta_3 \cdot \zeta_2}{x_3 - x_2} + V_{12345}
\frac{\zeta_4 \cdot \zeta_1}{x_4 - x_1} +
\nonumber \\ & & 
\hphantom{{}\times \biggl[- (2 \alpha') \biggl\{}  
+ V_{21345}
\frac{\zeta_4 \cdot \zeta_2}{x_4 - x_2} + V_{32145} \frac{\zeta_4
\cdot \zeta_3}{x_4 - x_3} + W_{1234} (\zeta_5 \cdot \zeta_1) +
W_{2134} (\zeta_5 \cdot \zeta_2) +
\nonumber \\ & &  
\hphantom{{}\times \biggl[- (2 \alpha') \biggl\{}  
+W_{3124} (\zeta_5 \cdot \zeta_3) + Z_{1234} (\zeta_5 \cdot \zeta_4)
  \biggr\} \theta_1 \theta_2 \theta_3   \biggr]\, ,
\label{A12345-3}
\end{eqnarray}   
where all the $S$, $T$, $U$, $V$, $W$ and $Z$ coefficients are bosonic
and known (they depend on $x_i$, $k_i$ and $\zeta_i$).
\end{enumerate}}

\subsection{Calculating the $\theta$ Grassmann integration and
deriving the final expression for $A(1, 2, 3, 4, 5)$}
\label{Grasstheta}

In this subsection we integrate the $\theta_i$ Grassmann variables and
we then fix $x_1$ and $x_4$, arriving to the final expression for
$A(1, 2, 3, 4, 5)$. As an intermediate step, to clear out any
confusion with the lengthy expression involved, we consider some
specific terms of the scattering amplitude, showing how their
coefficients are obtained.  
{\renewcommand{\theenumi}{\roman{enumi}}
\begin{enumerate} 
\item We start expanding the $\prod$ factor appearing
in~(\ref{A12345-3}) as
\begin{eqnarray}
\lefteqn{ \prod_{i>j}^4 |x_i - x_j - \theta_i \theta_j |^{2 \alpha'
k_i \cdot k_j} =}\qquad &&
\label{product}
\\ &&
\prod_{i>j}^4 (x_i - x_j )^{2 \alpha' k_i \cdot k_j} 
\left[   1 - 2 \alpha' 
\left\{ \frac{ (k_2 \cdot k_1) \theta_2 \theta_1}{x_2 - x_1} 
+ \frac{ (k_3 \cdot k_2) \theta_3 \theta_2}{x_3 - x_2}
 +\frac{(k_3 \cdot k_1)\theta_3\theta_1}{x_3-x_1}
 \right\}\right] .
\nonumber\end{eqnarray}
After substituting~(\ref{product}) in~(\ref{A12345-3}) and then
integrating over $\theta_1$, $\theta_2$ and $\theta_3$, we have that
\begin{eqnarray}
\lefteqn{ A(1, 2, 3, 4, 5)  = }&& 
\label{A12345-4}
\\ &&
2 g^3 (2 \alpha') (x_4 - x_1) \int_{x_1}^{x_4}
d x_3 \int_{x_1}^{x_3} d x_2 \prod_{i>j}^4 (x_i - x_j)^{2 \alpha' k_i
\cdot k_j} \times
\nonumber\\ &&{}\times\biggl[   \biggl\{  
T_{12345}+T_{21345}+T_{32145} +
\nonumber \\ & & 
\hphantom{{}\times\biggl[   \biggl\{  }
+(2 \alpha')  \left( \frac{(k_3
 \cdot k_2) S_{12345}}{x_3-x_2}
+ \frac{(k_3 \cdot k_1) S_{21345}}{x_3-x_1}
- \frac{(k_2 \cdot k_1) S_{32145}}{x_2-x_1} \right) \biggr\}+
\nonumber \\ & & 
\hphantom{{}\times\biggl[}
+ (2 \alpha') \biggl\{ U_{12345} \frac{\zeta_2 \cdot \zeta_1}{x_2 -x_1}
+ U_{13245} \frac{\zeta_3 \cdot \zeta_1}{x_3 - x_1}
+ U_{23145} \frac{\zeta_3 \cdot \zeta_2}{x_3 - x_2} + 
\nonumber\\ &&
\hphantom{{}\times\biggl[+ (2 \alpha') \biggl\{}
+V_{12345}
\frac{\zeta_4 \cdot \zeta_1}{x_4 - x_1} + V_{21345} \frac{\zeta_4
\cdot \zeta_2}{x_4 - x_2} + V_{32145} \frac{\zeta_4 \cdot \zeta_3}{x_4
- x_3}
\nonumber \\ & & 
\hphantom{{}\times\biggl[+ (2 \alpha') \biggl\{} 
 + W_{1234} (\zeta_5 \cdot \zeta_1)
+ W_{2134} (\zeta_5 \cdot \zeta_2) + W_{3124} (\zeta_5 \cdot \zeta_3)
+ Z_{1234} (\zeta_5 \cdot \zeta_4)   \biggr\} \biggr] .
\nonumber\end{eqnarray}

\item Now, using the expressions for the $S$, $T$, $U$, $V$, $W$ and
$Z$ known coefficients, given in the previous subsection, we can see
in~(\ref{A12345-4}) that
\begin{itemize}
\item The first curly bracket is the one responsible for terms of the
type
\begin{equation}
(\zeta_i \cdot \zeta_j)(\zeta_k \cdot \zeta_l)(\zeta_m \cdot k_n)
\times \{ \mbox{kinematic factor} \}\,,
\label{type1}
\end{equation}
in the scattering amplitude.
\item The second curly bracket is the one responsible for the
remaining terms of the scattering amplitude, which are of the type
\begin{equation}
(\zeta_i \cdot \zeta_j)(\zeta_k \cdot k_l)(\zeta_m \cdot k_n) (\zeta_p
\cdot k_q) \times \{ \mbox{kinematic factor} \}\,.
\label{type2}
\end{equation}
\end{itemize}
We will denote the resulting integrals of each curly bracket as
$I_{(\zeta \cdot \zeta)^2 (\zeta \cdot k)}$ and $I_{(\zeta \cdot
\zeta)(\zeta \cdot k)^3}$, respectively.  As an example, we will look
for the scattering amplitude terms which contribute as $(\zeta_1 \cdot
\zeta_2)(\zeta_3 \cdot \zeta_4)(\zeta_5 \cdot k_i)$.  They will be
found only in the first curly bracket, as part of the $I_{(\zeta \cdot
\zeta)^2 (\zeta \cdot k)}$ integral.  It turns out that these terms
only appear in four of the six $S$ and $T$ coefficients, in the
following way
\begin{eqnarray}
S_{12345} \rightarrow + \frac{(\zeta_1 \cdot \zeta_2)(\zeta_3 \cdot
\zeta_4)} {(x_2-x_1)(x_4-x_3)} \cdot (\zeta_5 \cdot k_1) 
\nonumber \\
S_{21345} \rightarrow - \frac{(\zeta_1 \cdot \zeta_2)(\zeta_3 \cdot
\zeta_4)} {(x_2-x_1)(x_4-x_3)} \cdot (\zeta_5 \cdot k_2) 
\nonumber \\
S_{32145} \rightarrow - \frac{(\zeta_1 \cdot \zeta_2)(\zeta_3 \cdot
\zeta_4)} {(x_2-x_1)(x_4-x_3)} \cdot (\zeta_5 \cdot k_3) 
\nonumber \\
T_{32145} \rightarrow - \frac{(\zeta_1 \cdot \zeta_2)(\zeta_3 \cdot
\zeta_4)} {(x_2-x_1)^2} \cdot \frac{(\zeta_5 \cdot k_3)}{x_4-x_3} \,.
\label{Sts}
\end{eqnarray} 
So their contribution to $A(1, 2, 3, 4, 5)$ is given by the integral
\begin{eqnarray}
\lefteqn{ I_{(\zeta_1 \cdot \zeta_2)(\zeta_3 \cdot \zeta_4)(\zeta_5
    \cdot k_i)}  = }&&
\nonumber\\ && 
2 g^3 (2 \alpha') (x_4 - x_1) \int_{x_1}^{x_4} d x_3
\int_{x_1}^{x_3} d x_2 \prod_{i>j}^4 (x_i - x_j)^{2 \alpha' k_i \cdot
k_j} \frac{(\zeta_1 \cdot \zeta_2)(\zeta_3 \cdot
\zeta_4)}{(x_2-x_1)(x_4-x_3)}\times
\nonumber \\&&{}\times  \biggl\{   (2 \alpha') (k_2 \cdot k_3)
 \frac{(\zeta_5 \cdot
k_1)}{x_3-x_2}
- (2 \alpha') (k_3 \cdot k_1) \frac{(\zeta_5 \cdot k_2)}{x_3-x_1}
+ (2 \alpha') (k_1 \cdot k_2) \frac{(\zeta_5 \cdot k_3)}{x_2-x_1}
\nonumber\\ & & 
\hphantom{{}\times  \biggl\{}
-  \frac{(\zeta_5 \cdot k_3)}{x_2-x_1} \biggr\}\,,
\label{I1}
\end{eqnarray}
which can then be written as
\begin{eqnarray}
 I_{(\zeta_1 \cdot \zeta_2)(\zeta_3 \cdot \zeta_4)(\zeta_5
    \cdot k_i)}  &=&  2 g^3 (2 \alpha') (x_4 - x_1) (\zeta_1 \cdot
    \zeta_2) (\zeta_3 \cdot \zeta_4)\times
\nonumber \\&&{}\times  \biggl[   2 \alpha' L_2 \cdot (k_2
 \cdot k_3) (\zeta_5
\cdot k_1) - 2 \alpha' L_3 \cdot (k_1 \cdot k_3) (\zeta_5 \cdot k_2) +
\nonumber\\&&
\hphantom{{}\times  \biggl[} 
+(2 \alpha' (k_1 \cdot k_2)-1) L_5 \cdot (\zeta_5 \cdot k_3)  
\biggr]\,. 
\label{termo1}
\end{eqnarray}
Here, $L_2$, $L_3$ and $L_5$ are part of a list of known kinematic
factors (double integrals which depend in $\alpha'$ and the momenta
$k_i$) given in subsection~\ref{kin1} of
appendix~\ref{kinematic}.\footnote{In subsection~\ref{kin1}, $x_1$ and
$x_4$ have already been fixed to $0$ and $1$, respectively.}  We now
reconsider eq.~(\ref{A12345-4}) and fix there $x_1=0$ and $x_4=1$.
Carrying over there the same procedure already done in this subsection
for terms of the type $(\zeta_1 \cdot \zeta_2) (\zeta_3 \cdot
\zeta_4)(\zeta_5 \cdot k_i) \times \{\mbox{kinematic factor}\}$ , but
for every term of the type~(\ref{type1}) and every term of the
type~(\ref{type2}), we finally arrive to five-point amplitude:
\begin{eqnarray}
\lefteqn{ A(1, 2, 3, 4, 5)  =} && 
\nonumber\\&&
2 g^3 (2 \alpha')^2 \biggl[
  (\zeta_1 \cdot \zeta_2)(\zeta_3 \cdot \zeta_4) 
      \Bigl\{   (\zeta_5 \cdot k_1)(k_2 \cdot k_3) L_2 
     - (\zeta_5 \cdot k_2)(k_1 \cdot k_3) L_3 +
\nonumber \\ & & 
\hphantom{2 g^3 (2 \alpha')^2 \Bigl[ (\zeta_1 \cdot \zeta_2)(\zeta_3
  \cdot \zeta_4) \Bigl\{}
+ (\zeta_5 \cdot k_3)
\left( (k_2 \cdot k_4){L_3}'+(k_2 \cdot k_3)L_2 \right)
  \Bigr\}+
\nonumber \\ & & 
\hphantom{2 g^3 (2 \alpha')^2 \Bigl[}
+ (\zeta_1 \cdot \zeta_3)(\zeta_2 \cdot \zeta_4) \Bigl\{
    -(\zeta_5 \cdot k_1)(k_2 \cdot k_3) L_7 +
\nonumber\\ && 
\hphantom{2 g^3 (2 \alpha')^2 \Bigl[+ (\zeta_1 \cdot \zeta_3)(\zeta_2
    \cdot \zeta_4) \Bigl\{} 
+  (\zeta_5 \cdot k_2) \left( (k_3 \cdot k_4){L_1}'-(k_2\cdot k_3)L_7
  \right) -
\nonumber \\
 & & 
\hphantom{2 g^3 (2 \alpha')^2 \Bigl[+ (\zeta_1 \cdot \zeta_3)(\zeta_2
    \cdot \zeta_4) \Bigl\{}  
- (\zeta_5 \cdot k_3) (k_1 \cdot k_2)L_1   \Bigr\}+
\nonumber \\ & & 
\hphantom{2 g^3 (2 \alpha')^2 \Bigl[}
+ (\zeta_1 \cdot \zeta_4)(\zeta_2 \cdot \zeta_3)
\Bigl\{   (\zeta_5 \cdot k_1)   \left(   (k_3
\cdot k_4){K_4}'-(k_1 \cdot k_3)K_5  
\right) -
\nonumber\\ & & 
\hphantom{2 g^3 (2 \alpha')^2 \Bigl[+ (\zeta_1 \cdot \zeta_4)(\zeta_2
    \cdot \zeta_3) \Bigl\{ }
- (\zeta_5 \cdot k_2)(k_1 \cdot k_3) K_5
+ (\zeta_5 \cdot k_3)(k_1 \cdot k_2) K_4   \Bigr\}+
\nonumber \\ & & 
\hphantom{2 g^3 (2 \alpha')^2 \Bigl[}
+ (\zeta_2 \cdot \zeta_3)
\Bigl\{   (\zeta_5 \cdot k_1)  \left(  
(\zeta_1 \cdot k_2)(\zeta_4 \cdot k_3) L_2
- (\zeta_1 \cdot k_3)(\zeta_4 \cdot k_2) L_7   \right)+
\nonumber \\ & & 
\hphantom{2 g^3 (2 \alpha')^2 \Bigl[+ (\zeta_2 \cdot \zeta_3)\Bigl\{ }
+ (\zeta_5 \cdot k_2)
\Bigl(   (\zeta_1 \cdot k_3)(\zeta_4 \cdot k_1) K_5 + (\zeta_1
\cdot k_3)(\zeta_4 \cdot k_3) {L_4}' + 
\nonumber\\ &&
\hphantom{2 g^3 (2 \alpha')^2 \Bigl[+ (\zeta_2 \cdot \zeta_3)\Bigl\{ +
    (\zeta_5 \cdot k_2)\Bigl(} +(\zeta_1 \cdot k_2)(\zeta_4 \cdot k_3)
    L_2 + (\zeta_1 \cdot k_4)(\zeta_4 \cdot k_3) {K_4}' \Bigr)-
\nonumber \\ & & 
\hphantom{2 g^3 (2 \alpha')^2 \Bigl[+ (\zeta_2 \cdot \zeta_3)\Bigl\{}
- (\zeta_5 \cdot k_3)
\Bigl(   (\zeta_1 \cdot k_2)(\zeta_4 \cdot k_1) K_4 + (\zeta_1
 \cdot k_3)(\zeta_4 \cdot k_2) L_7 + 
\nonumber\\ &&
\hphantom{2 g^3 (2 \alpha')^2 \Bigl[+ (\zeta_2 \cdot \zeta_3)\Bigl\{-
    (\zeta_5 \cdot k_3)\Bigl(} 
+(\zeta_1 \cdot k_2)(\zeta_4 \cdot
 k_2) L_4 +
(\zeta_1 \cdot k_4)(\zeta_4 \cdot k_2) {K_5}'
  \Bigr)     \Bigr\}+
\nonumber \\ & & 
\hphantom{2 g^3 (2 \alpha')^2 \Bigl[}
+ (\zeta_1 \cdot \zeta_4)
\Bigl\{   (\zeta_5 \cdot k_2)  \left(  
(\zeta_2 \cdot k_1)(\zeta_3 \cdot k_4) K_2 - (\zeta_2
\cdot k_4)(\zeta_3 \cdot k_1) K_3   \right)-
\nonumber \\ & & 
\hphantom{2 g^3 (2 \alpha')^2 \Bigl[+ (\zeta_1 \cdot \zeta_4)
\Bigl\{}
- (\zeta_5 \cdot k_1)
\Bigl(   (\zeta_2 \cdot k_3)(\zeta_3 \cdot k_4) {K_4}' 
 - (\zeta_2 \cdot k_4)(\zeta_3 \cdot k_2) {K_5}' + 
\nonumber\\ &&
\hphantom{2 g^3 (2 \alpha')^2 \Bigl[+ (\zeta_1 \cdot \zeta_4)
\Bigl\{- (\zeta_5 \cdot k_1)\Bigl(}
+(\zeta_2 \cdot
 k_4)(\zeta_3 \cdot k_4) {K_1}' -
(\zeta_2
 \cdot k_1)(\zeta_3 \cdot k_4) {K_2}
  \Bigr)+
\nonumber \\ & & 
\hphantom{2 g^3 (2 \alpha')^2 \Bigl[}
+ (\zeta_5 \cdot k_4)
\Bigl(   (\zeta_2 \cdot k_1)(\zeta_3 \cdot k_2) K_4   +
(\zeta_2 \cdot k_1)(\zeta_3 \cdot k_1) K_1-
\nonumber\\ &&
\hphantom{2 g^3 (2 \alpha')^2 \Bigl[+ (\zeta_5 \cdot k_4)
\Bigl(}
- (\zeta_2 \cdot k_3)(\zeta_3 \cdot k_1) K_5 -
(\zeta_2 \cdot k_4)(\zeta_3 \cdot k_1) {K_3}
  \Bigr)     \Bigr\}+
\nonumber \\ & & 
\hphantom{2 g^3 (2 \alpha')^2 \Bigl[}
+  \left( \mbox{cyclic permutations of
indexes (1,2,3,4,5)} \right)   \biggr]\, .
\label{amplitude}
\end{eqnarray}
The $K_i$, ${K_i}'$, $L_i$ and ${L_i}'$ kinematic factors in this
formula are all detailed in subsection~\ref{kin1} of
appendix~\ref{kinematic}. Once again, they depend on $\alpha'$ and the
momenta $k_i$ (they do not depend on the polarizations $\zeta_i$).\\
In deriving~(\ref{amplitude}) from~(\ref{A12345-4}), we have used the
cyclic property satisfied by the open superstring tree level
amplitudes, mentioned in section~\ref{review}.\\ The complete on-shell
five gluon tree amplitude, ${\cal A}^{(5)}$, would be obtained from
eq.~(\ref{amplitude}) by using the rule given in
eq.~(\ref{general-amplitude}).\\ Expression~(\ref{amplitude}) is exact
in the sense that there is no $\alpha'$ expansion in it: besides the
$(2 \alpha')^2$ factor in the beginning, the rest of the $\alpha'$
dependence comes all in the kinematic factors.  These factors may be
expanded in an $\alpha'$ series to any desired order.  In
appendix~\ref{kinematic}, an $\alpha'$ expansion of these factors is
done up to linear terms in $\alpha'$, which is enough information to
derive, afterwards, the tree level scattering amplitude up to ${\cal
O}(\alpha'^3)$ terms, as presented in subsection~\ref{5-point}, in
eq.~(\ref{A5final}).\\ At a first look to eqs.~(\ref{termo1})
and~(\ref{amplitude}) there may seem to be some trouble when comparing
the $(\zeta_1 \cdot \zeta_2) (\zeta_3 \cdot \zeta_4)(\zeta_5 \cdot
k_i)$ terms of each, because for them to be equal it would be
necessary to have
\begin{equation}
\left( 2 \alpha' (k_1 \cdot k_2)-1  \right) L_5 = 2
\alpha' \left(  (k_2 \cdot k_4){L_3}'+(k_2 \cdot k_3)L_2
  \right).
\end{equation} 
This last formula is valid indeed and it is part of a set of relations
among the kinematic factors contained in subsection~\ref{kin2} of
appendix~\ref{kinematic}.
\end{enumerate}}

\section{Final remarks and conclusions}

In this paper we have derived the open superstring five-point
amplitude of massless bosons (eq.~(\ref{amplitude})).  We have also
derived its $\alpha'$ expansion up to ${\cal O}({\alpha'}^3)$ terms
(eq.~(\ref{A5final})), finding explicit expressions for the first
three non zero contributions to this expan\-sion: $A^{(0)}(1, 2, 3, 4$,
$5)$, $A^{(2)}(1, 2, 3, 4, 5)$ and $A^{(3)}(1, 2, 3, 4,
5)$\footnote{The $A^{(1)}(1, 2, 3, 4, 5)$ term, that would give the
order one contribution in $\alpha'$, turns out to be exactly zero.}
(given respectively in eqs.~(\ref{A(0)}),~(\ref{A(2)})
and~(\ref{A(3)})).  As far as we know this is the first time that the
on-shell five-point amplitude has been completely calculated up to
order three in $\alpha'$.  We have used the different order
contributions of the five-point amplitude to analyze the gluon
effective lagrangian up to ${\cal O}({\alpha'}^3)$ terms. The
$A^{(0)}(1, 2, 3, 4, 5)$ and $A^{(2)}(1, 2, 3, 4, 5)$ contributions
confirm the gluon effective lagrangian up to ${\cal O}({\alpha'}^2)$
terms (given in eq.~(\ref{L2})): this has been considered in
appendix~\ref{appD} (see the final paragraph).

\looseness=1 The ${\cal O}({\alpha'}^3)$ contribution to the effective
lagrangian, ${\cal L}_{(3)}$, deserves some special attention, since
three non-equivalent versions of it have been published in the
literature (\cite{Kitazawa:1987xj,Refolli:2001df}
and~\cite{Koerber:2001uu}).\footnote{In~\cite{Koerber:2001uu,
Koerber:2001hk} a detailed comparison among the three versions has
been done, concluding their non equivalence.} The three versions agree
in the $D^2 F^4$ terms and only differ in the $F^5$
terms~\cite{Koerber:2001uu, Koerber:2001hk}.  Now, the $D^2 F^4$ terms
can be directly derived from the four-point amplitude ${\cal A}^{(4)}$
expanded up to ${\cal O}({\alpha'}^3)$ terms~\cite{Bilal:2001} (see
eqs.~(\ref{A4}) and~(\ref{gammas})). In fact, we have confirmed in
appendix~\ref{appD} that the $D^2 F^4$ terms of ${\cal L}_{(3)}$,
given in eq.~(\ref{L3}), give a four-point amplitude that agrees with
the corresponding one obtained in Open Superstring theory (see
eqs.~(\ref{AL4}) and~(\ref{A4f})). So, from a scattering amplitude
approach, the real importance in the computation of a five-point
amplitude lies in the determination of the $F^5$ terms.  $A^{(3)}(1,
2, 3, 4, 5)$ turns out to be of crucial importance in this last
purpose. Using its expression we have found complete agreement with
the five-point amplitude coming from the ${\cal L}_{(3)}$ lagrangian
in eq.~(\ref{L3}), due to the authors of~\cite{Koerber:2001uu} (see
appendix~\ref{appD}). It should be mentioned that this lagrangian had
also passed a test done in~\cite{Koerber:2001hk} by the same authors.

So, our main conclusion agrees completely with the one
in~\cite{Koerber:2001uu}, namely, that the bosonic terms of the
non-abelian Born-Infeld supersymmetric lagrangian, up to ${\cal
O}({\alpha'}^3)$, are given~by\footnote{There is a misprint in equation
(5.1) of~\cite{Koerber:2001hk}, which contains the complete effective
lagrangian up
to ${\cal O}(\alpha'^3)$ corrections: some terms have an additional
factor $1/4$.}
\begin{eqnarray}
{\cal L} &=&- \frac{1}{4} tr\left(F_{{\mu}_1 {\mu}_2} F^{{\mu}_1
{\mu}_2} \right) +
\nonumber \\&&
{}+ \left(2 \pi \alpha' \right)^2 g^2
\tr \biggl[ \frac{1}{24}
F_{{\mu}_1}^{\;\,{\mu}_2}\,F_{{\mu}_2}^{\;\,{\mu}_3}
\,F_{{\mu}_3}^{\;\,{\mu}_4}\,F_{{\mu}_4}^{\;\,{\mu}_1} +
\frac{1}{12}\,F_{{\mu}_1}^{\;\,{\mu}_2}\,F_{{\mu}_2}^{\;\,{\mu}_3}
\,F_{{\mu}_4}^{\;\,{\mu}_1}\,F_{{\mu}_3}^{\;\,{\mu}_4}-
\nonumber \\ & & 
\hphantom{{}+ \left(2 \pi \alpha' \right)^2 g^2\tr \biggl[}
-\frac{1}{48}\,F_{{\mu}_1}^{\;\,{\mu}_2}\,F_{{\mu}_2}^{\;\,{\mu}_1}
\,F_{{\mu}_3}^{\;\,{\mu}_4}\,F_{{\mu}_4}^{\;\,{\mu}_3} -
\frac{1}{96}\,F_{{\mu}_1}^{\;\,{\mu}_2}\,F_{{\mu}_3}^{\;\,{\mu}_4}
\,F_{{\mu}_2}^{\;\,{\mu}_1}\,F_{{\mu}_4}^{\;\,{\mu}_3}\biggr]-
 \nonumber \\&&
{}- (2\,\alpha')^3\,2\,\zeta(3)\tr
 \biggl[{{\;}\atop{\;}}
i\,g^3\,F_{{\mu}_1}^{\;\,{\mu}_2}\,F_{{\mu}_2}^{\;\,{\mu}_3}
\,F_{{\mu}_3}^{\;\,{\mu}_4}\,F_{{\mu}_5}^{\;\,{\mu}_1}\,
F_{{\mu}_4}^{\;\,{\mu}_5}+
i\,g^3\,F_{{\mu}_1}^{\;\,{\mu}_2}\,F_{{\mu}_4}^{\;\,{\mu}_5}
\,F_{{\mu}_2}^{\;\,{\mu}_3}\,F_{{\mu}_5}^{\;\,{\mu}_1}\,
F_{{\mu}_3}^{\;\,{\mu}_4}-
\nonumber \\ & & 
\hphantom{{}- (2\,\alpha')^3\,2\,\zeta(3)\tr \biggl[} 
-  \frac{i\,g^3}{2}
\,F_{{\mu}_1}^{\;\,{\mu}_2}\,F_{{\mu}_2}^{\;\,{\mu}_3}
\,F_{{\mu}_4}^{\;\,{\mu}_5}\,F_{{\mu}_3}^{\;\,{\mu}_1}\,
F_{{\mu}_5}^{\;\,{\mu}_4}-
\nonumber\\ &&
\hphantom{{}- (2\,\alpha')^3\,2\,\zeta(3)\tr \biggl[} 
-g^2\,F_{{\mu}_1}^{\;\,{\mu}_2}\left(D^{{\mu}_1}
\,F_{{\mu}_3}^{\;\,{\mu}_4}\right)\left(D_{{\mu}_5}
\,F_{{\mu}_2}^{\;\,{\mu}_3}\right)\,F_{{\mu}_4}^{\;\,{\mu}_5}+
\nonumber \\ & & 
\hphantom{{}- (2\,\alpha')^3\,2\,\zeta(3)\tr \biggl[} 
+  \frac{g^2}{2}
\left(D^{{\mu}_1}\,F_{{\mu}_2}^{\;\,{\mu}_3}\right)
\left(D_{{\mu}_1}\,F_{{\mu}_3}^{\;\,{\mu}_4}\right)
\,F_{{\mu}_5}^{\;\,{\mu}_2}\,F_{{\mu}_4}^{\;\,{\mu}_5}+
\nonumber\\ &&
\hphantom{{}- (2\,\alpha')^3\,2\,\zeta(3)\tr \biggl[} 
+\frac{g^2}{2}\left(D^{{\mu}_1}\,F_{{\mu}_2}^{\;\,{\mu}_3}\right)
\,F_{{\mu}_5}^{\;\,{\mu}_2}\left(D_{{\mu}_1}
\,F_{{\mu}_3}^{\;\,{\mu}_4}\right)\,F_{{\mu}_4}^{\;\,{\mu}_5}-
\nonumber \\ & & 
\hphantom{{}- (2\,\alpha')^3\,2\,\zeta(3)\tr \biggl[} 
-  \frac{g^2}{8}\left(D^{{\mu}_1}\,F_{{\mu}_2}^{\;\,{\mu}_3}\right)
\,F_{{\mu}_4}^{\;\,{\mu}_5}\left(D_{{\mu}_1}
\,F_{{\mu}_3}^{\;\,{\mu}_2}\right)\,F_{{\mu}_5}^{\;\,{\mu}_4} +
\nonumber\\ &&
\hphantom{{}- (2\,\alpha')^3\,2\,\zeta(3)\tr \biggl[} 
+ g^2\,
\left(D_{{\mu}_5}\,F_{{\mu}_1}^{\;\,{\mu}_2}\right)
\,F_{{\mu}_3}^{\;\,{\mu}_4}\left(D^{{\mu}_1}
\,F_{{\mu}_2}^{\;\,{\mu}_3}\right)\,F_{{\mu}_4}^{\;\,{\mu}_5}\biggr]\,.
\label{Ltotal}
\end{eqnarray}

\acknowledgments

R.M.~would like to thank Osvaldo Chandia for many useful discussions
about string scattering amplitudes, from which this work
began. R.M.~would also like to thank Jos Vermaseren for useful
conversations and guidance in the use of the $harmpol$ package (which
uses FORM).  R.M.~acknowledges funding from CLAF in the first steps of
this work.  F.T.B.~acknowledges funding from CNPq and
F.R.M.~acknowledges funding from FAPESP.

\appendix

\section{Kinematic factors}
\label{kinematic}

\subsection{List of kinematic factors}
\label{kin1}
The $K_i$, ${K_i}'$, $L_i$ and ${L_i}'$ kinematic factors, present in
eq.~(\ref{amplitude}), are all defined as double integrals of the form
\begin{equation} 
\int_0^1 d x_3
\int_0^{x_3} d x_2 x_3^{2 \alpha' \alpha_{13}} (1-x_3)^{2 \alpha'
  \alpha_{34}} {x_2}^{2 \alpha' \alpha_{12}} (1-x_2)^{2 \alpha'
  \alpha_{24}} (x_3-x_2)^{2 \alpha' \alpha_{23}} \cdot
\left\{ 
{\kappa(x_2,x_3) \atop \lambda(x_2,x_3)}
\right\}, 
\label{double-int}
\end{equation} 
where
\begin{equation}
\alpha_{ij} = k_i \cdot k_j
\label{alphaij}
\end{equation}
and where the specific functions $\kappa(x_2,x_3)$ and
$\lambda(x_2,x_3)$, associated to each of the kinematic factors, are
given in  table~\ref{table1}

{\setlength{\tabcolsep}{2pt}
\TABLE[b]{\begin{tabular}{|c|c|c|c|}\hline
$\kappa(x_2,x_3)$ & Kinematic Factor & $\lambda(x_2,x_3)$ &
Kinematic Factor 
\\\hline
 $1/\{x_2 \cdot x_3\}$ &
$K_1$ & $1/\{x_2 \cdot (1-x_2) \cdot x_3\}$ & $L_1$ 
\\
 $1/\{(1-x_2)
\cdot (1-x_3)\}$ & ${K_1}'$ & $1/\{(1-x_2) \cdot x_3 \cdot
(1-x_3)\}$ & ${L_1}'$ 
\\
 $1/\{x_2 \cdot (1-x_3)\}$ & ${K_2}$ &
$1/\{x_2 \cdot (1-x_3) \cdot (x_3-x_2)\}$ & $L_2$ 
\\
 $1/\{(1-x_2)
\cdot x_3\}$ & ${K_3}$ & $1/\{x_2 \cdot x_3 \cdot (1-x_3)\}$ & $L_3$
\\
 $1/\{x_2 \cdot (x_3-x_2)\}$ & ${K_4}$ & $1/\{x_2 \cdot (1-x_2)
\cdot (1-x_3)\}$ & ${L_3}'$ 
\\
 $1/\{(1-x_3) \cdot (x_3-x_2)\}$ &
${K_4}'$ & $1/\{x_2 \cdot (1-x_2) \cdot (x_3-x_2) \}$ & $L_4$ 
\\
$1/\{x_3 \cdot (x_3-x_2)\}$ & ${K_5}$ & $1/\{x_3 \cdot (1-x_3) \cdot
(x_3-x_2)\}$ & ${L_4}'$ 
\\
 $1/\{(1-x_2) \cdot (x_3-x_2)\}$ & ${K_5}'$
& $1/\{{x_2}^2 \cdot (1-x_3)\}$ & $L_5$ 
\\
 $1/\{{(x_3-x_2)}^2 \}$ &
${K_6}$ & $1/\{(1-x_2) \cdot {x_3}^2\}$ & $L_6$ 
\\
 & &$1/\{(1-x_2) \cdot x_3 \cdot (x_3-x_2)\}$ & $L_7$
\\\hline
\end{tabular}\caption{List of kinematic factors.}\label{table1}}}

\subsection{Relations among the kinematic factors}
\label{kin2}

These integrals are not all independent.  There is a \emph{duality}
operation $\star$ that interchanges some of the $\alpha_{ij}$ among
themselves as follows,
\begin{equation}
\alpha_{12} \leftrightarrow \alpha_{34}, \alpha_{13} \leftrightarrow
  \alpha_{24}\qquad
\mbox{and} \quad\alpha_{23} \leftrightarrow \alpha_{23}\,,
\label{duality1}
\end{equation}
and such that every \emph{primed} kinematic factor can be obtained
from the corresponding \emph{non primed} one by means of the
\emph{duality} operation:
\begin{equation} 
{K_i}' = \star \{K_i\} \,, \qquad{L_i}' = \star \{L_i\} \,.
\label{duality}
\end{equation}
The kinematic factors $K_2$, $K_3$, $K_6$, $L_2$ and $L_7$ are
self-dual:
\begin{equation}
{K_2} = \star \{K_2\} \,,\qquad {K_3} = \star \{K_3\} \,,\qquad {K_6}
= \star \{K_6\} \,,\qquad {L_2} = \star \{L_2\} \,, \qquad {L_7} = \star
\{L_7\} \,.
\label{self-duality}
\end{equation}
The formulas in eqs.~(\ref{duality}) and~(\ref{self-duality}) can
easily be proved by first changing the order of integration in the
double integrals and then doing the substitution $x_2 \rightarrow
1-{x_3}'$, $x_3 \rightarrow 1-{x_2}'$.  There are also some additional
relations which can be proved using integration by parts:
\begin{eqnarray}
\alpha_{34} K_2 & = & \alpha_{13} K_1 + \alpha_{23} K_4 
\nonumber \\
\alpha_{24} K_3 & = & \alpha_{12} K_1 - \alpha_{23} K_5 
\nonumber \\
(2 \alpha' \alpha_{23}-1) K_6 & = & 2 \alpha' \left(
\alpha_{34}{K_4}'-\alpha_{13}K_5 \right) 
\nonumber \\ \alpha_{13} L_1
& = & \alpha_{34} {L_3}' - \alpha_{23} {L_4}' 
\nonumber \\ (2 \alpha'
\alpha_{12}-1) L_5 & = & 2 \alpha' \left(
\alpha_{24}{L_3}'+\alpha_{23}L_2 \right) 
\nonumber \\ (2 \alpha'
\alpha_{13}-1) L_6 & = & 2 \alpha' \left(
\alpha_{34}{L_1}'-\alpha_{23}L_7 \right)\,.
\label{relations}
\end{eqnarray}

\subsection{Explicit expression for the kinematic factors
up to ${\cal O}(\alpha')$ terms}
\label{kin3}

A matter of quite interest is to have an $\alpha'$ series for all
these kinematic factors, since this allows to have the superstring
corrections to the Yang-Mills five-gluon tree amplitude, ${\cal
A}^{(5)}$, at different orders in $\alpha'$.  At least this is desired
to be done up to ${\cal O}(\alpha')$ terms, since this is enough to
obtain the effective lagrangian up to ${\cal O}(\alpha'^3)$.  When
looking for the $\alpha'$ expansion of the kinematic factors, some
care must be taken with some of the double integrals, because they
need to be regularized. In the list given in subsection~\ref{kin1}
this is necessary to be done with $K_6$ and $L_5$.  The third and
fifth relations in~(\ref{relations}) assume that this regularization
has been taken into account.  After this regularization has been done,
any of the kinematic factors listed in the tables may be calculated in
terms of Beta functions and the Hypergeometric function ${ }_3 F_{2}$,
as was done in~\cite{Kitazawa:1987xj}.
From this kind of result, in principle an $\alpha'$ series can be
obtained.
We prefer, instead, to do a formal $\alpha'$ expansion for each
kinematic factor in a different way. As an example we will show our
procedure for
\begin{equation}
L_1 = \int_0^1 d x_3 x_3^{2 \alpha' \alpha_{13}-1} (1-x_3)^{2 \alpha'
\alpha_{34}} \int_0^{x_3} d x_2
{x_2}^{2 \alpha' \alpha_{12}-1} (1-x_2)^{2 \alpha' \alpha_{24}-1}
  (x_3-x_2)^{2 \alpha' \alpha_{23}}
\label{L1-1}
\end{equation}
We start making the substitution $x_2 = u \cdot x_3$ in the inner
integral (so $x_3$ acts as a constant in it), leading to
\begin{equation}
L_1 = \int_0^1 d x_3 x_3^{2 \alpha' \rho-1} (1-x_3)^{2 \alpha'
\alpha_{34}} \int_0^{1} d u
u^{2 \alpha' \alpha_{12}-1} (1-u)^{2 \alpha' \alpha_{23}} (1-u x_3)^{2
 \alpha' \alpha_{24}-1} \,,
\label{L1-2}
\end{equation} 
where
\begin{equation}
\rho = \alpha_{12} + \alpha_{13} + \alpha_{23}\,.
\label{rho}
\end{equation}
Now, in~(\ref{L1-2}) $L_1$ may be written as
\begin{eqnarray}
L_1 = & &B( 2 \alpha' \rho, 2\alpha' \alpha_{34}+1) \cdot B( 2 \alpha'
\alpha_{12}, 2 \alpha' \alpha_{23}+1) +
\label{L1-3}
 \\&&{}+ \int_0^1 d x_3 x_3^{2 \alpha' \rho-1} (1 \underbrace{-x_3)^{2
 \alpha' \alpha_{34}}
\int_0^{1} d u u^{2 \alpha' \alpha_{12}-1} (1-}_{I} u)^{2 \alpha'
 \alpha_{23}}
\left\{ (1-u x_3)^{2 \alpha' \alpha_{24}-1} -1 \right\} 
\nonumber\end{eqnarray}
(where $B(\lambda, \mu)$ is the Euler Beta function), such that the
double integral $I$ does have a well defined power series in
$\alpha'$.  Since the $\alpha'$ expansion of the Beta functions that
appear in~(\ref{L1-3}) is known, the only thing that is missing is the
series in $\alpha'$ for the integral term, which can be written as
\begin{eqnarray}
\lefteqn{ I  =  \left\{ \int_0^1 d x_3 \int_0^1 \frac{du}{1-ux_3}
  \right\} +}&& 
\nonumber\\  &&
{}+(2 \alpha') \biggl[ \rho \left\{ \int_0^1 dx_3 ln(x_3) \int_0^1 du
\frac{1}{1-ux_3} \right\}+
 \alpha_{34} \left\{ \int_0^1 dx_3 ln(1-x_3) \int_0^1 du
\frac{1}{1-ux_3} \right\} + 
\nonumber\\ &&
\hphantom{{}+(2 \alpha') \biggl[}
+\alpha_{12} \left\{ \int_0^1 dx_3 \int_0^1
du \frac{ln(u)}{1-ux_3} \right\} +
  \alpha_{23} \left\{ \int_0^1 dx_3
\int_0^1 du \frac{ln(1-u)}{u(1-ux_3)} \right\} + 
\nonumber \\&&{}
\hphantom{{}+(2 \alpha') \biggl[}
+\alpha_{24} \left\{
\int_0^1 \frac{dx_3}{x_3} \int_0^1 du \frac{ln(1-ux_3)}{u(1-ux_3)}
\right\} \biggr] + {\cal O}({\alpha'}^2) \,. 
\nonumber \\
\label{L1-4}
\end{eqnarray} 
Each of the double integrals that appear in~(\ref{L1-4}) can be
calculated in terms of $\zeta(2) = \pi^2/6$ and $\zeta(3)$, where
$\zeta(z)$ is the Riemann Zeta function.\footnote{The interested
reader can find a very nice method to calculate all these integrals,
and more complicated ones, with the aid of \emph{Harmonic
Polylogarithms}~\cite{Remiddi:1999ew} and the \emph{harmpol} package,
that uses computer language FORM~\cite{Vermaseren:2000nd}.} The result
is the following:
\begin{equation} 
I = \frac{\pi^2}{6} + 2 \alpha' \left[   - \zeta(3) \rho -2
\zeta(3) \alpha_{34}
- \zeta(3) \alpha_{12} -2 \zeta(3) \alpha_{23}- 2 \zeta(3) \alpha_{24}
\right]
+ {\cal O}({\alpha'}^2) \,.
\label{L1-5}
\end{equation}
Now, using the $\alpha'$ expansion for $I$ and the one for the Beta
functions appearing in~(\ref{L1-3}), in that equation we finally have
that
\begin{eqnarray*}
L_1 & = & \frac{1}{(2 \alpha')^2} \left\{ \frac{1}{\rho \cdot
\alpha_{12}} \right\} + \frac{\pi^2}{6} \left\{ 1 -
\frac{\alpha_{34}}{\alpha_{12}} - \frac{\alpha_{23}}{\rho} \right\}+
\\&&+ 2 \zeta(3) \alpha' \left\{ {-}\rho {-}2 \alpha_{34} {-}
\alpha_{12} {-} 2 \alpha_{23} {-} 2 \alpha_{24} +
\frac{\alpha_{34}^2}{\alpha_{12}} {+} \frac{\alpha_{23}^2}{\rho} {+}
\frac{\alpha_{23} \cdot \alpha_{12}}{\rho} {+} \frac{\rho \cdot
\alpha_{34}}{\alpha_{12}} \right\} {+} {\cal O}(\alpha'^2).  
\end{eqnarray*}
Doing analogous procedures to this one and using the relations among
the kinematic factors, given in subsection~\ref{kin2}, we obtain the
following list:
\begin{eqnarray} 
K_1 & = & \frac{1}{(2 \alpha')^2} \left\{ \frac{1}{\rho \cdot
\alpha_{12}} \right\}
- \frac{\pi^2}{6}
\left\{\frac{\alpha_{34}}{\alpha_{12}} + \frac{\alpha_{23}}{\rho}
\right\} +
\nonumber \\&&{}+ 2 \zeta(3) \alpha' \left\{ - \alpha_{24} +
\frac{\alpha_{34} \cdot \rho}{\alpha_{12}} + \frac{\alpha_{12} \cdot
\alpha_{23}}{\rho} + \frac{\alpha_{34}^2}{\alpha_{12}} +
\frac{\alpha_{23}^2}{\rho} \right\} + {\cal O}(\alpha'^2) \,,
\label{K1final}
\\
{K_1}' & = & \frac{1}{(2 \alpha')^2} \left\{ \frac{1}{\phi \cdot
\alpha_{34}} \right\}
- \frac{\pi^2}{6}
\left\{\frac{\alpha_{12}}{\alpha_{34}} + \frac{\alpha_{23}}{\phi}
\right\} +
\nonumber \\&&{}+ 2 \zeta(3) \alpha' \left\{ - \alpha_{13} +
\frac{\alpha_{12} \cdot \phi}{\alpha_{34}} + \frac{\alpha_{34} \cdot
\alpha_{23}}{\phi} + \frac{\alpha_{12}^2}{\alpha_{34}} +
\frac{\alpha_{23}^2}{\phi} \right\} + {\cal O}(\alpha'^2) \,,
\label{K1primefinal}
\\
K_2 & = & \frac{1}{(2 \alpha')^2} \left\{ \frac{1}{\alpha_{12} \cdot
\alpha_{34}} \right\}
- \frac{\pi^2}{6}
\left\{\frac{\rho}{\alpha_{12}} + \frac{\alpha_{23}}{\alpha_{34}} +
\frac{\alpha_{24}}{\alpha_{34}} \right\} +
\nonumber \\&&{}+ 2 \zeta(3)
\alpha' \left\{ 2 \alpha_{24} {+} \frac{\alpha_{34} \cdot
\rho}{\alpha_{12}} {+} \frac{\rho^2}{\alpha_{12}} {+}
\frac{\alpha_{23}^2}{\alpha_{34}} {+ }\frac{\alpha_{24}^2}{\alpha_{34}}
{+} 2 \frac{\alpha_{23} \cdot \alpha_{24}}{\alpha_{34}} {+}
\frac{\alpha_{12} \cdot \alpha_{23}}{\alpha_{34}} {+} \frac{\alpha_{12}
\cdot \alpha_{24}}{\alpha_{34}} \right\} + 
\nonumber\\ &&
{}+{\cal O}(\alpha'^2) \,,
\label{K3final}
\\
K_3 & = & \frac{\pi^2}{6} - 2 \zeta(3) \alpha' \left\{ \rho + 2
\alpha_{23} + 2 \alpha_{34} + \alpha_{12} + \alpha_{24} \right\} +
{\cal O}(\alpha'^2) \,,
\label{K4final}
\\
K_4 & = & \frac{1}{(2 \alpha')^2} \left\{ \frac{1}{\rho \cdot
\alpha_{12}} + \frac{1}{\rho \cdot \alpha_{23}} \right\}
- \frac{\pi^2}{6}
\left\{\frac{\alpha_{24}}{\alpha_{23}} +
\frac{\alpha_{34}}{\alpha_{23}} + \frac{\alpha_{12}}{\rho} +
\frac{\alpha_{23}}{\rho} + \frac{\alpha_{34}}{\alpha_{12}} \right\}+
\nonumber \\&&
{}+ 2 \zeta(3) \alpha' \biggl\{ \alpha_{24} +
\frac{\alpha_{24}^2}{\alpha_{23}} + 2 \frac{\alpha_{34} \cdot
\alpha_{24}}{\alpha_{23}} + \frac{\alpha_{34} \cdot \rho}{\alpha_{12}}
+ 2 \frac{\alpha_{12} \cdot \alpha_{23}}{\rho} +
\frac{\alpha_{34}^2}{\alpha_{23}} {+} \frac{\alpha_{12}^2}{\rho} {+}
\frac{\rho \cdot \alpha_{24}}{\alpha_{23}} {+}
\nonumber \\&&
\hphantom{{}+ 2 \zeta(3) \alpha' \biggl\{}
+  \frac{\rho \cdot \alpha_{34}}{\alpha_{23}}
+ \frac{\alpha_{23}^2}{\rho} + \frac{\alpha_{34}^2}{\alpha_{12}}
\biggr\} + {\cal O}(\alpha'^2) \,,
\label{K5final}
\\
{K_4}' & = & \frac{1}{(2 \alpha')^2} \left\{ \frac{1}{\phi \cdot
\alpha_{34}} + \frac{1}{\phi \cdot \alpha_{23}} \right\}
- \frac{\pi^2}{6}
\left\{\frac{\alpha_{13}}{\alpha_{23}} +
\frac{\alpha_{12}}{\alpha_{23}} + \frac{\alpha_{34}}{\phi} +
\frac{\alpha_{23}}{\phi} + \frac{\alpha_{12}}{\alpha_{34}} \right\}+
\nonumber \\&&{}+ 2 \zeta(3) \alpha' \biggl\{ \alpha_{13} +
\frac{\alpha_{13}^2}{\alpha_{23}} + 2 \frac{\alpha_{12} \cdot
\alpha_{13}}{\alpha_{23}} + \frac{\alpha_{12} \cdot \phi}{\alpha_{34}}
+ 2 \frac{\alpha_{34} \cdot \alpha_{23}}{\phi} {+}
\frac{\alpha_{12}^2}{\alpha_{23}} + \frac{\alpha_{34}^2}{\phi}{ +}
\frac{\phi \cdot \alpha_{13}}{\alpha_{23}} {+}
\nonumber \\&&
\hphantom{{}+ 2 \zeta(3) \alpha' \biggl\{}
+ \frac{\phi \cdot \alpha_{12}}{\alpha_{23}}
+ \frac{\alpha_{23}^2}{\phi} + \frac{\alpha_{12}^2}{\alpha_{34}}
\biggr\} + {\cal O}(\alpha'^2) \,,
\label{K5primefinal}
\\
K_5 & = & \frac{1}{(2 \alpha')^2} \left\{ \frac{1}{\alpha_{23} \cdot
\rho} \right\}
- \frac{\pi^2}{6}
\left\{\frac{\alpha_{34}}{\alpha_{23}} + \frac{\alpha_{12}}{\rho} +
\frac{\alpha_{24}}{\alpha_{23}} \right\} +
\nonumber \\&&{}+ 2 \zeta(3)
\alpha' \left\{ 2 \alpha_{24} + \frac{\alpha_{12}^2}{\rho} +
\frac{\alpha_{34}^2}{\alpha_{23}} + \frac{\rho \cdot
\alpha_{34}}{\alpha_{23}} + \frac{\rho \cdot \alpha_{24}}{\alpha_{23}}
+ 2 \frac{\alpha_{34} \cdot \alpha_{24}}{\alpha_{23}} {+}
\frac{\alpha_{12} \cdot \alpha_{23}}{\rho} {+}
\frac{\alpha_{24}^2}{\alpha_{23}} \right\} {+ }
\nonumber\\ &&
{}+{\cal O}(\alpha'^2) \,,
\label{K6final}
\\
{K_5}' & = & \frac{1}{(2 \alpha')^2} \left\{ \frac{1}{\alpha_{23}
\cdot \phi} \right\}
- \frac{\pi^2}{6}
\left\{\frac{\alpha_{12}}{\alpha_{23}} + \frac{\alpha_{34}}{\phi} +
\frac{\alpha_{13}}{\alpha_{23}} \right\} +
\nonumber \\&&{}+ 2 \zeta(3)
\alpha' \left\{ 2 \alpha_{13} + \frac{\alpha_{34}^2}{\phi} +
\frac{\alpha_{12}^2}{\alpha_{23}} + \frac{\phi \cdot
\alpha_{12}}{\alpha_{23}} + \frac{\phi \cdot \alpha_{13}}{\alpha_{23}}
+ 2 \frac{\alpha_{12} \cdot \alpha_{13}}{\alpha_{23}} {+}
\frac{\alpha_{34} \cdot \alpha_{23}}{\phi} {+}
\frac{\alpha_{13}^2}{\alpha_{23}} \right\} {+} 
\nonumber\\ &&
{}+{\cal O}(\alpha'^2) \,,
\label{K6primefinal}
\\
L_1 & = & \frac{1}{(2 \alpha')^2} \left\{ \frac{1}{\rho \cdot
\alpha_{12}} \right\} + \frac{\pi^2}{6} \left\{ 1 -
\frac{\alpha_{34}}{\alpha_{12}} - \frac{\alpha_{23}}{\rho} \right\}+
\nonumber \\&&{}+ 2 \zeta(3) \alpha' \left\{ -\rho -2 \alpha_{34} -
\alpha_{12} - 2 \alpha_{23} - 2 \alpha_{24} +
\frac{\alpha_{34}^2}{\alpha_{12}} + \frac{\alpha_{23}^2}{\rho} {+}
\frac{\alpha_{23} \cdot \alpha_{12}}{\rho} + \frac{\rho \cdot
\alpha_{34}}{\alpha_{12}} \right\} {+} 
\nonumber\\ &&
{}+{\cal O}(\alpha'^2) \,, 
\label{L1final}
\\
{L_1}' & = & \frac{1}{(2 \alpha')^2} \left\{ \frac{1}{\phi \cdot
\alpha_{34}} \right\} + \frac{\pi^2}{6} \left\{ 1 -
\frac{\alpha_{12}}{\alpha_{34}} - \frac{\alpha_{23}}{\phi} \right\}+
\nonumber \\&&{}+ 2 \zeta(3) \alpha' \left\{ -\phi -2 \alpha_{12} -
\alpha_{34} - 2 \alpha_{23} - 2 \alpha_{13} +
\frac{\alpha_{12}^2}{\alpha_{34}} + \frac{\alpha_{23}^2}{\phi} +
\frac{\alpha_{23} \cdot \alpha_{34}}{\phi} {+} \frac{\phi \cdot
\alpha_{12}}{\alpha_{34}} \right\} {+} 
\nonumber\\ &&
{}+{\cal O}(\alpha'^2) \,, 
\label{L1primefinal}
\\
L_2 & = & \frac{1}{(2 \alpha')^2} \left\{ \frac{1}{\rho \cdot
\alpha_{12}} + \frac{1}{\phi \cdot \alpha_{34}} + \frac{1}{\rho \cdot
\alpha_{23}} + \frac{1}{\phi \cdot \alpha_{23}}
+ \frac{1}{\alpha_{12} \cdot \alpha_{34}} \right\} +
\frac{\pi^2}{6}
\biggl\{ 1 - \frac{\alpha_{12}}{\rho} - \frac{\alpha_{34}}{\phi}-
\nonumber \\&&{}
- \frac{\alpha_{12}}{\alpha_{34}}- \frac{\alpha_{34}}{\alpha_{12}}
- \frac{\alpha_{34}}{\alpha_{23}}- \frac{\rho}{\alpha_{12}}
- \frac{\alpha_{23}}{\phi}- \frac{\rho}{\alpha_{23}}
- \frac{\alpha_{23}}{\rho} -\frac{\alpha_{23}}{\alpha_{34}}
- \frac{\alpha_{24}}{\alpha_{34}}- \frac{\alpha_{24}}{\alpha_{23}}
  \biggr\}+
\nonumber \\&&{}+ 2 \zeta(3) \alpha'
\biggl\{ -\phi + 3 \alpha_{24} + \frac{\alpha_{12}^2}{\alpha_{34}} +
\frac{\alpha_{23}^2}{\rho} + \frac{\rho^2}{\alpha_{12}} +
\frac{\alpha_{34}^2}{\alpha_{12}} + \frac{\alpha_{34}^2}{\alpha_{23}}
+ \frac{\alpha_{34}^2}{\phi} + \frac{\alpha_{12}^2}{\rho} +
\frac{\alpha_{23}^2}{\alpha_{34}} + 
\nonumber\\ &&
\hphantom{{}+ 2 \zeta(3) \alpha' \biggl\{}
+\frac{\alpha_{23}^2}{\phi} +
\frac{\alpha_{24}^2}{\alpha_{34}} + \frac{\alpha_{24}^2}{\alpha_{23}}
+ \frac{\rho^2}{\alpha_{23}}
+ 2 \frac{\rho \cdot \alpha_{34}}{\alpha_{23}} + 2 \frac{\rho \cdot
\alpha_{24}}{\alpha_{23}} + 2 \frac{\alpha_{24} \cdot
\alpha_{34}}{\alpha_{23}} + 
\nonumber\\ &&
\hphantom{{}+ 2 \zeta(3) \alpha' \biggl\{}
+2 \frac{\alpha_{12} \cdot
\alpha_{23}}{\rho} + 2 \frac{\alpha_{12} \cdot
\alpha_{23}}{\alpha_{34}} + 2 \frac{\rho \cdot
\alpha_{34}}{\alpha_{12}} + 2 \frac{\alpha_{12} \cdot
\alpha_{24}}{\alpha_{34}}+
\nonumber \\&&
\hphantom{{}+ 2 \zeta(3) \alpha' \biggl\{}
+  2 \frac{\alpha_{24} \cdot
 \alpha_{23}}{\alpha_{34}}
+ 2 \frac{\alpha_{34} \cdot \alpha_{23}}{\phi} \biggr\} + {\cal
O}(\alpha'^2) \,,
\label{L2final}
\\
L_3 & = & \frac{1}{(2 \alpha')^2} \left\{ \frac{1}{\rho \cdot
\alpha_{12}} + \frac{1}{\alpha_{34} \cdot \alpha_{12}} \right\}
- \frac{\pi^2}{6}
\left\{\frac{\alpha_{23}}{\alpha_{34}} + \frac{\alpha_{23}}{\rho} +
\frac{\rho}{\alpha_{12}} + \frac{\alpha_{34}}{\alpha_{12}} +
\frac{\alpha_{24}}{\alpha_{34}} \right\} +
\nonumber \\&&{}+ 2 \zeta(3)
\alpha' \biggl\{ \alpha_{24} + \frac{\alpha_{24}^2}{\alpha_{34}} + 2
\frac{\alpha_{24} \cdot \alpha_{23}}{\alpha_{34}} + 2 \frac{\rho \cdot
\alpha_{34}}{\alpha_{12}} + \frac{\alpha_{12} \cdot \alpha_{23}}{\rho}
+ \frac{\alpha_{34}^2}{\alpha_{12}} + \frac{\alpha_{23}^2}{\rho} +
\nonumber\\ &&
\hphantom{{}+ 2 \zeta(3)\alpha' \biggl\{ } 
\frac{\alpha_{12} \cdot \alpha_{23}}{\alpha_{34}} 
+  \frac{\alpha_{12} \cdot
 \alpha_{24}}{\alpha_{34}}
+ \frac{\alpha_{23}^2}{\alpha_{34}} + \frac{\rho^2}{\alpha_{12}}
\biggr\} + {\cal O}(\alpha'^2) \,,
\label{L3final}
\\
{L_3}' & = & \frac{1}{(2 \alpha')^2} \left\{ \frac{1}{\phi \cdot
\alpha_{34}} + \frac{1}{\alpha_{12} \cdot \alpha_{34}} \right\}
- \frac{\pi^2}{6}
\left\{\frac{\alpha_{23}}{\alpha_{12}} + \frac{\alpha_{23}}{\phi} +
\frac{\phi}{\alpha_{34}} + \frac{\alpha_{12}}{\alpha_{34}} +
\frac{\alpha_{13}}{\alpha_{12}} \right\} +
\nonumber \\&&{}+ 2 \zeta(3)
\alpha' \biggl\{ \alpha_{13} + \frac{\alpha_{13}^2}{\alpha_{12}} + 2
\frac{\alpha_{13} \cdot \alpha_{23}}{\alpha_{12}} + 2 \frac{\phi \cdot
\alpha_{12}}{\alpha_{34}} + \frac{\alpha_{34} \cdot \alpha_{23}}{\phi}
+ \frac{\alpha_{12}^2}{\alpha_{34}} + \frac{\alpha_{23}^2}{\phi} +
\nonumber\\ &&
\hphantom{{}+ 2 \zeta(3)\alpha' \biggl\{}
+\frac{\alpha_{34} \cdot \alpha_{23}}{\alpha_{12}} 
+ \frac{\alpha_{34} \cdot
 \alpha_{13}}{\alpha_{12}}
+ \frac{\alpha_{23}^2}{\alpha_{12}} + \frac{\phi^2}{\alpha_{34}}
\biggr\} + {\cal O}(\alpha'^2) \,,
\label{L3primefinal}
\\
L_4 & = & \frac{1}{(2 \alpha')^2} \left\{ \frac{1}{\alpha_{23} \cdot
\phi} + \frac{1}{\alpha_{23} \cdot \alpha_{12}}
-\frac{\alpha_{13}}{\alpha_{23} \cdot \rho \cdot \alpha_{12}} \right\}+
\label{L4final}\\ &&
{}+ \frac{\pi^2}{6} \left\{\frac{\alpha_{13}}{\rho} -
\frac{\alpha_{34}}{\phi}
- \frac{\alpha_{34}}{\alpha_{12}} - \frac{\alpha_{13}}{\alpha_{23}}
- \frac{\alpha_{12}}{\alpha_{23}} - \frac{\phi}{\alpha_{23}}
\right\} +
\nonumber \\&&{}+ 2 \zeta(3) \alpha' \biggl\{ 2 \alpha_{13} +
\frac{\alpha_{34} \cdot \alpha_{23}}{\phi}
- \frac{\alpha_{13} \cdot \alpha_{12}}{\rho}
+ \frac{\alpha_{34} \cdot \alpha_{23}}{\alpha_{12}} +
\frac{\alpha_{13} \cdot \rho}{\alpha_{23}} + \frac{\alpha_{13} \cdot
\alpha_{12}}{\alpha_{23}}+
\nonumber \\&&
\hphantom{{}+ 2 \zeta(3) \alpha' \biggl\{}
- \frac{\alpha_{23} \cdot \alpha_{13}}{\rho} + 2 \frac{\alpha_{34}
  \cdot \alpha_{13}}{\alpha_{12}} + 3 \frac{\alpha_{34} \cdot
  \alpha_{13}}{\alpha_{23}} + 2 \frac{\alpha_{13} \cdot
  \alpha_{24}}{\alpha_{23}} + 2 \frac{\phi \cdot
  \alpha_{12}}{\alpha_{23}} + 
\nonumber\\ &&
\hphantom{{}+ 2 \zeta(3) \alpha' \biggl\{}
+\frac{\phi^2}{\alpha_{23}} +
\frac{\alpha_{34}^2}{\phi} + \frac{\alpha_{34}^2}{\alpha_{12}} -
\frac{\alpha_{13} \cdot \rho \cdot \alpha_{34}}{\alpha_{23} \cdot 
  \alpha_{12}} + \frac{\alpha_{34} \cdot \alpha_{13}^2}{\alpha_{23}
  \cdot \alpha_{12}}
+  \frac{\alpha_{12}^2}{\alpha_{23}} \biggr\}
+ {\cal O}(\alpha'^2) \,,
\nonumber
\\
{L_4}' & = & \frac{1}{(2 \alpha')^2} \left\{ \frac{1}{\alpha_{23}
\cdot \rho} + \frac{1}{\alpha_{23} \cdot \alpha_{34}}
-\frac{\alpha_{24}}{\alpha_{23} \cdot \phi \cdot \alpha_{34}} \right\}+
\nonumber\\ &&
{}+ \frac{\pi^2}{6} \left\{\frac{\alpha_{24}}{\phi} -
\frac{\alpha_{12}}{\rho}
- \frac{\alpha_{12}}{\alpha_{34}} - \frac{\alpha_{24}}{\alpha_{23}}
- \frac{\alpha_{34}}{\alpha_{23}} - \frac{\rho}{\alpha_{23}}
\right\} +
\nonumber \\&&{}+ 2 \zeta(3) \alpha' \biggl\{ 2 \alpha_{24} +
\frac{\alpha_{12} \cdot \alpha_{23}}{\rho}
- \frac{\alpha_{24} \cdot \alpha_{34}}{\phi}
+ \frac{\alpha_{12} \cdot \alpha_{23}}{\alpha_{34}} +
\frac{\alpha_{24} \cdot \phi}{\alpha_{23}} + \frac{\alpha_{24} \cdot
\alpha_{34}}{\alpha_{23}}-
\nonumber \\&&
\hphantom{{}+ 2 \zeta(3) \alpha' \biggl\{} 
- \frac{\alpha_{23} \cdot \alpha_{24}}{\phi} 
+  2 \frac{\alpha_{12} \cdot
 \alpha_{24}}{\alpha_{34}}
+ 3 \frac{\alpha_{12} \cdot \alpha_{24}}{\alpha_{23}} + 2
\frac{\alpha_{24} \cdot \alpha_{13}}{\alpha_{23}} + 2 \frac{\rho \cdot
\alpha_{34}}{\alpha_{23}} + \frac{\rho^2}{\alpha_{23}} +
\nonumber\\&&
\hphantom{{}+ 2 \zeta(3) \alpha' \biggl\{} 
+\frac{\alpha_{12}^2}{\rho} +
  \frac{\alpha_{12}^2}{\alpha_{34}}
- \frac{\alpha_{24} \cdot \phi \cdot \alpha_{12}}{\alpha_{23} \cdot
 \alpha_{34}}
+ \frac{\alpha_{12} \cdot \alpha_{24}^2}{\alpha_{23} \cdot
\alpha_{34}} 
+ \frac{\alpha_{34}^2}{\alpha_{23}} \biggr\}
+ {\cal O}(\alpha'^2) \,,
\label{L4primefinal}
\\
L_7 & = & \frac{1}{(2 \alpha')^2} \left\{ \frac{1}{\rho \cdot
\alpha_{23}} + \frac{1}{\phi \cdot \alpha_{23}} \right\}
- \frac{\pi^2}{6}
\left\{\frac{\alpha_{34}}{\alpha_{23}} + \frac{\rho}{\alpha_{23}} +
\frac{\alpha_{34}}{\phi} + \frac{\alpha_{12}}{\rho} +
\frac{\alpha_{24}}{\alpha_{23}} \right\} +
\nonumber \\&&
{}+ 2 \zeta(3) \alpha' \biggl\{ 2 \alpha_{34} - 2 \alpha_{12} + 2
\alpha_{24} + 2 \rho 
+ \frac{\alpha_{34} \cdot \alpha_{23}}{\phi} + 2
\frac{\alpha_{34}^2}{\alpha_{23}} + \frac{\rho^2}{\alpha_{23}} +
\frac{\alpha_{34}^2}{\phi} + \frac{\phi \cdot
\alpha_{12}}{\alpha_{23}}-
\nonumber\\ &&
\hphantom{{}+ 2 \zeta(3) \alpha' \biggl\{}
- \frac{\alpha_{34} \cdot \phi}{\alpha_{23}}
+ 2 \frac{\rho \cdot \alpha_{24}}{\alpha_{23}}
+  2 \frac{\rho \cdot \alpha_{34}}{\alpha_{23}}
- \frac{\alpha_{12} \cdot \alpha_{24}}{\alpha_{23}} -\frac{\alpha_{34}
 \cdot \alpha_{12}}{\alpha_{23}}
+ 3 \frac{\alpha_{34} \cdot \alpha_{24}}{\alpha_{23}} +
\nonumber\\ &&
\hphantom{{}+ 2 \zeta(3) \alpha' \biggl\{}
+\frac{\alpha_{23} \cdot \alpha_{12}}{\rho} +
\frac{\alpha_{12}^2}{\rho} + \frac{\alpha_{24}^2}{\alpha_{23}}
\biggr\} 
+ {\cal O}(\alpha'^2) \,.
\label{L9final}
\end{eqnarray}
Besides the $\alpha_{ij}$ and $\rho$ variables, respectively defined
in eqs.~(\ref{alphaij}) and~(\ref{rho}), the $\phi$ variable
appearing in many of the kinematic factors is defined as
\begin{equation}
\phi = \alpha_{23} + \alpha_{24} + \alpha_{34}\,.
\label{phi}
\end{equation} 
On this list, the expressions for $K_6$, $L_5$ and $L_6$ have been
omitted since, using the relations in~(\ref{relations}), they can be
directly obtained in terms of kinematic factors that do appear on the
list.  In fact, using these relations, $K_6$, $L_5$ and $L_6$ have
been already eliminated in the expression for $A(1,2,3,4,5)$ that
comes in~(\ref{amplitude}) .  A remarkable fact, present in every
kinematic factor, is that there is no $1/\alpha'$ contribution to it.

\section{Contributions to the five gluon tree amplitude, at higher orders in
$\alpha'$}\label{A0A2A3} 

Using the explicit expressions of the kinematic factors (up to ${\cal
O}({\alpha'})$ terms) given in the previous subsection, and
substituting them in eq.~(\ref{amplitude}), leads directly to the
expressions of $A^{(0)}(1,2,3,4,5)$, $A^{(2)}(1,2,3,4,5)$ and
$A^{(3)}(1,2,3,4,5)$ of eq.~(\ref{A5final}). Since the expression for
$A^{(0)}(1,2,3,4,5)$ was already presented in eq.~(\ref{A(0)}), here
we just give the following expressions for $A^{(2)}(1,2,3,4,5)$ and
$A^{(3)}(1,2,3,4,5)$:
\newcommand{\qLtwo}{\biggl\{-{\frac {\alpha_{{2 4}}}{\alpha_{{3 4}}}}-
{\frac {\alpha_{{2 3}}}{\alpha_{{3 4}}}}- {\frac {\rho}{\alpha_{{1
2}}}}- {\frac {\alpha_{{1 2}}}{\alpha_{{3 4}}}} -
\nnbb & &
\hspace{7cm}
-{\frac {\alpha_{{2 3}}}{\phi}}+
1-{\frac {\alpha_{{3 4}}}{\phi}}- {\frac {\alpha_{{3 4}}}{\alpha_{{1
2}}}}- {\frac {\alpha_{{2 3}}}{\rho}} -
\nnbb & &
\hspace{7cm}
-{\frac {\alpha_{{3 4}}}{\alpha_{{2 3}}}}-
{\frac {\rho}{\alpha_{{2 3}}}}- {\frac {\alpha_{{2 4}}}{\alpha_{{2
3}}}}- {\frac {\alpha_{{1 2}}}{\rho}}\biggr\}}
\newcommand{\qLthree}{\left\{-{\frac {\alpha_{{2 3}}}{\alpha_{{3 4}}}}-
{\frac {\alpha_{{2 3}}}{\rho}}- {\frac {\rho}{\alpha_{{1 2}}}}- {\frac
{\alpha_{{3 4}}}{\alpha_{{1 2}}}}
-{\frac {\alpha_{{2 4}}}{\alpha_{{3 4}}}}\right\}}
\newcommand{\qLthreep}{\left\{-{\frac {\phi}{\alpha_{{3 4}}}}- {\frac
{\alpha_{{1 2}}}{\alpha_{{3 4}}}}- {\frac {\alpha_{{2 3}}}{\alpha_{{1
2}}}}- {\frac {\alpha_{{2 3}}}{\phi}}
-{\frac {\alpha_{{1 3}}}{\alpha_{{1 2}}}}\right\}}
\newcommand{\qLnine}{\left\{-{\frac {\alpha_{{3 4}}}{\phi}}- {\frac
{\alpha_{{3 4}}}{\alpha_{{2 3}}}}- {\frac {\rho}{\alpha_{{2
3}}}}-{\frac {\alpha_{{2 4}}}{\alpha_{{2 3}}}}
-{\frac {\alpha_{{1 2}}}{\rho}}\right\}}
\newcommand{\qLonep}{\left\{1-{\frac {\alpha_{{1 2}}}{\alpha_{{3 4}}}}-
{\frac {\alpha_{{2 3}}}{\phi}}\right\}} 
\newcommand{\qLone}{\left\{1-{\frac
{\alpha_{{3 4}}}{\alpha_{{1 2}}}}- {\frac {\alpha_{{2 3}}}{\rho}}\right\}}
\newcommand{\qKfivep}{\left\{-{\frac {\alpha_{{1 3}}}{\alpha_{{2 3}}}}-
{\frac {\alpha_{{1 2}}}{\alpha_{{2 3}}}}- {\frac {\alpha_{{3
4}}}{\phi}}- {\frac {\alpha_{{2 3}}}{\phi}}
-{\frac {\alpha_{{1 2}}}{\alpha_{{3 4}}}}\right\}}
\newcommand{\qKsix}{\left\{-{\frac {\alpha_{{3 4}}}{\alpha_{{2 3}}}}-
{\frac {\alpha_{{1 2}}}{\rho}}- {\frac {\alpha_{{2 4}}}{\alpha_{{2
3}}}}\right\}} 
\newcommand{\qKfive}{\left\{-{\frac {\alpha_{{2 4}}}{\alpha_{{2
3}}}}- {\frac {\alpha_{{3 4}}}{\alpha_{{2 3}}}}- {\frac {\alpha_{{1
2}}}{\rho}}- {\frac {\alpha_{{2 3}}}{\rho}}
-{\frac {\alpha_{{3 4}}}{\alpha_{{1 2}}}}\right\}}
\newcommand{\qLfourp}{\left\{-{\frac {\alpha_{{1 2}}}{\alpha_{{3 4}}}}+
{\frac {\alpha_{{2 4}}}{\phi}}- {\frac {\rho}{\alpha_{{2 3}}}}- {\frac
{\alpha_{{3 4}}}{\alpha_{{2 3}}}}
-{\frac {\alpha_{{2 4}}}{\alpha_{{2 3}}}}-
{\frac {\alpha_{{1 2}}}{\rho}}\right\}} 
\newcommand{\qLfour}{\left\{-{\frac
{\alpha_{{3 4}}}{\alpha_{{1 2}}}}- {\frac {\alpha_{{3 4}}}{\phi}}+
{\frac {\alpha_{{1 3}}}{\rho}}- {\frac {\phi}{\alpha_{{2 3}}}}
-{\frac {\alpha_{{1 3}}}{\alpha_{{2 3}}}}-
{\frac {\alpha_{{1 2}}}{\alpha_{{2 3}}}}\right\}}
\newcommand{\qKsixp}{\left\{-{\frac {\alpha_{{1 2}}}{\alpha_{{2 3}}}}-
{\frac {\alpha_{{3 4}}}{\phi}}- {\frac {\alpha_{{1 3}}}{\alpha_{{2
3}}}}\right\}} 
\newcommand{\qKthree}{\left\{-{\frac {\rho}{\alpha_{{1 2}}}}-
{\frac {\alpha_{{2 3}}}{\alpha_{{3 4}}}}- {\frac {\alpha_{{2
4}}}{\alpha_{{3 4}}}}\right\}} 
\newcommand{\qKonep}{\left\{-{\frac {\alpha_{{1
2}}}{\alpha_{{3 4}}}}- {\frac {\alpha_{{2 3}}}{\phi}}\right\}}
\newcommand{\qKone}{\left\{-{\frac {\alpha_{{3 4}}}{\alpha_{{1 2}}}}-
{\frac {\alpha_{{2 3}}}{\rho}}\right\}} 
\newcommand{\cLtwo}{\biggl\{{\frac
{{\,\alpha_{{34}}}^{2}}{\,\alpha_{{12}}}}+ 3\,\,\alpha_{{24}}+ {\frac
{{\,\alpha_{{12}}}^{2}}{\,\alpha_{{34}}}}+ {\frac
{2\,\rho\,\,\alpha_{{24}}}{\,\alpha_{{23}}}}+ \nnbb & &
\hspace{5.8cm}+{\frac
{2\,\,\alpha_{{12}}\,\alpha_{{24}}}{\,\alpha_{{34}}}}+ {\frac
{2\,\rho\,\,\alpha_{{34}}}{\,\alpha_{{23}}}}+ {\frac
{2\,\,\alpha_{{34}}\,\alpha_{{24}}}{\,\alpha_{{23}}}}+ {\frac
{{\rho}^{2}}{\,\alpha_{{12}}}} +\nnbb & &
\hspace{5.8cm}
+{\frac
{{\,\alpha_{{34}}}^{2}}{\phi}}+ {\frac {{\,\alpha_{{23}}}^{2}}{\rho}}+
{\frac {2\,\,\alpha_{{23}}\,\alpha_{{24}}}{\,\alpha_{{34}}}}+ {\frac
{2\,\,\alpha_{{12}}\,\alpha_{{23}}}{\,\alpha_{{34}}}}+ \nnbb & &
\hspace{5.8cm}
+{\frac
{2\,\,\alpha_{{34}}\,\alpha_{{23}}}{\phi}}+ {\frac
{2\,\,\alpha_{{12}}\,\alpha_{{23}}}{\rho}}+ {\frac
{2\,\rho\,\,\alpha_{{34}}}{\,\alpha_{{12}}}}- \phi+ \nnbb & &
\hspace{5.8cm}
+{\frac
{{\rho}^{2}}{\,\alpha_{{23}}}}+ {\frac {{\,\alpha_{{12}}}^{2}}{\rho}}+
{\frac {{\,\alpha_{{23}}}^{2}}{\,\alpha_{{34}}}}+ {\frac
{{\,\alpha_{{24}}}^{2}}{\,\alpha_{{34}}}}+ \nnbb & &
\hspace{5.8cm}
+ {\frac
{{\,\alpha_{{34}}}^{2}}{\,\alpha_{{23}}}}+ {\frac
{{\,\alpha_{{24}}}^{2}}{\,\alpha_{{23}}}}+ {\frac
{{\,\alpha_{{23}}}^{2}}{\phi}}\biggr\}}
\newcommand{\cLthree}{\biggl\{{\frac
{\,\alpha_{{12}}\,\alpha_{{23}}}{\,\alpha_{{34}}}}+ {\frac
{\,\alpha_{{12}}\,\alpha_{{23}}}{\rho}}+ {\frac
{2\,\,\alpha_{{34}}\rho}{\,\alpha_{{12}}}}+ {\frac
{{\,\alpha_{{23}}}^{2}}{\,\alpha_{{34}}}}+ \nnbb & &
\hspace{5.8cm}
+{\frac
{{\,\alpha_{{23}}}^{2}}{\rho}}+ {\frac
{{\,\alpha_{{34}}}^{2}}{\,\alpha_{{12}}}}+ {\frac
{{\rho}^{2}}{\,\alpha_{{12}}}}+ {\frac
{2\,\,\alpha_{{23}}\,\alpha_{{24}}}{\,\alpha_{{34}}}}+ \nnbb & &
\hspace{5.8cm}
+ {\frac
{{\,\alpha_{{24}}}^{2}}{\,\alpha_{{34}}}}+ {\frac
{\,\alpha_{{12}}\,\alpha_{{24}}}{\,\alpha_{{34}}}}+ \,\alpha_{{24}}\biggr\}}
\newcommand{\cLthreep}{\biggl\{{\frac {{\phi}^{2}}{\,\alpha_{{34}}}}+ {\frac
{\,\alpha_{{34}}\,\alpha_{{23}}}{\,\alpha_{{12}}}}+ {\frac
{\,\alpha_{{34}}\,\alpha_{{23}}}{\phi}}+ {\frac
{2\,\phi\,\,\alpha_{{12}}}{\,\alpha_{{34}}}}+ \nnbb & &
\hspace{5.8cm}
+{\frac
{{\,\alpha_{{23}}}^{2}}{\,\alpha_{{12}}}}+ {\frac
{{\,\alpha_{{23}}}^{2}}{\phi}}+ {\frac
{{\,\alpha_{{12}}}^{2}}{\,\alpha_{{34}}}}+ {\frac
{2\,\,\alpha_{{23}}\,\alpha_{{13}}}{\,\alpha_{{12}}}}+ \nnbb & &
\hspace{5.8cm}
+{\frac
{{\,\alpha_{{13}}}^{2}}{\,\alpha_{{12}}}}+ {\frac
{\,\alpha_{{34}}\,\alpha_{{13}}}{\,\alpha_{{12}}}}+ \,\alpha_{{13}}\biggr\}}
\newcommand{\cLnine}{\biggl\{{\frac {\,\alpha_{{12}}\,\alpha_{{23}}}{\rho}}-
{\frac {\,\alpha_{{34}}\phi}{\,\alpha_{{23}}}}- {\frac
{\,\alpha_{{34}}\,\alpha_{{12}}}{\,\alpha_{{23}}}}- {\frac
{\,\alpha_{{12}}\,\alpha_{{24}}}{\,\alpha_{{23}}}} +\nnbb & &
\hspace{5.8cm}
+{\frac
{\phi\,\,\alpha_{{12}}}{\,\alpha_{{23}}}}+ {\frac
{2\,\rho\,\,\alpha_{{24}}}{\,\alpha_{{23}}}}+ {\frac
{2\,\rho\,\,\alpha_{{34}}}{\,\alpha_{{23}}}}+ {\frac
{3\,\,\alpha_{{34}}\,\alpha_{{24}}}{\,\alpha_{{23}}}} +\nnbb & &
\hspace{5.8cm}
+2\,\,\alpha_{{24}}-
2\,\,\alpha_{{12}}+ {\frac {{\,\alpha_{{24}}}^{2}}{\,\alpha_{{23}}}}+
{\frac {2\,{\,\alpha_{{34}}}^{2}}{\,\alpha_{{23}}}} +\nnbb & &
\hspace{5.8cm}
+{\frac
{{\,\alpha_{{12}}}^{2}}{\rho}}+ {\frac {{\,\alpha_{{34}}}^{2}}{\phi}}+
{\frac {{\rho}^{2}}{\,\alpha_{{23}}}}+ {\frac
{\,\alpha_{{34}}\,\alpha_{{23}}}{\phi}}+ 2\,\,\alpha_{{34}}+
2\,\rho\biggr\}} 
\newcommand{\cLonep}{\biggl\{{\frac
{{\,\alpha_{{12}}}^{2}}{\,\alpha_{{34}}}}+ {\frac
{{\,\alpha_{{23}}}^{2}}{\phi}}+ {\frac
{\,\alpha_{{34}}\,\alpha_{{23}}}{\phi}}+ {\frac
{\phi\,\,\alpha_{{12}}}{\,\alpha_{{34}}}} -\nnbb & &
\hspace{5.8cm}
-\phi-
2\,\,\alpha_{{12}}- \,\alpha_{{34}}- 2\,\,\alpha_{{23}}-
2\,\,\alpha_{{13}}\biggr\}} 
\newcommand{\cLone}{\biggl\{{\frac
{{\,\alpha_{{34}}}^{2}}{\,\alpha_{{12}}}}+ {\frac
{{\,\alpha_{{23}}}^{2}}{\rho}}+ {\frac
{\,\alpha_{{12}}\,\alpha_{{23}}}{\rho}}+ {\frac
{\rho\,\,\alpha_{{34}}}{\,\alpha_{{12}}}} -\nnbb & &
\hspace{5.8cm}
-\rho-2\,\,\alpha_{{34}}-
\,\alpha_{{12}}- 2\,\,\alpha_{{23}}- 2\,\,\alpha_{{24}}\biggr\}}
\newcommand{\cKfivep}{\biggl\{{\frac
{{\,\alpha_{{13}}}^{2}}{\,\alpha_{{23}}}}+ {\frac
{2\,\,\alpha_{{12}}\,\alpha_{{13}}}{\,\alpha_{{23}}}}+ {\frac
{\phi\,\,\alpha_{{12}}}{\,\alpha_{{34}}}}+ {\frac
{2\,\,\alpha_{{34}}\,\alpha_{{23}}}{\phi}}+ \nnbb & &
\hspace{5.8cm}
+{\frac
{{\,\alpha_{{12}}}^{2}}{\,\alpha_{{23}}}}+ {\frac
{{\,\alpha_{{34}}}^{2}}{\phi}}+ {\frac
{\phi\,\,\alpha_{{13}}}{\,\alpha_{{23}}}}+ \nnbb & &
\hspace{5.8cm}
+{\frac
{\phi\,\,\alpha_{{12}}}{\,\alpha_{{23}}}}+ {\frac
{{\,\alpha_{{23}}}^{2}}{\phi}}+ \,\alpha_{{13}}+ {\frac
{{\,\alpha_{{12}}}^{2}}{\,\alpha_{{34}}}}\biggr\}}
\newcommand{\cKsix}{\biggl\{{\frac {{\,\alpha_{{12}}}^{2}}{\rho}}+ {\frac
{{\,\alpha_{{34}}}^{2}}{\,\alpha_{{23}}}}+ {\frac
{\rho\,\,\alpha_{{34}}}{\,\alpha_{{23}}}}+ {\frac
{\rho\,\,\alpha_{{24}}}{\,\alpha_{{23}}}}+ \nnbb & &
\hspace{5.8cm}
+{\frac
{2\,\,\alpha_{{34}}\,\alpha_{{24}}}{\,\alpha_{{23}}}}+ {\frac
{\,\alpha_{{12}}\,\alpha_{{23}}}{\rho}}+ {\frac
{{\,\alpha_{{24}}}^{2}}{\,\alpha_{{23}}}}+ 2\,\,\alpha_{{24}}\biggr\}}
\newcommand{\cKfive}{\biggl\{{\frac
{{\,\alpha_{{24}}}^{2}}{\,\alpha_{{23}}}}+ {\frac
{2\,\,\alpha_{{34}}\,\alpha_{{24}}}{\,\alpha_{{23}}}}+ {\frac
{\rho\,\,\alpha_{{34}}}{\,\alpha_{{12}}}}+ {\frac
{2\,\,\alpha_{{12}}\,\alpha_{{23}}}{\rho}} +\nnbb & &
\hspace{5.8cm}
+{\frac
{{\,\alpha_{{34}}}^{2}}{\,\alpha_{{23}}}}+ {\frac
{{\,\alpha_{{12}}}^{2}}{\rho}}+ {\frac
{\rho\,\,\alpha_{{24}}}{\,\alpha_{{23}}}}+ {\frac
{\rho\,\,\alpha_{{34}}}{\,\alpha_{{23}}}} +\nnbb & &
\hspace{5.8cm}
+{\frac
{{\,\alpha_{{23}}}^{2}}{\rho}}+ \,\alpha_{{24}}+ {\frac
{{\,\alpha_{{34}}}^{2}}{\,\alpha_{{12}}}}\biggr\}}
\newcommand{\cLfourp}{\biggl\{{\frac
{{\,\alpha_{{12}}}^{2}}{\,\alpha_{{34}}}}+ {\frac
{\,\alpha_{{12}}\,\alpha_{{23}}}{\rho}}+ {\frac
{\,\alpha_{{34}}\,\alpha_{{24}}}{\,\alpha_{{23}}}}+ {\frac
{2\,\rho\,\,\alpha_{{34}}}{\,\alpha_{{23}}}}+ \nnbb & &
\hspace{5.8cm}
+2\,\,\alpha_{{24}}+ {\frac
{{\,\alpha_{{12}}}^{2}}{\rho}}+ {\frac
{{\,\alpha_{{34}}}^{2}}{\,\alpha_{{23}}}}+ {\frac
{{\rho}^{2}}{\,\alpha_{{23}}}} +\nnbb & &
\hspace{5.8cm}
+{\frac
{2\,\,\alpha_{{13}}\,\alpha_{{24}}}{\,\alpha_{{23}}}}+ {\frac
{\,\alpha_{{24}}\phi}{\,\alpha_{{23}}}}+ {\frac
{\,\alpha_{{12}}\,\alpha_{{23}}}{\,\alpha_{{34}}}}+ {\frac
{3\,\,\alpha_{{12}}\,\alpha_{{24}}}{\,\alpha_{{23}}}} -\nnbb & &
\hspace{5.8cm}
-{\frac {\,\alpha_{{23}}\,\alpha_{{24}}}{\phi}}-
{\frac {\,\alpha_{{34}}\,\alpha_{{24}}}{\phi}}-
{\frac{\,\alpha_{{24}}\phi\,\,\alpha_{{12}}}
{\,\alpha_{{23}}\,\alpha_{{34}}}}  
+\nnbb & & \hspace{5.8cm}
+{\frac
{\,\alpha_{{12}}{\,\alpha_{{24}}}^{2}}{\,\alpha_{{23}}\,\alpha_{{34}}}}+
{\frac {2\,\,\alpha_{{12}}\,\alpha_{{24}}}{\,\alpha_{{34}}}}\biggr\}}
\newcommand{\cLfour}{\biggl\{2 \alpha_{13}+ \frac{\alpha_{34}
\alpha_{23}}{\phi}
- \frac{\alpha_{13} \alpha_{12}}{\rho}
+ \frac{\alpha_{34} \alpha_{23}}{\alpha_{12}} + \frac{\alpha_{13}
\rho}{\alpha_{23}} +\nnbb & & 
\hspace{5.8cm}
+ \frac{\alpha_{13} \alpha_{12}}{\alpha_{23}}
- \frac{\alpha_{23}\alpha_{13}}{\rho}
+ \frac{2\alpha_{34} \alpha_{13}}{\alpha_{12}} + \frac{3\alpha_{34}
\alpha_{13}}{\alpha_{23}} +\nnbb & &
\hspace{5.8cm}
+ \frac{2\alpha_{13}
\alpha_{24}}{\alpha_{23}} + \frac{\phi \alpha_{12}}{\alpha_{23}} +
\frac{\phi^2}{\alpha_{23}} + \frac{\alpha_{34}^2}{\phi} +\nnbb & &
\hspace{5.8cm}
+
\frac{\alpha_{34}^2}{\alpha_{12}}
- \frac{\alpha_{13} \rho \cdot \alpha_{34}}{\alpha_{23} \alpha_{12}}
+ \frac{\alpha_{34} \alpha_{13}^2}{\alpha_{23} \alpha_{12}}
+\frac{\alpha_{12}^2}{\alpha_{23}}\biggr\}} 
\newcommand{\cKsixp}{\biggl\{{\frac
{{\,\alpha_{{34}}}^{2}}{\phi}}+ {\frac
{{\,\alpha_{{12}}}^{2}}{\,\alpha_{{23}}}}+ {\frac
{\phi\,\,\alpha_{{12}}}{\,\alpha_{{23}}}}+ {\frac
{\phi\,\,\alpha_{{13}}}{\,\alpha_{{23}}}} +\nnbb & &
\hspace{5.8cm}
+{\frac
{2\,\,\alpha_{{12}}\,\alpha_{{13}}}{\,\alpha_{{23}}}}+ {\frac
{\,\alpha_{{34}}\,\alpha_{{23}}}{\phi}}+ {\frac
{{\,\alpha_{{13}}}^{2}}{\,\alpha_{{23}}}}+ 2\,\,\alpha_{{13}}\biggr\}}
\newcommand{\cKthree}{\biggl\{2\,\,\alpha_{{24}}+ {\frac
{\,\alpha_{{34}}\rho}{\,\alpha_{{12}}}}+ {\frac
{{\rho}^{2}}{\,\alpha_{{12}}}}+ {\frac
{{\,\alpha_{{23}}}^{2}}{\,\alpha_{{34}}}} +\nnbb & &
\hspace{5.8cm}
+{\frac
{{\,\alpha_{{24}}}^{2}}{\,\alpha_{{34}}}}+ {\frac
{2\,\,\alpha_{{23}}\,\alpha_{{24}}}{\,\alpha_{{34}}}}+ {\frac
{\,\alpha_{{23}}\,\alpha_{{12}}}{\,\alpha_{{34}}}}+ {\frac
{\,\alpha_{{12}}\,\alpha_{{24}}}{\,\alpha_{{34}}}}\biggr\}}
\newcommand{\cKfour}{\biggl\{-\rho- 2\,\,\alpha_{{23}}- 2\,\,\alpha_{{34}}-
\,\alpha_{{12}}- \,\alpha_{{24}}\biggr\}}
\newcommand{\cKonep}{\biggl\{-\,\alpha_{{13}}+ {\frac
{\phi\,\,\alpha_{{12}}}{\,\alpha_{{34}}}}+ {\frac
{\,\alpha_{{34}}\,\alpha_{{23}}}{\phi}}+ {\frac
{{\,\alpha_{{12}}}^{2}}{\,\alpha_{{34}}}}+ {\frac
{{\,\alpha_{{23}}}^{2}}{\phi}}\biggr\}}
\newcommand{\cKone}{\biggl\{-\,\alpha_{{24}}+ {\frac
{\,\alpha_{{34}}\rho}{\,\alpha_{{12}}}}+ {\frac
{\,\alpha_{{12}}\,\alpha_{{23}}}{\rho}}+ {\frac
{{\,\alpha_{{34}}}^{2}}{\,\alpha_{{12}}}}+ {\frac
{{\,\alpha_{{23}}}^{2}}{\rho}}\biggr\}}
{\small\begin{eqnarray}
\lefteqn{ A^{(2)}(1, 2, 3, 4, 5)  =  {4\pi^2 g^3\over 3}\times }&&
\nonumber\\ &&
\times\biggl[(\zeta_1 \cdot \zeta_2)(\zeta_3 \cdot \zeta_4)
{{}\atop{}} \biggl\{(\zeta_5 \cdot k_1) \alpha_{23} {{}\atop{}}
\qLtwo- \nnbb & &
\hphantom{\times\biggl[(\zeta_1 \cdot \zeta_2)(\zeta_3 \cdot \zeta_4)
{{}\atop{}} \biggl\{}
      -(\zeta_5 \cdot k_2) \alpha_{13} \qLthree+
\nnbb & & 
\hphantom{\times\biggl[(\zeta_1 \cdot \zeta_2)(\zeta_3 \cdot \zeta_4)
{{}\atop{}} \biggl\{}
+(\zeta_5 \cdot k_3)
      \alpha_{24} \qLthreep+
\nnbb & &  
\hphantom{\times\biggl[(\zeta_1 \cdot \zeta_2)(\zeta_3 \cdot \zeta_4)
{{}\atop{}} \biggl\{}
+(\zeta_5 \cdot k_3)
    \alpha_{23} \qLtwo 
\biggr\}+
\nnbb & & 
\hphantom{\times\biggl[} 
+ (\zeta_1 \cdot \zeta_3)(\zeta_2 \cdot \zeta_4)
\biggl\{-(\zeta_5 \cdot k_1) \alpha_{23}  \qLnine+
\nnbb & & 
\hphantom{\times\biggl[+ (\zeta_1 \cdot \zeta_3)(\zeta_2 \cdot
    \zeta_4)\biggl\{}  
+(\zeta_5 \cdot k_2)
      \alpha_{34} \qLonep-
\nnbb & & 
\hphantom{\times\biggl[+ (\zeta_1 \cdot \zeta_3)(\zeta_2 \cdot
    \zeta_4)\biggl\{}  
      -(\zeta_5 \cdot k_2) \alpha_{23} \qLnine -
\nnbb & &
\hphantom{\times\biggl[+ (\zeta_1 \cdot \zeta_3)(\zeta_2 \cdot
    \zeta_4)\biggl\{}  
     -(\zeta_5 \cdot k_3) \alpha_{12} \qLone 
\biggr\}+
\nnbb & & 
\hphantom{\times\biggl[} 
+ (\zeta_1 \cdot \zeta_4)(\zeta_2 \cdot \zeta_3)
\biggl\{(\zeta_5 \cdot k_1) \alpha_{34} \qKfivep-
\nnbb & & 
\hphantom{\times\biggl[+ (\zeta_1 \cdot \zeta_4)(\zeta_2 \cdot
    \zeta_3)\biggl\{}  
      -(\zeta_5 \cdot k_1) \alpha_{13} \qKsix -
\nnbb & &
\hphantom{\times\biggl[+ (\zeta_1 \cdot \zeta_4)(\zeta_2 \cdot
    \zeta_3)\biggl\{}  
      -(\zeta_5 \cdot k_2) \alpha_{13} \qKsix+
\nnbb & & 
\hphantom{\times\biggl[+ (\zeta_1 \cdot \zeta_4)(\zeta_2 \cdot
    \zeta_3)\biggl\{}  
+(\zeta_5 \cdot k_3)
   \alpha_{12}\qKfive
\biggr\}\biggr]+
\nnbb & & 
+\biggl[(\zeta_2 \cdot \zeta_3)\biggl\{  (\zeta_5\cdot k_2)\biggl( 
(\zeta_1 \cdot k_3)(\zeta_4 \cdot k_1)
\qKsix +
\nnbb & & 
\hphantom{+\biggl[(\zeta_2 \cdot \zeta_3)\biggl\{  (\zeta_5\cdot
    k_2)\biggl(}  
+(\zeta_1 \cdot k_3)(\zeta_4 \cdot k_3) \qLfourp+
\nnbb & & 
\hphantom{+\biggl[(\zeta_2 \cdot \zeta_3)\biggl\{  (\zeta_5\cdot
    k_2)\biggl(}  
+(\zeta_1 \cdot k_2)(\zeta_4 \cdot k_3) \qLtwo +
\nnbb & &
\hphantom{+\biggl[(\zeta_2 \cdot \zeta_3)\biggl\{  (\zeta_5\cdot
    k_2)\biggl(}  
+(\zeta_1 \cdot k_4)(\zeta_4 \cdot k_3) \qKfivep
\biggr) -
\nnbb & &
\hphantom{+\biggl[(\zeta_2 \cdot \zeta_3)\biggl\{}
-(\zeta_5\cdot k_3)\biggl(
(\zeta_1 \cdot k_2)(\zeta_4 \cdot k_1) \qKfive 
\nnbb & & 
\hphantom{+\biggl[(\zeta_2 \cdot \zeta_3)\biggl\{-(\zeta_5\cdot k_3)\biggl(}
+(\zeta_1
\cdot k_3)(\zeta_4 \cdot k_2) \qLnine +
\nnbb & & 
\hphantom{+\biggl[(\zeta_2 \cdot \zeta_3)\biggl\{-(\zeta_5\cdot k_3)\biggl(}
+(\zeta_1 \cdot
k_2)(\zeta_4 \cdot k_2) \qLfour +
\nnbb & & 
\hphantom{+\biggl[(\zeta_2 \cdot \zeta_3)\biggl\{-(\zeta_5\cdot k_3)\biggl(}
+(\zeta_1 \cdot k_4)(\zeta_4
\cdot k_2) \qKsixp 
\biggr) +
\nnbb & & 
\hphantom{+\biggl[(\zeta_2 \cdot \zeta_3)\biggl\{}
+(\zeta_5\cdot k_1)\biggl(  (\zeta_1 \cdot k_2)(\zeta_4 \cdot k_3)
\qLtwo -
\nnbb & &
\hphantom{+\biggl[(\zeta_2 \cdot \zeta_3)\biggl\{+(\zeta_5\cdot k_1)\biggl(}
-(\zeta_1 \cdot k_3)(\zeta_4 \cdot k_2) \qLnine
\biggr)\biggr\} +
\nnbb & & 
\hphantom{+\biggl[}
+(\zeta_1 \cdot \zeta_4)\biggl\{
-(\zeta_5\cdot k_1)\biggl(
(\zeta_2 \cdot k_3)(\zeta_3 \cdot k_4) \qKfivep -
\nnbb & &
\hphantom{+\biggl[+(\zeta_1 \cdot \zeta_4)\biggl\{-(\zeta_5\cdot
    k_1)\biggl(} 
-(\zeta_2 \cdot k_4)(\zeta_3 \cdot k_2) \qKsixp+
\nnbb & & 
\hphantom{+\biggl[+(\zeta_1 \cdot \zeta_4)\biggl\{-(\zeta_5\cdot
    k_1)\biggl(} 
+(\zeta_2 \cdot k_4)(\zeta_3 \cdot k_4) \qKonep -
\nnbb & &
\hphantom{+\biggl[+(\zeta_1 \cdot \zeta_4)\biggl\{-(\zeta_5\cdot
    k_1)\biggl(} 
-(\zeta_2 \cdot k_1)(\zeta_3 \cdot k_4) \qKthree
\biggr)-
\nnbb & & 
\hphantom{+\biggl[+(\zeta_1 \cdot \zeta_4)\biggl\{}
-(\zeta_5\cdot k_4)\biggl(  (\zeta_2 \cdot
k_1)(\zeta_3 \cdot k_2) \qKfive +
\nnbb & & 
\hphantom{+\biggl[+(\zeta_1 \cdot \zeta_4)\biggl\{-(\zeta_5\cdot
    k_4)\biggl(} 
+(\zeta_2 \cdot k_1)(\zeta_3
\cdot k_1) \qKone -
\nnbb & &
\hphantom{+\biggl[+(\zeta_1 \cdot \zeta_4)\biggl\{-(\zeta_5\cdot
    k_4)\biggl(} 
-(\zeta_2 \cdot k_3)(\zeta_3 \cdot k_1) \qKsix-
\nnbb & &
\hphantom{+\biggl[+(\zeta_1 \cdot \zeta_4)\biggl\{-(\zeta_5\cdot
    k_4)\biggl(} 
-(\zeta_2 \cdot k_4)(\zeta_3 \cdot k_1) \biggr)+
\nnbb & & 
\hphantom{+\biggl[+(\zeta_1 \cdot \zeta_4)\biggl\{} 
+(\zeta_5\cdot k_2)\biggl(  (\zeta_2 \cdot
k_1)(\zeta_3 \cdot k_4) \qKthree -
\nnbb & &
\hphantom{+\biggl[+(\zeta_1 \cdot \zeta_4)\biggl\{+(\zeta_5\cdot
    k_2)\biggl(}  
-(\zeta_2 \cdot k_4)(\zeta_3 \cdot k_1)\biggr)
\biggr\}\biggr] +
\nnbb & & 
+\left( \mbox{cyclic
permutations of indexes (1,2,3,4,5)} \right)
\label{A(2)}\end{eqnarray}}
and
{\small\begin{eqnarray}
\lefteqn{ A^{(3)}(1, 2, 3, 4, 5)  =  {16\zeta(3)g^3}\times}\!\!&&
\nonumber\\ &&
{}\times \biggl[(\zeta_1 \cdot
\zeta_2)(\zeta_3 \cdot \zeta_4)  \biggl\{(\zeta_5
\cdot k_1) \alpha_{23}  \cLtwo -
\nnbb & &
\hphantom{{}\times \biggl[(\zeta_1 \cdot
\zeta_2)(\zeta_3 \cdot \zeta_4)  \biggl\{} 
      -(\zeta_5 \cdot k_2) \alpha_{13} \cLthree+
\nnbb &&
\hphantom{{}\times \biggl[(\zeta_1 \cdot
\zeta_2)(\zeta_3 \cdot \zeta_4)  \biggl\{} 
+(\zeta_5 \cdot k_3)
      \alpha_{24} \cLthreep+
\nnbb &&
\hphantom{{}\times \biggl[(\zeta_1 \cdot
\zeta_2)(\zeta_3 \cdot \zeta_4)  \biggl\{} 
+(\zeta_5 \cdot k_3)
    \alpha_{23} \cLtwo\biggr\}+
\nnbb &&
\hphantom{{}\times \biggl[}
+ (\zeta_1 \cdot \zeta_3)(\zeta_2 \cdot \zeta_4)
\biggl\{-(\zeta_5 \cdot k_1) \alpha_{23} \cLnine+
\nnbb &&
\hphantom{{}\times \biggl[+ (\zeta_1 \cdot \zeta_3)(\zeta_2 \cdot
    \zeta_4)\biggl\{} 
+(\zeta_5 \cdot k_2)
      \alpha_{34} \cLonep-
\nnbb &&
\hphantom{{}\times \biggl[+ (\zeta_1 \cdot \zeta_3)(\zeta_2 \cdot
    \zeta_4)\biggl\{} 
      -(\zeta_5 \cdot k_2) \alpha_{23} \cLnine- \nnbb & &
\hphantom{{}\times \biggl[+ (\zeta_1 \cdot \zeta_3)(\zeta_2 \cdot
    \zeta_4)\biggl\{} 
     -(\zeta_5 \cdot k_3) \alpha_{12} \cLone \biggr\}+
\nnbb &&
\hphantom{{}\times \biggl[}
+ (\zeta_1 \cdot \zeta_4)(\zeta_2 \cdot \zeta_3)
\biggl\{(\zeta_5 \cdot k_1) \alpha_{34} \cKfivep-
\nnbb &&
\hphantom{{}\times \biggl[+ (\zeta_1 \cdot \zeta_4)(\zeta_2 \cdot
    \zeta_3)\biggl\{} 
      -(\zeta_5 \cdot k_1) \alpha_{13} \cKsix -\nnbb & &
\hphantom{{}\times \biggl[+ (\zeta_1 \cdot \zeta_4)(\zeta_2 \cdot
    \zeta_3)\biggl\{} 
      -(\zeta_5 \cdot k_2) \alpha_{13} \cKsix+
\nnbb &&
\hphantom{{}\times \biggl[+ (\zeta_1 \cdot \zeta_4)(\zeta_2 \cdot
    \zeta_3)\biggl\{} 
+(\zeta_5 \cdot k_3)
   \alpha_{12}\cKfive\biggr\}\biggr]+
\nnbb &&
{}+\biggl[(\zeta_2 \cdot
\zeta_3)\biggl\{  (\zeta_5\cdot
k_2)\biggl(  (\zeta_1 \cdot k_3)(\zeta_4 \cdot k_1)
\cKsix +
\nnbb &&
\hphantom{{}+\biggl[(\zeta_2 \cdot \zeta_3)\biggl\{ (\zeta_5\cdot
k_2)\biggl(}
+(\zeta_1 \cdot k_3)(\zeta_4 \cdot k_3) \cLfourp+
\nnbb &&
\hphantom{{}+\biggl[(\zeta_2 \cdot \zeta_3)\biggl\{ (\zeta_5\cdot
k_2)\biggl(}
+(\zeta_1 \cdot k_2)(\zeta_4 \cdot k_3) \cLtwo 
\nnbb & &
\hphantom{{}+\biggl[(\zeta_2 \cdot \zeta_3)\biggl\{ (\zeta_5\cdot
k_2)\biggl(}
+(\zeta_1 \cdot k_4)(\zeta_4 \cdot k_3) \cKfivep
\biggr) 
\nnbb & &
\hphantom{{}+\biggl[(\zeta_2 \cdot \zeta_3)\biggl\{}
-(\zeta_5\cdot k_3)\biggl(
(\zeta_1 \cdot k_2)(\zeta_4 \cdot k_1) \cKfive +
\nnbb &&
\hphantom{{}+\biggl[(\zeta_2 \cdot \zeta_3)\biggl\{-(\zeta_5\cdot
    k_3)\biggl(} 
+(\zeta_1
\cdot k_3)(\zeta_4 \cdot k_2) \cLnine +
\nnbb &&
\hphantom{{}+\biggl[(\zeta_2 \cdot \zeta_3)\biggl\{-(\zeta_5\cdot
    k_3)\biggl(} 
+(\zeta_1 \cdot
k_2)(\zeta_4 \cdot k_2) \cLfour +
\nnbb &&
\hphantom{{}+\biggl[(\zeta_2 \cdot \zeta_3)\biggl\{-(\zeta_5\cdot
    k_3)\biggl(} 
+(\zeta_1 \cdot k_4)(\zeta_4
\cdot k_2) \cKsixp \biggr) +
\nnbb &&
\hphantom{{}+\biggl[(\zeta_2 \cdot \zeta_3)\biggl\{}
+(\zeta_5\cdot
k_1)\biggl(  (\zeta_1 \cdot k_2)(\zeta_4 \cdot k_3)
\cLtwo -
\nnbb & &
\hphantom{{}+\biggl[(\zeta_2 \cdot \zeta_3)\biggl\{+(\zeta_5\cdot
k_1)\biggl(}
-(\zeta_1 \cdot k_3)(\zeta_4 \cdot k_2) \cLnine
\biggr)\biggr\} +
\nnbb &&
\hphantom{{}+\biggl[}
+(\zeta_1 \cdot
\zeta_4)\biggl\{
-(\zeta_5\cdot k_1)\biggl(
(\zeta_2 \cdot k_3)(\zeta_3 \cdot k_4) \cKfivep -
\nnbb & &
\hphantom{{}+\biggl[+(\zeta_1 \cdot \zeta_4)\biggl\{ -(\zeta_5\cdot
k_1)\biggl(}
-(\zeta_2 \cdot k_4)(\zeta_3 \cdot k_2) \cKsixp+
\nnbb &&
+(\zeta_2 \cdot k_4)(\zeta_3 \cdot k_4) \cKonep -
\hphantom{{}+\biggl[+(\zeta_1 \cdot \zeta_4)\biggl\{ -(\zeta_5\cdot
k_1)\biggl(}
\nnbb & &
\hphantom{{}+\biggl[+(\zeta_1 \cdot \zeta_4)\biggl\{ -(\zeta_5\cdot
k_1)\biggl(}
-(\zeta_2 \cdot k_1)(\zeta_3 \cdot k_4) \cKthree
\biggr)-
\nnbb &&
\hphantom{{}+\biggl[+(\zeta_1 \cdot \zeta_4)\biggl\{ }
-(\zeta_5\cdot k_4)\biggl(  (\zeta_2 \cdot
k_1)(\zeta_3 \cdot k_2) \cKfive +
\nnbb &&
\hphantom{{}+\biggl[+(\zeta_1 \cdot \zeta_4)\biggl\{-(\zeta_5\cdot
    k_4)\biggl( }
+(\zeta_2 \cdot k_1)(\zeta_3
\cdot k_1) \cKone -
\nnbb & &
\hphantom{{}+\biggl[+(\zeta_1 \cdot \zeta_4)\biggl\{-(\zeta_5\cdot
    k_4)\biggl( }
-(\zeta_2 \cdot k_3)(\zeta_3 \cdot k_1) \cKsix-
\nnbb & &
\hphantom{{}+\biggl[+(\zeta_1 \cdot \zeta_4)\biggl\{-(\zeta_5\cdot
    k_4)\biggl( }
-(\zeta_2 \cdot k_4)(\zeta_3 \cdot k_1) \cKfour
\biggr)+
\nnbb &&
\hphantom{{}+\biggl[+(\zeta_1 \cdot \zeta_4)\biggl\{}
+(\zeta_5\cdot k_2)\biggl(  (\zeta_2 \cdot
k_1)(\zeta_3 \cdot k_4) \cKthree -
\nnbb & &
\hphantom{{}+\biggl[+(\zeta_1 \cdot \zeta_4)\biggl\{+(\zeta_5\cdot
    k_2)\biggl(} 
-(\zeta_2 \cdot k_4)(\zeta_3 \cdot k_1) \cKfour
\biggr)\biggr\}\biggr]+
\nnbb &&
{}+\left( \mbox{cyclic
permutations of indexes (1,2,3,4,5)} \right).
\label{A(3)}\end{eqnarray}}

In the appendix~\ref{appD} we compute the same amplitudes, using the
standard Feynman diagram technique, from the action~(\ref{Ltotal}) and
verify explicitly that both approaches give the same result.

\section{The Feynman rules}\label{appC}

In this appendix we derive the Feynman rules from the lagrangian
${\cal L}={\cal L}_{(0,2)} + {\cal L}_{(3)}$, where ${\cal L}_{(0,2)}$
and ${\cal L}_{(3)}$ are given in~(\ref{L2}) and~(\ref{L3}),
respectively. We will follow the standard procedure employed in
non-abelian gauge theories.  The gauge fields are matrices in the \un
internal space, so that $A_\mu = A^a_\mu\, \lambda^a$, where the
hermitian generators $\lambda^a$ satisfy the usual relations
\be\label{colmat} \tr \lambda^a \, \lambda^b =
\delta^{ab}\,;\qquad \left[\lambda^a, \lambda^b\right] = i\, f^{abc} \,
\lambda^c\,.
\ee
Adding the gauge fixing term
\be
{\mathcal L}_{GF} = -{1\over 2\alpha} \tr (\pd_\mu\, A^\mu)^2
\ee
and choosing the \emph{Feynman gauge} condition $\alpha = 1$, we
obtain from eq.~(\ref{L2}) the propagator
\begin{equation}
\begin{array}{c}
\includegraphics*{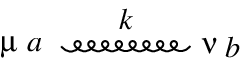}
\end{array}
 :  \displaystyle{\delta^{a b}\; {\eta_{\mu\nu}\over k^2}}\,,
\label{AA}
\end{equation}
where $k^2 = -k_0^2 + \vec k^2$, in agreement with the convention for
$\eta_{\mu \nu}$, given in~(\ref{metric}).  Since we are only
interested in the tree level amplitudes, it will not be necessary to
include \emph{ghost fields}.

The three point-vertex comes only from the standard $F^2$ term in
eq.~(\ref{L2}) and is given~by
\begin{eqnarray} 
\lefteqn{ \begin{array}{c}
\includegraphics*{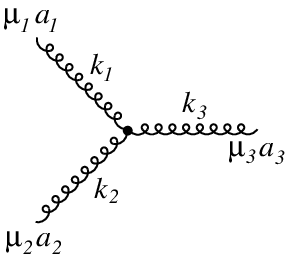}
\end{array} : 
{\mathcal V}_{\mu_1\mu_2\mu_3}^{a_1a_2a_3}(k_1,k_2,k_3)
 =}\qquad\qquad\qquad\qquad\qquad\qquad&&
\nonumber\\[-25pt]&&  
-i\,g\,f^{a_1 a_2 a_3}\, \Bigl[
(k_1-k_2)_{\mu_3}\, \eta_{\mu_1\mu_2}  +
\nonumber\\&&
\hphantom{-i\,g\,f^{a_1 a_2 a_3}\, \Bigl[}
+  
(k_2-k_3)_{\mu_1}\, \eta_{\mu_2\mu_3} +
(k_3-k_1)_{\mu_2}\, \eta_{\mu_1\mu_3} \Bigr] 
\nonumber\\ 
& \equiv & g\,f^{a_1 a_2 a_3}\, V_{\mu_1\mu_2\mu_3}(k_1,k_2,k_3)\,.
\label{AAA}
\end{eqnarray}
Our momentum conventions in the previous vertex as well as in the
following ones are such that the momenta are all inwards. Also, when
computing an $N$ gluon amplitude, factors
$-i\,(2\pi)^{10}\,\delta^{10}(k_1+k_2+\cdots+k_N)$ and $i/(2\pi)^{10}$
are to be included for each interaction vertex and internal lines,
respectively.  In some steps of the calculation, it will be convenient
to employ the relations~(\ref{colmat}) so that we can
write~(\ref{AAA}) as follows
\begin{equation}\label{v30}
{\mathcal V}_{\mu_1\mu_2\mu_3}^{a_1a_2a_3}(k_1,k_2,k_3)  =
- g\,{\sumpp}\left(\lambda^{a_1}\lambda^{a_2} \lambda^{a_3}\right)
\left[
\left({k_1}_{\mu_3}\, \eta_{\mu_1\mu_2}- {k_1}_{\mu_2}\,
      \eta_{\mu_1\mu_3}\right)
+ {\rm cyclic\; perm} \right]
\end{equation}
where the meaning of $\sum_{\rm perm^{\prime}}$ is similar to the one
in eq.~(\ref{general-amplitude}) (except that here we have Lorentz
indices instead of polarizations) and the terms in \emph{cyclic perm}
are obtained adding the cyclic permutations of the Lorentz indices and
momenta.  The four and five gluon vertices can be expressed similarly
as follows
\begin{equation}
\begin{array}{c}
\includegraphics*{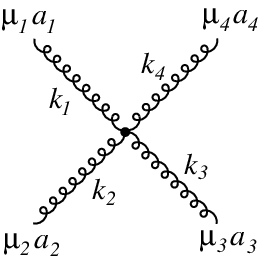}
\end{array}
 \! :\!  {\mathcal
V}_{\mu_1\mu_2\mu_3\mu_4}^{a_1a_2a_3a_4}(k_1,k_2,k_3,k_4)\!=\! g^2
{\sumpp} (\lambda^{a_1}\lambda^{a_2}
\lambda^{a_3}\lambda^{a_4})
V_{\mu_1\mu_2\mu_3\mu_4}(k_1,k_2,k_3,k_4) \label{AAAA}
\end{equation}
and
\begin{eqnarray} 
\lefteqn{ \begin{array}{c}
\includegraphics*{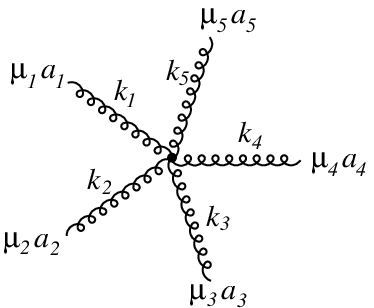}
\end{array}  : 
{\mathcal V}_{\mu_1\ldots \mu_5}^{a_1a_2\ldots a_5}
(k_1,\ldots ,k_5) =}
\qquad\qquad\qquad\qquad\qquad\qquad&&
\nonumber\\[-25pt] &&
g^3\,{\sumpp} (\lambda^{a_1}\lambda^{a_2}\ldots\lambda^{a_5})
V_{\mu_1\mu_2\ldots \mu_5}(k_1,k_2,\ldots,k_5)\,,\qquad
\label{AAAAA}
\end{eqnarray} 
where the Lorentz factors can be derived from~(\ref{L2})
and~(\ref{L3}) in a straightforward way. In the following we will
present the results for the terms of order zero, two and three in
$\alp$.  Let us begin with the four gluon vertex. The zeroth order
contribution can be easily obtained from the first term in~(\ref{L2})
and is given by
\be\label{v40} V^{(0)}_{\mu_1\mu_2\mu_3\mu_4}
= -\frac 1 2 \left(\eta_{\mu_1\mu_2}\,\eta_{\mu_3\mu_4}-
                     \eta_{\mu_1\mu_3}\,\eta_{\mu_2\mu_4}\right)
+ {\rm cyclic\; perm} .
\ee
The terms proportional to ${\alp}^2$ are also straightforward, but
much more involved. We have employed computer algebra to extract this
and all higher order vertex contributions. Exploring the tensor
symmetry we obtain the following result
\begin{eqnarray}
\lefteqn{  \left({6\over \pi^2 {\alp}^2}\right)
V^{(2)}_{\mu_1\ldots\mu_4}(k_1\ldots k_4)  = }\quad&& 
\nonumber\\ &&
\Bigl\{\Bigl[
\left({k_2}\cdot{k_3}\,\eta_{{\mu_3\mu_4}}-
{k_2}_{{\mu_3}}{k_3}_{{\mu_4}}\right)
\left({k_1}\cdot{k_4}\,\eta_{{\mu_1\mu_2}}-
{k_4}_{{\mu_1}}{k_1}_{{\mu_2}}\right)+
\nnbb&&
\hphantom{\Bigl\{\Bigl[} 
+   (k_1,\mu_1)\,(k_4,\mu_4)
\longleftrightarrow (\mu_1,k_1)\,(\mu_4,k_4)\Bigr] +
\nnbb&&
\hphantom{\Bigl\{}
+\Bigl[ {(k_1,\mu_1)\,(k_2,\mu_2)\,(k_3,\mu_3)\,(k_4,\mu_4)}
\longleftrightarrow
{(\mu_1,k_1)\,(\mu_2,k_2)\,(\mu_3,k_3)\,(\mu_4,k_4)}\Bigr]\Bigr\}+
\nnbb&&
{}+ 2\times \{{\rm previous\; curly\; bracket\; with}\;
(\mu_3,k_3)\longleftrightarrow (\mu_4,k_4)\} -
\nnbb&&
{}- 2 \left\{
\left({k_1}\cdot{k_2}\,\eta_{\mu_1\mu_2}-
{k_1}_{{\mu_2}}{k_2}_{{\mu_1}}\right)
\left({k_3}\cdot{k_4}\,\eta_{\mu_3\mu_4}-
{k_3}_{{\mu_4}}{k_4}_{{\mu_3}}\right)
\right\} +
\nnbb&&
{}+ \frac 1 2\times \{{\rm previous\; curly\;
bracket\; with}\; (\mu_2,k_2)\longleftrightarrow (\mu_3,k_3)\} +
\nnbb &&
{}+ {\rm cyclic\; perm},\label{v42}
\end{eqnarray}
where an interchange like $(k_1,\mu_1)\,(k_2,\mu_2)\longleftrightarrow
(\mu_1,k_1)\,(\mu_2,k_2)$ yields, for instance, $k_1\cdot
k_3\longleftrightarrow {k_3}_{\mu_1}$ and \hbox{$k_1\cdot
k_2\longleftrightarrow\eta_{\mu_1\mu_2}$}.  In writing the above
expression, we have organized the terms in such a way that there is a
direct correspondence with the terms of order $\alp^2$ in
eq.~(\ref{L2}). The second and the third term of eq.~(\ref{L2})
generates the first and the second curly brackets of eq.~(\ref{v42}),
respectively. The fourth and the fifth terms of eq.~(\ref{L2})
generates the the last two curly brackets eq.~(\ref{v42}).

Let us now consider the contributions to
$V_{\mu_1\ldots\mu_4}(k_1\ldots k_4)$ which are of order $\alp^3$.
These are generated from the last five terms of eq.~(\ref{L3}) with
the covariant derivative replaced by the usual one. These terms can
also be written, exploring the tensor symmetry, as follows
\begin{eqnarray}
\lefteqn{ \left({1\over \zeta(3) {\alp}^3}\right)
V^{(3)}_{\mu_1\ldots\mu_4}(k_1\ldots k_4)  = }&&
\nonumber\\ && 
\Bigl\{16\Bigl[
\left({k_2}_{\mu_1}k_1\cdot k_3-{k_3}_{\mu_1}k_1\cdot k_2\right)\left(
\eta_{\mu_2\mu_3} \left(k_3\cdot k_4{k_2}_{\mu_4}-k_2\cdot
k_4{k_3}_{\mu_4}\right)
- \mu_2 \longleftrightarrow k_2 \right)+
\nnbb & & 
\hphantom{\Bigl\{16\Bigl[} 
 +  \left(\eta_{\mu_1\mu_3}k_1\cdot
k_2-{k_2}_{\mu_1}{k_1}_{\mu_3}\right) \left({k_3}_{\mu_4}
\left({k_4}_{\mu_2}k_2\cdot k_3-{k_3}_{\mu_2}k_2\cdot k_4\right)
- \mu_4 \longleftrightarrow k_4 \right)\Bigr]\Bigr\}+
\nnbb&&{}+ 
\Bigl\{{8}\,{k_1\cdot k_2}\Bigl[
\left(\eta_{\mu_3\mu_4}\,k_2\cdot
k_4-{k_2}_{\mu_4}\,{k_4}_{\mu_3}\right)
\left(\eta_{\mu_1\mu_2}\,k_1\cdot
k_3-{k_3}_{\mu_1}\,{k_1}_{\mu_2}\right)+
\nnbb & & 
\hphantom{{}+ \Bigl\{{8}\,{k_1\cdot k_2}\Bigl[}
 + (k_1,\mu_1)\,(k_3,\mu_3)
   \longleftrightarrow (\mu_1,k_1)\,(\mu_3,k_3)\Bigr]+
\nnbb & & 
\hphantom{{}+ \Bigl\{}
+ \left[ {(k_1,\mu_1)\,(k_2,\mu_2)\,(k_3,\mu_3)\,(k_4,\mu_4)}
\longleftrightarrow
{(\mu_1,k_1)\,(\mu_2,k_2)\,(\mu_3,k_3)\,(\mu_4,k_4)} \right]\Bigr\}+
\nnbb&&
{}+ \left\{{\rm previous\; curly\; bracket\; with}\;
(\mu_2,k_2)\longleftrightarrow (\mu_3,k_3)\right\} -
\nnbb&&{}- 
\left\{8\,k_1\cdot k_3\left[
\left({k_2}\cdot{k_4}\eta_{\mu_2\mu_4}-{k_4}_{\mu_2}\,{k_2}_{\mu_4}\right)
\left({k_1}\cdot{k_3}\eta_{\mu_1\mu_3}-{k_1}_{\mu_3}\,{k_3}_{\mu_1}\right)
\right]\right\} +
\nnbb & &
{}+ \left\{16\, k_1\cdot k_4
\left(\eta_{\mu_1\mu_3}k_1\cdot k_3-{k_1}_{\mu_3}{k_3}_{\mu_1}\right)
\left(\eta_{\mu_2\mu_4}k_2\cdot k_3-{k_3}_{\mu_2}{k_2}_{\mu_4}\right)
- \mu_4 \longleftrightarrow k_4 \right\}+
\nnbb&&
{}+ {\rm cyclic\; perm}\,.
\label{v43}
\end{eqnarray}
Similarly to the eq.~(\ref{v42}), each one of the five brackets in the
previous expression is in direct correspondence with the last five
terms in eq.~(\ref{L3}). The compact and symmetric form of these
expressions makes then quite useful when performing the calculations
of scattering amplitudes. Everything can be expressed in terms of the
product of two tensors of the form of the zeroth order four-gluon
vertex given by~(\ref{v40}).

An important property of the four-gluon vertices in eq.~(\ref{v42})
and~(\ref{v43}) is that they obey simple Ward identities like
\begin{eqnarray} \label{wi4} 
k_1^{\mu_1}\, V^{(2,3)}_{\mu_1\ldots\mu_4}(k_1\ldots
k_4)&=& k_2^{\mu_2}\, V^{(2,3)}_{\mu_1\ldots\mu_4}(k_1\ldots k_4)=
k_3^{\mu_3}\, V^{(2,3)}_{\mu_1\ldots\mu_4}(k_1\ldots k_4)=
\nonumber\\ 
&=&k_4^{\mu_4}\, V^{(2,3)}_{\mu_1\ldots\mu_4}(k_1\ldots k_4)=0\,.
\end{eqnarray}
These identities are a direct consequence of the {invariance} of the
Born-Infeld action under a non-abelian gauge transformation
\be\label{gt} 
A_\mu \longrightarrow A_\mu - \frac 1 g \pd_\mu\omega +
i\left[A_\mu,\omega \right],
\ee
where $\omega$ is an infinitesimal matrix. The \emph{zero} on the
right hand side of eq.~(\ref{wi4}) reflects the absence of $A^3$ terms
of order $\alp^2$ or higher. In fact, since each individual term in
the lagrangians~(\ref{L2}) and~(\ref{L3}) is gauge invariant by
itself, the identity~(\ref{wi4}) is fulfilled by each curly bracket in
eqs.~(\ref{v42}) and~(\ref{v43}).

The contribution of order $\alp^2$ to the five-gluon vertex can also
be written in a very compact form as follows
\begin{eqnarray}\label{v52}
\lefteqn{ \left(6\over4
  \pi^2\alpha'^2\right)V_{\mu_1\ldots\mu_5}^{(2)}(k_1\ldots
  k_5)=}&&
\nonumber\\ &&
\Bigl\{\Bigr[\left(k_1\cdot
  k_5\eta_{\mu_1\mu_2}-{k_1}_{\mu_2}{k_5}_{\mu_1}\right)
\left({k_2}_{\mu_3}\eta_{\mu_4\mu_5}-{k_2}_{\mu_4}\eta_{\mu_3\mu_5}\right)+
\nonumber  \\&&
\hphantom{\Bigl\{\Bigr[} 
{}+(\mu_2,k_2)\leftrightarrow (\mu_5,k_5)\Bigr]
  -\mu_5\leftrightarrow k_5 \Bigr\} +
\nonumber \\
&&{}+ \Bigl\{\left[\left(k_1\cdot
  k_4\eta_{\mu_1\mu_2}-{k_1}_{\mu_2}{k_4}_{\mu_1}\right)
\left({k_5}_{\mu_4}\eta_{\mu_3\mu_5}-{k_5}_{\mu_3}\eta_{\mu_4\mu_5}\right)
-(\mu_2,k_2)\leftrightarrow (\mu_3,k_3)\right]+
\nonumber \\&&
\hphantom{{}+ \Bigl\{}
{}+  \left[\left(\eta_{\mu_1\mu_4}\eta_{\mu_2\mu_3}-\eta_{\mu_1\mu_3}
\eta_{\mu_2\mu_4}\right)
\left({k_3}_{\mu_5}k_4\cdot k_5-{k_4}_{\mu_5}k_3\cdot
  k_5\right)-\mu_3\leftrightarrow k_3\right]+
\nonumber \\&&
\hphantom{{}+ \Bigl\{}
- \mu_4\leftrightarrow k_4\Bigr\} +
\nonumber \\&&
{}+ \left\{\left({k_1}_{\mu_3}\eta_{\mu_1\mu_2}-
{k_1}_{\mu_2}\eta_{\mu_1\mu_3}\right)\left(k_4\cdot
  k_5\eta_{\mu_4\mu_5}-{k_4}_{\mu_5}{k_5}_{\mu_4}\right)-
(\mu_1,k_1)\leftrightarrow
  (\mu_3,k_3)\right\} +
\nonumber \\&&
{}+ \left\{\left(k_1\cdot
  k_3\eta_{\mu_1\mu_3}-{k_1}_{\mu_3}{k_3}_{\mu_1}\right)
\left({k_2}_{\mu_5}\eta_{\mu_2\mu_4}-{k_2}_{\mu_4}\eta_{\mu_2\mu_5}\right)
\right\}+
\nnbb&&
{}+ {\rm cyclic\; perm}.
\end{eqnarray}
Notice that after including all the cyclic permutations this
expression generates $300$ terms.

We have verified that~(\ref{v52}) and~(\ref{v42}) are related by the
following Ward identity
\be\label{wi542}
{k_1}^{\mu_1}\,V^{(2)}_{\mu_1\ldots\mu_5}(k_1,\ldots,k_5) =
V^{(2)}_{\mu_2\mu_3\mu_4\mu_5}(k_1+k_2,k_3,k_4,k_5)
\nnbb -
V^{(2)}_{\mu_2\mu_3\mu_4\mu_5}(k_2,k_3,k_4,k_1+k_5),
\ee
which is again a consequence of the gauge invariance of the effective
action~(\ref{L2}) under the gauge transformation~(\ref{gt}).

Finally, the contributions of order $\alp^3$ to the five gluon vertex
are given by
{\small
\begin{eqnarray}\label{v53}
\lefteqn{ \left(1\over 16\zeta(3)
  \alpha'^3\right)V_{\mu_1\ldots\mu_5}^{(3)}(k_1\ldots k_5)=}&&
\nnbb&&
\Big\{\left[({k_1}_{\mu_4}\eta_{\mu_1\mu_2}-
{k_1}_{\mu_2}\eta_{\mu_1\mu_4})(k_4\cdot
  k_5{k_3}_{\mu_5}-k_3\cdot
  k_5{k_4}_{\mu_5}){k_2}_{\mu_3}+(\mu_4,k_4)(\mu_5,k_5)\leftrightarrow
(k_4,\mu_4)(k_5,\mu_5)\right]{+}
\nnbb&&
\hphantom{\Big\{} 
+\left[({k_3}_{\mu_5}\eta_{\mu_2\mu_3}{-}
{k_3}_{\mu_2}\eta_{\mu_3\mu_5})(k_4\cdot
  k_5\eta_{\mu_1\mu_4}{-}{k_4}_{\mu_1}{k_5}_{\mu_4})k_1\cdot
  k_2+(\mu_1,k_1)(\mu_4,k_4)\leftrightarrow
  (k_1,\mu_1)(k_4,\mu_4)\right]\!{+}
\nonumber\\&& 
\hphantom{\Big\{} 
+(\mu_3,k_3)(\mu_5,k_5)\leftrightarrow
  (k_3,\mu_3)(k_5,\mu_5)\Bigr\}_{(1)}{+}
\nnbb&& 
{}+\Bigl\{{\rm previous\; curly\; bracket\; with\;}
  (\mu_1,k_1)(\mu_2,k_2)(\mu_3,k_3)(\mu_4,k_4)(\mu_5,k_5)\leftrightarrow
\nnbb&& 
\hphantom{{}+\Bigl\{} 
\leftrightarrow 
  (\mu_1,k_1)(\mu_3,k_3)(\mu_5,k_5)(\mu_4,k_4)(\mu_2,k_2)\Bigr\}_{(2)}{+}
\nnbb&& 
{}+\left\{\left[\left ({ k_1}\cdot{ k_2}\eta_{{{
  \mu_1}{ \mu_4}}}{-}{ k_1}_{{{ \mu_4}}} { k_2}_{{{ \mu_1}}}\right
  )\!\left ({ k_3}\cdot{ k_5} \eta_{{{ \mu_3}{ \mu_5}}}{-}{ k_3}_{{{
  \mu_5}}}{ k_5}_{{{ \mu_3}}} \right ){ k_4}_{{{
  \mu_2}}}-(\mu_2,k_2)\leftrightarrow(\mu_4,k_4)\right]-\mu_4\leftrightarrow
  k_4\right\}_{(3)} \!\!\!{+}
\nnbb&& 
{}+\Bigl\{\left[\left ({ k_1}\cdot { k_2}\eta_{{{
\mu_1}{ \mu_4}}}-{ k_1}_{{{ \mu_4}}}{ k_2}_{{{ \mu_1}}} \right )\left
({ k_2}\cdot { k_4}{ k_5}_{{{ \mu_2}}}-{ k_2}\cdot { k_5}{ k_4}_{{{
\mu_2}}}
\right )\eta_{{{ \mu_3}{ \mu_5}}}-\mu_5\leftrightarrow
  k_5\right]{+}
\nonumber\\
&& \hphantom{{}+\Bigl\{}
+\left[\left ({ k_5}_{{{ \mu_4}}}\eta_{{{ \mu_2}{
 \mu_5}}}-\eta_{{{ \mu_4}{ \mu_5}}}{ k_5}_{{{ \mu_2}}}\right )\left ({
 k_2}\cdot { k_3}\eta_{{{ \mu_1}{ \mu_3}}}-{ k_2}_{{{ \mu_3}}}{
k_3}_{{{ \mu_1}}}\right ){ k_1}\cdot { k_2}-\mu_1\leftrightarrow
  k_1\right]{+}
\nonumber\\
&& \hphantom{{}+\Bigl\{}
+\left[\left[\left ({ k_5}_{{{ \mu_3}}}\eta_{{{
 \mu_2}{ \mu_5}}}-\eta_{{{ \mu_3}{ \mu_5}}}{ k_5}_{{{
 \mu_2}}}\right ){ k_2}_{{{ \mu_4}}}-\mu_2\leftrightarrow
  k_2\right]\left({ k_1}\cdot { k_2}{ k_4}_{{{ \mu_1}}}-{ k_1}\cdot {
  k_4}{ k_2}_{{{ \mu_1}}}\right)\right]{+}
\nonumber\\
&& \hphantom{{}+\Bigl\{}
+\left[\left ({ k_2}\cdot { k_5}\eta_{{{ \mu_4}{
\mu_5}}}-{ k_2}_{{{ \mu_5}}}{ k_5}_{{{ \mu_4}}}\right )\left({
k_1}\cdot { k_2}\eta_{{{ \mu_1}{ \mu_3}}}-{ k_1}_{{{ \mu_3}}}{
k_2}_{{{
\mu_1}}}\right ){ k_3}_{{{ \mu_2}}}-\mu_3\leftrightarrow
  k_3\right]{+}
\nonumber\\
&& \hphantom{{}+\Bigl\{}
+\left[\left[\left ({ k_1}\cdot { k_4}\eta_{{{
\mu_1}{ \mu_2}}}{-}{ k_1}_{{{ \mu_2}}}{ k_4}_{{{ \mu_1}}}
\right )\left ({ k_4}_{{{ \mu_5}}}{ k_5}_{{{ \mu_3}}}{-}\eta_{{{ \mu_3}{
 \mu_5}}}{ k_4}\cdot {
 k_5}\right ){ k_3}_{{{ \mu_4}}}{-}\mu_3\leftrightarrow
  k_3\right]{-}(\mu_2,k_2)\leftrightarrow (\mu_3,k_3)\right]\!{+}
\nonumber\\
&& \hphantom{{}+\Bigl\{}
+\left[\left[\left ({ k_1}_{{{ \mu_4}}}\eta_{{{
 \mu_1}{ \mu_3}}}{-}{ k_1}_{{{ \mu_3}}}\eta_{{{ \mu_1}{ \mu_4}}}\right
 )\left ({ k_4}_{{{ \mu_2}}}{ k_2}_{{{ \mu_5}}}{-}\eta_{{{ \mu_2}{
 \mu_5}}}{
 k_2}\cdot { k_4}\right ){ k_4}\cdot { k_5}{-}\mu_5\leftrightarrow
  k_5\right]{-}(\mu_2,k_2){\leftrightarrow} (\mu_3,k_3)\right]\!{+}
\nonumber\\
&& \hphantom{{}+\Bigl\{}
+\left[\left[\left ({ k_1}\cdot { k_2}\eta_{{{
\mu_1}{ \mu_3}}}{-}{ k_1}_{{{ \mu_3}}}{ k_2}_{{{ \mu_1}}} \right )\left
({ k_3}_{{{ \mu_2}}}{ k_2}_{{{ \mu_5}}}{-}{ k_2}\cdot { k_3}\eta_{{{
\mu_2}{
\mu_5}}}\right ){-}\mu_3\leftrightarrow
  k_3\right]{k_3}_{\mu_4}{-}(\mu_4,k_4)\leftrightarrow
  (\mu_5,k_5)\right]\!{-}
\nonumber\\
&& \hphantom{{}+\Bigl\{}
-\left[\left ({ k_2}_{{{ \mu_5}}}\eta_{{{ \mu_2}{
 \mu_3}}}-{ k_2}_{{{ \mu_3}}}\eta_{{{ \mu_2}{ \mu_5}}}\right )\left ({
 k_1}_{{{ \mu_4}}}{ k_2}_{{{ \mu_1}}}-{ k_1}\cdot { k_2}\eta_{{{
\mu_1}{ \mu_4}}}\right ){ k_5}\cdot\left( { k_4}+{
  k_3}\right)-\mu_5\leftrightarrow k_5\right]{+}
\nonumber\\
&& \hphantom{{}+\Bigl\{}
+\left[\left[\left ({ k_4}\cdot { k_5}\eta_{{{
\mu_2}{ \mu_5}}}{-}{ k_4}_{{{ \mu_5}}}{ k_5}_{{{ \mu_2}}} \right )\!\left
({ k_1}\cdot { k_4}\eta_{{{ \mu_3}{ \mu_4}}}{-}{ k_1}_{{{ \mu_4}}}{
k_4}_{{{
\mu_3}}}\right ){-}(\mu_3,k_3)\leftrightarrow
  (\mu_2,k_2)\right]{k_2}_{\mu_1}-\mu_1{\leftrightarrow}
  k_1\right]{+}
\nonumber\\
&& \hphantom{{}+\Bigl\{}
+\left[\left[\left (\eta_{{{ \mu_3}{ \mu_5}}}{
k_4}\cdot { k_5}{-}{ k_4}_{{{ \mu_5}}}{ k_5}_{{{ \mu_3}}}
\right )\!\left ({ k_4}_{{{ \mu_2}}}{ k_3}_{{{ \mu_4}}}{-}\eta_{{{ \mu_2}{
 \mu_4}}}{ k_3}\cdot {
 k_4}\right ){-}\mu_3{\leftrightarrow}
  k_3\right]{k_3}_{\mu_1}-(\mu_1,k_1)\leftrightarrow
  (\mu_2,k_2)\right]{+}
\nonumber\\
&& \hphantom{{}+\Bigl\{}
+\left[\left ({ k_1}\cdot { k_2}\eta_{{{ \mu_1}{
\mu_3}}}-{ k_1}_{{{ \mu_3}}}{ k_2}_{{{ \mu_1}}} \right )\left ({
k_2}\cdot { k_5}\eta_{{{ \mu_2}{ \mu_4}}}-{ k_2}_{{{ \mu_4}}}{
k_5}_{{{
\mu_2}}}\right ){ k_3}_{{{ \mu_5}}}-\mu_5\leftrightarrow
  k_5\right]{+}
\nonumber\\
&& \hphantom{{}+\Bigl\{}
+\left[\left ({ k_2}\cdot { k_5}{ k_4}_{{{
  \mu_5}}}-{ k_4}\cdot { k_5}{ k_2}_{{{ \mu_5}}}\right )\eta_{{{
  \mu_2}{ \mu_4}}}-\mu_2\leftrightarrow k_2\right]\left ({ k_1}\cdot {
  k_2}\eta_{{{ \mu_1}{ \mu_3}}}-{ k_1}_{{{ \mu_3}}}{ k_2}_{{{
  \mu_1}}}\right ){+}
\nonumber\\
&& \hphantom{{}+\Bigl\{}
+\left[\left ({ k_4}_{{{ \mu_1}}}{ k_1}\cdot {
k_3}-{ k_3}_{{{ \mu_1}}}{ k_1}\cdot { k_4}\right )
\eta_{{{ \mu_2}{ \mu_4}}}-\mu_4\leftrightarrow k_4\right]\left ({
  k_4}\cdot { k_5}\eta_{{{ \mu_3}{ \mu_5}}}-{ k_4}_{{{ \mu_5}}}{
  k_5}_{{{ \mu_3}}}\right ){+}
\nonumber\\
&& \hphantom{{}+\Bigl\{}
+\left[\left ({ k_4}\cdot { k_5}\eta_{{{
\mu_2}{ \mu_5}}}-{ k_4}_{{{ \mu_5}}}{ k_5}_{{{ \mu_2}}} \right )\left
({ k_1}{\cdot k_4}\eta_{{{ \mu_3}{ \mu_4}}}-{ k_1}_{{{ \mu_4}}}{
k_4}_{{{
\mu_3}}}\right ){ k_3}_{{{ \mu_1}}}-\mu_1\leftrightarrow
  k_1\right]\Bigr\}_{(4)}{+}
\nnbb&& 
{}+{1\over 2}\Bigl\{\bigl[\bigl[\left[\left ({
 k_3}_{{{ \mu_2}}}\eta_{{{ \mu_1}{ \mu_3}}}-\eta_{{{ \mu_2}{ \mu_3}}}{
 k_3}_{{{ \mu_1}}}\right )\left ({ k_2}_{{{ \mu_4}}}{ k_1}_{{{
 \mu_5}}}-{ k_2}_{{{ \mu_5}}}{ k_1}_
{{{ \mu_4}}}\right )-(\mu_3,k_3)\leftrightarrow
  (\mu_5,k_5)\right]{-}
\nonumber\\ &&
\hphantom{{}+{1\over 2}\Bigl\{\bigl[\bigl[} 
-\mu_2\leftrightarrow k_2\bigr]k_2\cdot
  k_3{-}\mu_1\leftrightarrow k_1\bigr]{+}
\nonumber\\
&& 
\hphantom{{}+{1\over 2}\Bigl\{}
+\left[\left[\left (\eta_{{{ \mu_2}{
 \mu_4}}}\eta_{{{ \mu_3}{ \mu_5}}}{-}\eta_{{{ \mu_2}{ \mu_5}}}\eta_{{{
 \mu_3}{ \mu_4}}}\right )\!\left ({ k_1}{\cdot} { k_4}{ k_5}_{{{
 \mu_1}}}{-}{ k_1}{\cdot} { k_5}{
k_4}_{{{ \mu_1}}}\right )\!{-}\mu_4\leftrightarrow k_4\right]\!k_4\cdot
  \left(k_2+k_3\right){-}\mu_5\leftrightarrow k_5\right]\!{+}
\nonumber\\
&& \hphantom{{}+{1\over 2}\Bigl\{}
+\bigl[\left[\left (\eta_{{{ \mu_2}{ \mu_5}}}{
 k_5}_{{{ \mu_3}}}-{ k_5}_{{{ \mu_2}}}\eta_{{{ \mu_3}{ \mu_5}}}\right
 )\left ({ k_3}\cdot { k_4}\eta_{{{ \mu_1}{ \mu_4}}}-{ k_3}_{{{
 \mu_4}}}{
k_4}_{{{ \mu_1}}}\right )+(\mu_4,k_4)\leftrightarrow
  (\mu_5,k_5)\right]k_3\cdot k_4{-}
\nonumber\\ &&
\hphantom{{}+{1\over 2}\Bigl\{+\bigl[} 
-(\mu_1,k_1)\leftrightarrow
  (\mu_2,k_2)\bigr]{+}
\nonumber\\
&& \hphantom{{}+{1\over 2}\Bigl\{}
+\left[\left[\left ({ k_5}_{{{ \mu_2}}}\eta_{{{
 \mu_1}{ \mu_5}}}-\eta_{{{ \mu_2}{ \mu_5}}}{ k_5}_{{{ \mu_1}}}\right
 )\left ({ k_1}_{{{ \mu_4}}}{ k_2}_{{{ \mu_3}}}-{ k_1}_{{{ \mu_3}}}{
 k_2}_
{{{ \mu_4}}}\right )-\mu_2\leftrightarrow k_2\right]k_2\cdot
  k_4-\mu_1\leftrightarrow k_1\right]{+}
\nonumber\\
&& \hphantom{{}+{1\over 2}\Bigl\{}
+\left[\left[\left ({ k_2}\cdot { k_3}{ k_1}_{{{
    \mu_3}}}-{ k_2}_{{{ \mu_3}}}{ k_1}\cdot { k_3}\right )
\left (\eta_{{{ \mu_2}{ \mu_5}}}{ k_5}_{{{ \mu_1}}}-{ k_5}_{{{
 \mu_2}}}\eta_{{{ \mu_1}{
 \mu_5}}}\right )-\mu_2\leftrightarrow
  k_2\right]{k_2}_{\mu_4}-\mu_1\leftrightarrow k_1\right]{+}
\nonumber\\
&& \hphantom{{}+{1\over 2}\Bigl\{}
+\left[\left[\left ({ k_5}_{{{ \mu_1}}}\eta_{{{
 \mu_3}{ \mu_5}}}-\eta_{{{ \mu_1}{ \mu_5}}}{ k_5}_{{{ \mu_3}}}\right
 )\left ({ k_3}_{{{ \mu_4}}}{ k_1}\cdot { k_4}-{ k_3}\cdot { k_4}{
 k_1}_{{{
 \mu_4}}}\right )-\mu_3\leftrightarrow
  k_3\right]{k_4}_{\mu_2}-\mu_1\leftrightarrow k_1\right]{+}
\nonumber\\
&& \hphantom{{}+{1\over 2}\Bigl\{}
+\left[\left[\left ({ k_4}_{{{
 \mu_1}}}\eta_{{{ \mu_2}{ \mu_4}}}{-}\eta_{{{ \mu_1}{ \mu_4}}}{ k_4}_{{{
 \mu_2}}}\right )\left ({ k_1}\cdot { k_5}{ k_2}_{{{ \mu_5}}}{-}{
 k_1}_{{{ \mu_5}}}{ k_2}\cdot {
 k_5}\right ){-}\mu_2{\leftrightarrow}
  k_2\right]{\left(k_2-k_4\right)}_{\mu_3}{-}\mu_1{\leftrightarrow}
  k_1\right]\Bigr\}_{(5)}\!\!\!{+}
\nonumber\\
&& {}+\Bigl\{\left[\left[\left (\eta_{{{ \mu_2}{
 \mu_5}}}\eta_{{{ \mu_1}{ \mu_4}}}{-}\eta_{{{ \mu_2}{ \mu_4}}}\eta_{{{
 \mu_1}{ \mu_5}}}\right )\left ({ k_1}\cdot{ k_3}{ k_2}_{{{ \mu_3}}}{-}{
 k_1}_{{{ \mu_3}}}{
 k_2}\cdot{ k_3}\right )-\mu_2\leftrightarrow k_2\right]
  k_2\cdot\left(k_4+k_5\right)-\mu_1\leftrightarrow
  k_1\right]{+}
\nonumber\\
&& \hphantom{ {}+\Bigl\{}
+\left[\left[\left ({ k_1}\cdot{ k_5}\eta_{{{
\mu_1}{ \mu_4}}}{-}{ k_1}_{{{ \mu_4}}}{ k_5}_{{{ \mu_1}}} \right )\left
({ k_2}_{{{ \mu_3}}}\eta_{{{ \mu_2}{ \mu_5}}}{-}{ k_2}_{{{
\mu_5}}}\eta_{{{
\mu_2}{ \mu_3}}}\right ){-}\mu_5{\leftrightarrow} k_5\right] k_2\cdot
  k_5{-}(\mu_3,k_3){\leftrightarrow} (\mu_4,k_4)\right]{+}
\nonumber\\
&& \hphantom{ {}+\Bigl\{}
+\left[\left[\left ({ k_1}_{{{ \mu_5}}}{ k_3}_{{{
\mu_1}}}-{ k_1}\cdot{ k_3}\eta_{{{ \mu_1}{ \mu_5}}}
\right )\left ({ k_5}_{{{ \mu_2}}}{ k_2}_{{{ \mu_3}}}-\eta_{{{ \mu_2}{
 \mu_3}}}{ k_2}\cdot{
 k_5}\right )-\mu_3\leftrightarrow k_3\right]-\mu_5\leftrightarrow
  k_5\right]{k_2}_{\mu_4}{-}
\nonumber\\
&& \hphantom{ {}+\Bigl\{}
-\bigl[{\rm previous\; bracket\; with\;}
  (\mu_1,k_1)(\mu_2,k_2)(\mu_3,k_3)(\mu_4,k_4)(\mu_5,k_5)\leftrightarrow
\nonumber\\ &&
\hphantom{ {}+\Bigl\{-\bigl[}
\leftrightarrow
  (\mu_4,k_4)(\mu_3,k_3)(\mu_1,k_1)(\mu_5,k_5)(\mu_2,k_2)\bigr]
{k_2}_{\mu_5}\Bigr\}_{(6)}{+}
\nonumber \\
&& {}+\Bigl\{\left[\left ({ k_2}\cdot{k_4}\eta_{{{
\mu_2}{ \mu_4}}}{-}{ k_2}_{{{ \mu_4}}}{ k_4}_{{{ \mu_2}}}
\right ){ k_2}_{{{
 \mu_5}}}{-}(\mu_4,k_4)\leftrightarrow(\mu_5,k_5)+\left ({ k_2}_{{{
 \mu_5}}}\eta_{{{ \mu_2}{ \mu_4}}}{-}{ k_2}_{{{ \mu_4}}}\eta_{{{ \mu_2}{
 \mu_5}}}\right )({ k_4}{+}{ k_5})\cdot {k_2}\right]\!{\times}
\nonumber \\
&& \hphantom{{}+\Bigl\{}
\times \left ({ k_1}\cdot{ k_3}\eta_{{{ \mu_1}{
\mu_3}}}-{ k_1}_{{{ \mu_3}}}{ k_3}_{{{ \mu_1}}}\right )+\left ({
k_2}\cdot{ k_5}\eta_{{{ \mu_2}{ \mu_5}}}-{ k_2}_{{{ \mu_5}}}{ k_5}_{{{
\mu_2}}}\right )\left ({ k_1}_{{{ \mu_4}}}\eta_{{{
\mu_1}{ \mu_3}}}-{ k_1}_{{{ \mu_3}}}\eta_{{{ \mu_1}{ \mu_4}}}\right ){
  k_2}\cdot{ k_5}\Bigr\}_{(7)}\!\!\!\!{+}
\nnbb
&& {}+\Bigl\{\bigl[\bigl[\bigl[\left[\left ({ k_3}_{{{
 \mu_5}}}\eta_{{{ \mu_1}{ \mu_3}}}{-}\eta_{{{ \mu_3}{ \mu_5}}}{ k_3}_{{{
 \mu_1}}}\right )\!\left ({ k_2}_{{{ \mu_4}}}{ k_4}_{{{ \mu_2}}}{-}{
 k_2}\cdot{ k_4}\eta_{{{
\mu_2},{ \mu_4}}}\right ){+}(\mu_4,k_4)(\mu_5,k_5)\leftrightarrow
  (\mu_4,\mu_5)(k_5,\mu_5)\right]\!{-}
\nnbb
&&\hphantom{ {}+\Bigl\{\bigl[\bigl[\bigl[}  
-(\mu_4,k_4)\leftrightarrow
  (\mu_5,k_5)\bigr]k_1\cdot k_2{+}\left ({ k_3}\cdot{ k_5}\eta_{{{
  \mu_3}{ \mu_5}}}-{ k_3}_{{{ \mu_5}}}{ k_5}_{{{ \mu_3}}}
\right )\left ({ k_4}\cdot{ k_5}\eta_{{{ \mu_1}{ \mu_4}}}{-}{ k_5}_{{{
\mu_4}}}{ k_4}_{{{
\mu_1}}}\right ){ k_1}_{{{ \mu_2}}}\bigr]{-}
\nonumber\\ &&
\hphantom{ {}+\Bigl\{\bigl[\bigl[\bigl[}
-\mu_1\leftrightarrow k_1
  \bigr]{+}
\nnbb
&& \hphantom{ {}+\Bigl\{}
+\bigl[\bigl[\left ({ k_5}_{{{ \mu_2}}}\eta_{{{
\mu_3}{ \mu_5}}}-\eta_{{{ \mu_2}{ \mu_5}}}{ k_5}
_{{{ \mu_3}}}\right )\left[\left ({ k_1}\cdot { k_2}\eta_{{{ \mu_1}{
 \mu_4}}}-{ k_1}_{{{ \mu_4}}}{ k_2}_{{{ \mu_1}}}\right )+\left ({
 k_1}\cdot { k_3}\eta_{{{ \mu_1}{ \mu_4}}}-{ k_1}_{{{ \mu_4}}}{ k_3}_
{{{ \mu_1}}}\right )\right]{+}
\nnbb &&\hphantom{ {}+\Bigl\{}
+\left ({ k_1}_{{{ \mu_4}}}\eta_{{{ \mu_1}{ \mu_3}}}-{
k_1}_{{{ \mu_3}} }\eta_{{{ \mu_1}{ \mu_4}}}\right )\left ({ k_2}\cdot
{ k_5}\eta_{{{ \mu_2}{ \mu_5}}}-{ k_2}_
{{{ \mu_5}}}{ k_5}_{{{ \mu_2}}}\right )\bigr]{ k_4}\cdot {
  k_5}-\mu_4\leftrightarrow k_4 \bigr]{+}
\nnbb
&& \hphantom{ {}+\Bigl\{}
+2\!\left[\left ({ k_2}{\cdot} { k_5}\eta_{{{
\mu_2}{ \mu_5}}}{-}{ k_2}_{{{ \mu_5}}}{ k_5}_{{{ \mu_2}}} \right )\!\left
(\left ({ k_1}{\cdot} { k_2}\eta_{{{ \mu_1}{ \mu_4}}}{-}{ k_1}_{{{
\mu_4}}}{ k_2}_
{{{ \mu_1}}}\right ){ k_5}_{{{ \mu_3}}}\!{-}\!\left ({ k_1}{\cdot} {
 k_2}\eta_{{{ \mu_1}{ \mu_3}}}\!{-}{ k_1}_{{{ \mu_3}}}{ k_2}_{{{
 \mu_1}}}\right ){ k_5}_{{{ \mu_4}}}\right
)\right]\!\Bigr\}_{(8)}\!\!\!\!\! {+}
\nnbb && {}+ \mbox{cyclic perm.}
\end{eqnarray}}
Notice that we have numbered the curly brackets in the previous
expression in correspondence with the terms in the
lagrangian~(\ref{L3}).  The $F^5$ terms in~(\ref{L3}) yields the first
three curly brackets in eq.~(\ref{v53}). Since there are no four-gluon
vertex associated with these terms, the corresponding Ward identities
are quite simple insofar as its right hand side is identically
zero. This simple property has been explicitly verified for the first
three curly brackets in eq.~(\ref{v53}). Each of the curly brackets
numbered from (4) to
(8) satisfy an Ward identity similar to the one in
eq.~(\ref{wi542}). Therefore the complete vertex
$V^{(3)}_{\mu_1,\ldots,\mu_5}(k_1,\ldots,k_5)$ satisfy
\be\label{wi543}
{k_1}^{\mu_1}\,V^{(3)}_{\mu_1\ldots\mu_5}(k_1,\ldots,k_5) =
V^{(3)}_{\mu_2\mu_3\mu_4\mu_5}(k_1+k_2,k_3,k_4,k_5) -
V^{(3)}_{\mu_2\mu_3\mu_4\mu_5}(k_2,k_3,k_4,k_1+k_5)\,.
\ee

\section{The three, four and five-point amplitudes}\label{appD}

We will now employ the Feynman rules presented in appendix~\ref{appC}
in order to compute\linebreak the scattering amplitudes ${\cal
A}^{(N)}$ of $N$ gluons with colors $a_1,a_2,\ldots, a_N$,
polarizations\linebreak $\zeta_1,\zeta_2,\ldots,\zeta_N$ and external
momenta $k_1,k_2,\ldots, k_N$, satisfying the physical conditions\linebreak
in~(\ref{onshell}).  In what follows we will present and discuss the
explicit results for $N=3,4,5$ up to order~$\alp^3$.

The three-gluon amplitude ${\cal A}^{(3)}$ can be easily obtained from
eq.~(\ref{v30}). Contracting with
${\zeta_1}_{\mu_1}\,{\zeta_2}_{\mu_2}\,{\zeta_3}_{\mu_3}$ and
inserting the factor $-i\,(2\pi)^{10}\,\delta^{10}(k_1+k_2+k_3)$ one
easily obtains
\be\label{A3} {\mathcal A}^{(3)} = i\,g
\,(2\pi)^{10}\,\delta^{10}(k_1+k_2+k_3)
{\sump}\left(\lambda^{a_1}\lambda^{a_2} \lambda^{a_3}\right)
\left[\left(\zeta_3\cdot k_1\right)\,\left(\zeta_1\cdot
      \zeta_2\right)- \left(\zeta_2\cdot
      k_1\right)\,\left(\zeta_1\cdot \zeta_3\right)\right].
\ee
Using the cyclic invariance of the trace we can write eq.~(\ref{A3})
as follows
\be\label{A3a} {\mathcal A}^{(3)} =
i\,(2\pi)^{10}\,\delta^{10}(k_1+k_2+k_3)
\,{\sumpp}\left(\lambda^{a_1}\lambda^{a_2} \lambda^{a_3}\right)
A(1,2,3)\,,
\ee
where
\be\label{A123n} A(1,2,3) = g\,
\left[\left(\zeta_3\cdot k_1\right)\left(\zeta_1\cdot
      \zeta_2\right)- \left(\zeta_2\cdot
      k_1\right)\left(\zeta_1\cdot \zeta_3\right)\right]
+({\rm cyclic\; perm})
\ee
As we can see, this is the simplest example were general structure of
the amplitudes given in eq.~(\ref{general-amplitude}) arises.
Inserting eq.~(\ref{A123n}) into eq.~(\ref{A3a}) one can easily obtain
the result given in eq.~(\ref{Athree}).  There are some simple
properties of~(\ref{A3a}) which is shared by any other
amplitude. First of all, it exhibits the Bose symmetry under
permutations of the external gluons. Secondly, it is invariant under
Lorentz transformations. And last but not least, it is invariant under
independent gauge transformations $\zeta_i\rightarrow \zeta_i +
\,\alpha_i \, k_i$, ($i=1,2,3$) of the 3 external fields. One should
also remark that~(\ref{A3a}) is an exact tree amplitude independent of
the parameter $\alp$. That would not be so, if the lagrangian
contained a term like $\alp\, F^3$.

Let us make some other observations about the three gluon amplitude.
Making use of the gauge freedom $\zeta_i^{\mu_i}\rightarrow
\zeta_i^{\mu_i} + \,\alpha_i \, k_i^{\mu_i}$, ($i=1,2,3$), one can
always choose a gauge such that all the time components of
$\zeta_i^{\mu_i}$, ($i=1,2,3$) are equal to zero. In this case, the
transversality condition implies that $\vec \zeta_i \perp \vec k_i$,
($i=1,2,3$).  On the other hand, the physical conditions on the
external momenta implies that $\vec k_1\, ||\, \vec k_2$ and $\vec
k_1\, ||\, \vec k_3$.  Therefore, the polarizations of three the
external gluons are all perpendicular to $\vec k_1\, ,\vec k_2$ and
$\vec k_3$ and the amplitude~(\ref{A3a}) vanishes for this choice of
gauge. Since it is gauge invariant, it is always zero. Physically this
is just a consequence of the conservation of the total (spin and
orbital) angular momentum of the three on-shell transverse massless
particles.  Our strategy to the computation of the four and five gluon
amplitudes will be first to derive the
relation~(\ref{general-amplitude}) and then to compute the Lorentz
scalars $A(1,2,3,4)$ and $A(1,2,3,4,5)$.  As we will see, these
Lorentz scalars can be expressed in terms combinations of contractions
of the Lorentz scalars in the Feynman rules.  All we need to do that
are a few simple relations involving the color factors associated with
the diagrams shown in figures~\ref{fA4} and~\ref{fA5}.

The color factor associated with the diagram (a) of figure~\ref{fA4}
can be expressed in terms of traces of the group generators as
\be\label{C4} C_{(4)}^{a_1 a_2 a_3 a_4}\equiv g^2\,f^{a_1 a_2 d} \,
f^{d a_3 a_4} =
- i\, g^2\tr \left(\lambda^{a_1}\,[\lambda^{a_2},
 \lambda^{d}]\right)  f^{d a_3 a_4} = - g^2\tr
 \left(\lambda^{a_1}\,
[\lambda^{a_2}, [\lambda^{a_3}, \lambda^{a_4}]]\right),
\ee
where we have employed eq.~(\ref{colmat}).  Similarly, the diagram (a)
of figure~\ref{fA5} has a color factor
\begin{eqnarray}\label{C5a}
C_{(5a)}^{a_1 a_2 a_3 a_4 a_5}\equiv g^3\, f^{a_1 a_2 d} \, f^{d e
a_5}\, f^{a_3 a_4 e} & = &
- i\, g^3\, \tr \left(\lambda^{a_1}\,[\lambda^{a_2},
 \lambda^{d}]\right) \, f^{d e a_5}\,f^{a_3 a_4 e} 
\nnbb & = &
- g^3\, \tr \left(\lambda^{a_1}\,[\lambda^{a_2},
[\lambda^{e},\lambda^{a_5}]]
\right)\,f^{a_3 a_4 e} 
\nnbb & = & i\, g^3\, \tr
\left(\lambda^{a_1}\,[\lambda^{a_2},
[[\lambda^{a_3},\lambda^{a_4}],\lambda^{a_5}]]\right).
\end{eqnarray}

\FIGURE[t]{\centerline{\epsfig{file=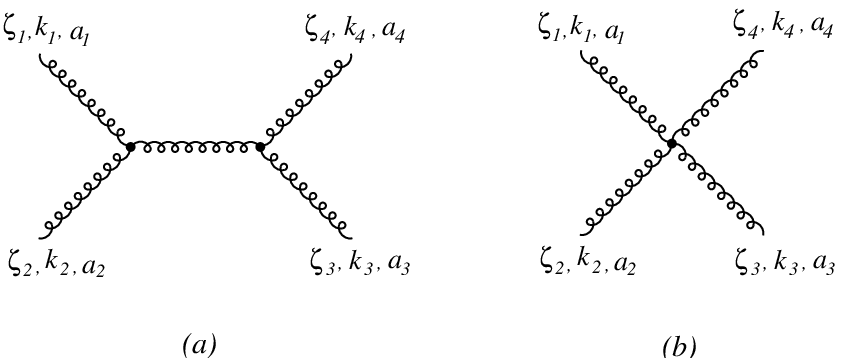}\caption{Basic tree
diagrams which contributes to the four gluon scattering
amplitude. There are three permutations of the diagram (a) which can
be obtained fixing the external gluon ($\zeta_1\,k_1\,a_1$) and cyclic
permuting the other three.}\label{fA4}}}

The diagram ($b$) of figure~\ref{fA5} can be expressed in terms of six
different color factors, each one corresponding to the different
traces in the four gluon vertex given by eq.~(\ref{AAAA}). For
instance, one these factors is
\be\label{C5b} C_{(5b)}^{a_1 a_2 a_3 a_4 a_5} \equiv g^3\,
\tr\left(\lambda^{a_1}\,\lambda^{a_2}\,\lambda^{a_3}\,\lambda^{e}\right)
f^{e a_4 a_5} = -i\, g^3\, \tr \left(\lambda^{a_1}\,\lambda^{a_2}\,
\lambda^{a_3}\,[\lambda^{a_4},\lambda^{a_5}]\right)
\ee
and the other five are obtained by doing permutations of $\{{a_2},
{a_3}, {e}\}$ and replacing $\lambda^{e}$ by
$[\lambda^{a_4},\lambda^{a_5}]$.  We have now the basic ingredients
and definitions in order to generate the Lorentz scalars $A(1,2,3,4)$
and $A(1,2,3,4,5)$.

Using the relation~(\ref{C4}) in the expression associated with the
diagram ($a$) of figure~\ref{fA4} and performing the permutations as
indicated in the figure, we obtain for the Lorentz factor of
$\tr\left(\lambda^{a_1}\,\lambda^{a_2}\,\lambda^{a_3}\,\lambda^{a_4}\right)$
the following expression
\begin{equation}\label{A4gen}
-A(1,2,3,4) = V_{33}(1,2,3,4) - V_{33}(1,4,2,3) - V_{4}(1,2,3,4)\,.
\end{equation}
The explicit form of the Lorentz scalar $V_{33}$ is given by
\be\label{v33} V_{33}(1,2,3,4) =
\zeta_1^{\mu_1}\,\zeta_2^{\mu_2}\,\zeta_3^{\mu_3}\,\zeta_4^{\mu_4}\,
\left.
{V_{\mu_1\mu_2\sigma}(k_1,k_2,-k_1-k_2)
 V_{\;\;\mu_3\mu_4}^{\sigma}(k_1+k_2,k_3,k_4)\over 2\, k_1\cdot k_2}
\right|_{\rm phys},
\ee
where $V$ is the Lorentz factor of the three gluon vertex as defined
in eq.~(\ref{AAA}) and the subscript ``phys'' means that the
conditions~(\ref{onshell}) are being used.  The scalar $V_4$ is given
by
\begin{eqnarray} 
V_{4}(1,2,3,4) &=& 
\zeta_1^{\mu_1}\,\zeta_2^{\mu_2}\,\zeta_3^{\mu_3}\,\zeta_4^{\mu_4}
\Bigl( V^{(0)}_{\mu_1\ldots \mu_4}(k_1, \ldots ,k_4) +
V^{(2)}_{\mu_1\ldots \mu_4}(k_1, \ldots ,k_4) +
\nonumber\\ &&
\hphantom{\zeta_1^{\mu_1}\,\zeta_2^{\mu_2}\,\zeta_3^{\mu_3}\,\zeta_4^{\mu_4}
\Bigl(}
 +V^{(3)}_{\mu_1\ldots
\mu_4}(k_1, \ldots ,k_4) \Bigr)\Bigr|_{\rm phys}\,.
\label{v4} 
\end{eqnarray}
Making a systematic use of the physical conditions~(\ref{onshell}), we
were able to simplify the result to the following form:
\begin{eqnarray}\label{AL4}
-A(1,2,3,4) & = & g^2\left(-\frac{2}{(k_1\cdot k_2)\,(k_1\cdot k_4)}
+ \frac{4\,\pi^2}{3}\,\alp^2 +16\,(k_1\cdot k_3)
\zeta(3)\,\alp^3\right) \times
\nnbb&&{}\times  K(\zeta_1, k_1; \zeta_2, k_2;
 \zeta_3, k_3; \zeta_4, k_4)
\end{eqnarray}
where the kinematic factor $K(\zeta_1, k_1; \zeta_2, k_2; \zeta_3,
k_3;\zeta_4, k_4)$ agrees with~(\ref{K}). Finally, inserting the
factors $-i\, (2\pi)^{10}\delta^{10}(k_1+k_2+k_3+k_4)$, the color
factor $\lambda^{a_1}\lambda^{a_2}\lambda^{a_3}\lambda^{a_4}$ and
adding the remaining non-cyclic permutations, we obtain
\be\label{A4f} {\mathcal A}^{(4)}=
i\,(2\pi)^{10}\,\delta^{10}(k_1+k_2+k_3+k_4)
{\sumpp}\left(\lambda^{a_1}\lambda^{a_2}\lambda^{a_3}\lambda^{a_4}\right)
{A}(1,2,3,4)\,.
\ee
As we can see, eq.~(\ref{A4f}) exhibits the structure of the general
amplitude given by eq. ({\ref{general-amplitude}).  The on-shell gauge
invariance of ${\mathcal A}^{(4)}$ can be argued by the following two
nice properties satisfied by the kinematic factor $K$ present
in~(\ref{AL4}):\footnote{See~\cite[section 7.4.3]{Green:1987sp}.}
it has $total$ symmetry in the four external particles (so $K$ may be
written as a common factor in~(\ref{A4f})) and it vanishes whenever
any $\zeta_i$ is substituted by the corresponding $k_i$, after using
the physical conditions in~(\ref{onshell}) (so $K$ is, itself,
on-shell gauge invariant).  We remark that the $\alp$ corrections are
associated only with the quartic vertex in figure~\ref{fA4} ($b$).

\FIGURE[t]{\centerline{\epsfig{file=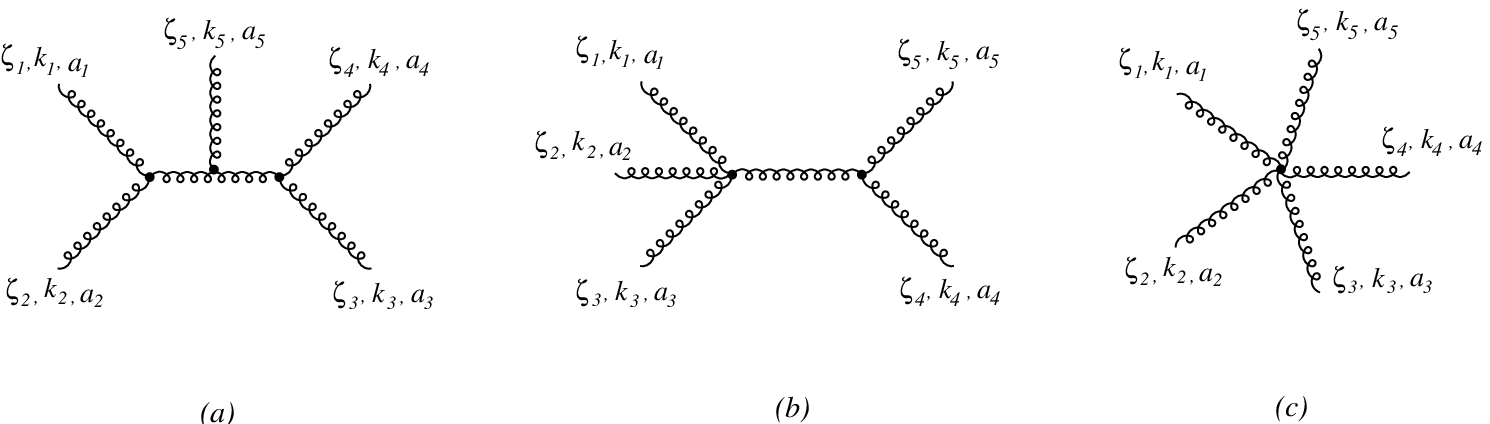}\caption{Basic tree
diagrams which contributes to the five gluon scattering amplitude. The
$15$ permutations of the diagram ($a$) can be grouped in $5$ sets of
three permutations. The first set is obtained fixing the external
gluons ($\zeta_1\,k_1\,a_1$) and ($\zeta_5\,k_5\,a_5$) and cyclic
permuting the other three. The other four sets are obtained from the
first performing $5\leftrightarrow 1$, $5\leftrightarrow 2$,
$5\leftrightarrow 3$ and $5\leftrightarrow 4$.  There are $10$
permutations of diagram ($b$) which correspond to the $10$ distinct
possibilities [$(4,5)$, $(3,5)$, $(2,5)$, $(1,5)$, $(3,4)$, $(2,4)$,
$(1,4)$, $(2,3)$, $(1,3)$, $(1,2)$] for the external gluons in the
cubic vertex}\label{fA5}}}

Let us now consider the five gluon amplitude shown in
figure~\ref{fA5}.  Using eqs.~(\ref{C5a}) and~(\ref{C5b}), performing
the permutations indicated in the figure~\ref{fA5} and collecting the
coefficient of
$\tr(\lambda^{a_1}\lambda^{a_2}\lambda^{a_3}\lambda^{a_4}\lambda^{a_5})$,
we obtain the following expression for the Lorentz scalar
$A(1,2,3,4,5)$
\begin{eqnarray}\label{A5gen}
A(1,2,3,4,5) & = & i\Bigr[ V_{333}(1,2,3,4,5) +
V_{333}(1,5,3,4,2)  +
\nnbb&&\hphantom{i\Bigl[}+ V_{333}(1,2,5,4,3) +
V_{333}(5,4,2,3,1)
- V_{333}(1,5,2,3,4) -
\nnbb&&\hphantom{i\Bigl[}- 
V_{43;1}(1,2,3,4,5) - V_{43;1}(3,4,5,1,2) -
\nnbb&&\hphantom{i\Bigl[}- 
V_{43;3}(1,4,5,2,3) - V_{43;6}(1,2,5,3,4) + V_{43;3}(4,2,3,1,5) +
\nnbb
&&\hphantom{i\Bigl[}+   V_5(1,2,3,4,5)\Bigr]_{{\rm phys}}\,,
\end{eqnarray}
where
\begin{eqnarray}\label{v333}
\lefteqn{ V_{333}(1,2,3,4,5)  =}\qquad &&
\nonumber\\ &&
{V_{\mu_1\mu_2}^{\;\;\;\;\;\rho}(k_1,k_2,-k_1-k_2)
V_{\;\;\mu_3\mu_4}^{\sigma}(-k_3-k_4,k_3,k_4)
V_{\mu_5\rho\sigma}(k_5,k_1+k_2,k_3+k_4) \over (2\, k_1\cdot k_2)(2\,
k_3\cdot k_4)} \times
\nnbb&&{}\times 
\zeta_1^{\mu_1}\,\zeta_2^{\mu_2}\,
\zeta_3^{\mu_3}\,\zeta_4^{\mu_4}\,\zeta_5^{\mu_5}
\Bigr|_{\rm phys},
\\
\label{v43i}
\lefteqn{ V_{43;i}(1,2,3,4,5) =} \qquad&&
\nonumber\\ &&
\Biggl[{V_{(i)\,\mu_1\mu_2\mu_3}^{(0)\;\;\;\;\;\;\;\;\;\;\rho}
(k_1,k_2,k_3,k_4+k_5)V_{\rho\mu_4\mu_5}(-k_4-k_5,k_4,k_5) \over (2\,
k_4\cdot k_5)}+
\nnbb&&\hphantom{\Biggl[} + 
{V_{(i)\,\mu_1\mu_2\mu_3}^{(2)\;\;\;\;\;\;\;\;\;\;\rho}
(k_1,k_2,k_3,k_4+k_5)V_{\rho\mu_4\mu_5}(-k_4-k_5,k_4,k_5) \over (2\,
k_4\cdot k_5)}+
\nnbb&&\hphantom{\Biggl[}+
{V_{(i)\,\mu_1\mu_2\mu_3}^{(3)\;\;\;\;\;\;\;\;\;\;\rho}
(k_1,k_2,k_3,k_4+k_5)V_{\rho\mu_4\mu_5}(-k_4-k_5,k_4,k_5) \over (2\,
k_4\cdot k_5)}\Biggr] \times
\nnbb&&{}\times 
\zeta_1^{\mu_1}\,\zeta_2^{\mu_2}\,\zeta_3^{\mu_3}\,
\zeta_4^{\mu_4}\,\zeta_5^{\mu_5}
\Bigr|_{\rm phys};\qquad (i=1,\ldots, 6)
\end{eqnarray}
and
\be\label{v5} V_{5}(1,2,3,4,5) = 
\zeta_1^{\mu_1}\,\zeta_2^{\mu_2}\,\zeta_3^{\mu_3}\,\zeta_4^{\mu_4}
\,\zeta_5^{\mu_5} \left( V^{(2)}_{\mu_1\ldots \mu_5}(k_1, \ldots ,k_5)
+ V^{(3)}_{\mu_1\ldots \mu_5}(k_1, \ldots ,k_5) \right)\Bigr|_{\rm
phys}.
\ee
The quantities $V_{(i)\,\mu_1\mu_2\mu_3,\mu_4}^{(0),(2),(3)}$
($i=1\dots 6$) are the Lorentz factors of the independent traces (see
eq.~(\ref{AAAA}))
\begin{eqnarray*} 
&&\tr({\lambda}^{a_1}{\lambda}^{a_2}{\lambda}^{a_3}{\lambda}^{a_4})\,,
\\&&
\tr({\lambda}^{a_1}{\lambda}^{a_3}{\lambda}^{a_4}{\lambda}^{a_2})\,,
\\&&
\tr({\lambda}^{a_1}{\lambda}^{a_4}{\lambda}^{a_2}{\lambda}^{a_3})\,,
\\&&
\tr({\lambda}^{a_1}{\lambda}^{a_4}{\lambda}^{a_3}{\lambda}^{a_2})\,,
\\&&
\tr({\lambda}^{a_1}{\lambda}^{a_3}{\lambda}^{a_2}{\lambda}^{a_4})
\end{eqnarray*}
and
$$\tr({\lambda}^{a_1}{\lambda}^{a_2}{\lambda}^{a_4}{\lambda}^{a_3})\,,$$
respectively.

\pagebreak[3]

The eqs.~(\ref{A5gen}) to~(\ref{v5}), together with the results of the
appendix~\ref{appC}, constitute a well defined set of inputs to a
computer aided calculation. The computation of these Lorentz scalars,
as well as the ones associated with the four gluon amplitude, was
performed using the \emph{Maple} version of the computer algebra
package HIP~\cite{hsieh:1992ti}. Although the result contains a large
number of terms (the full expression, including the cyclic
permutations, contains $562$, $1272$ and $8992$ terms respectively for
the powers $0$, $2$ and $3$ of $\alp$), we were able to verify that
eq.~(\ref{A5gen}) reproduces exactly the expression given in
eq.~(\ref{amplitude}). We have also explicitly verified the gauge
invariance of the final amplitude
\be\label{A5f} {\mathcal A}^{(5)}=
i\,(2\pi)^{10}\,\delta^{10}(k_1+k_2+k_3+k_4+k_5)
{\sumpp}\left(\lambda^{a_1}\lambda^{a_2}\lambda^{a_3}\lambda^{a_4}
\lambda^{a_5}\right){A}(1,2,3,4,5)\,.
\ee

\end{document}